%% file: book_arxiv.tex
\documentclass[singlecolor,letterpaper,oneside,openany,12pt]{book}
\usepackage[margin=1.25in]{geometry}
\usepackage{lastpage}
\usepackage[DIV=14,BCOR=2mm,headinclude=true,footinclude=false]{typearea}
\usepackage{fancyhdr}
\fancypagestyle{plain}{\fancyhf{}} 
\usepackage{graphicx}
\usepackage{fixltx2e,fix-cm}
\usepackage[english]{babel}
\usepackage{epigrafica}
\usepackage{epigraph}
\usepackage[LGR,OT1]{fontenc}
\usepackage{lmodern}
\usepackage{makeidx}
\usepackage{multicol}
\usepackage[square, numbers, comma, sort&compress, sectionbib]{natbib}
\usepackage{chapterbib}
\usepackage{subcaption}
\usepackage{color}
\usepackage[backref=section, breaklinks, plainpages=false, colorlinks=true, anchorcolor=cyan, linkcolor=cyan, citecolor=cyan, filecolor=magenta, urlcolor=magenta, linktoc=all]{hyperref}
\usepackage{xspace}
\usepackage[skins]{tcolorbox}
\usepackage{rotating}
\usepackage{amsmath,amstext,amssymb}
\usepackage{enumitem}
\usepackage[mathscr]{eucal}
\usepackage{aas_macros}
\usepackage{bm}
\usepackage{changepage}

\usepackage{algorithm,algcompatible}

\title{The Nanohertz Gravitational Wave Astronomer}
\author{Stephen~R.~Taylor}

\begin{document}

\maketitle
\cleardoublepage
\tableofcontents
\include{Preface}
\include{Author}
\include{01}

\include{02}
\include{03}
\include{04}

\include{05}
\include{06}
\include{07}
\include{08}

\end{document}

%% file: Preface.tex
\chapter*{Preface}

Gravitational waves are a radically new way to peer into the darkest depths of the cosmos. Almost a century passed from their first prediction by Albert Einstein (as a consequence of his dynamic, warping space-time description of gravity) until their direction detection by the LIGO experiment. This was a century filled with theoretical and experimental leaps, culminating in the measurement of two black holes inspiraling and merging, releasing huge quantities of energy in the form of gravitational waves. 
	
However, there were signs along the way to this first detection. Pulsars are rapidly rotating neutron stars that emit radiation along their magnetic field axes, which may be askew from their rotation axis. This creates a lighthouse effect when the radiation is swept into our line of sight. Through tireless observations of the radio pulse arrival times, we are able to construct detailed models of the pulsar's rotation, binary dynamics, and interstellar environment. The first hint of gravitational waves came from the measured orbital decay of a binary star system that contained a pulsar. The decay was in extraordinary agreement with predictions based on gravitational-wave emission. 
	
In fact, pulsars can be used to make direct detections of gravitational waves using a similar principle as LIGO. When a gravitational wave passes between a pulsar and the Earth, it stretches and squeezes the intermediate space-time, leading to deviations of the measured pulse arrival times away from model expectations. Combining the data from many Galactic pulsars can corroborate such a signal, and enhance its detection significance. This technique is known as a Pulsar Timing Array (PTA). PTAs in North America, Europe, and Australia have been active for the last couple of decades, monitoring almost one hundred ultra-stable pulsars, with the goal of measuring gravitational waves entering the Galaxy from inspiraling supermassive black-hole binary systems at cosmological distances. These black-hole systems are the most massive compact objects in the Universe, having masses almost a billions times as big as our Sun, and typically lurk in the hearts of massive galaxies. They only form binary systems when their host galaxies collide together.
	
In this book, I provide an overview of PTAs as a precision gravitational-wave detection instrument, then review the types of signal and noise processes that we encounter in typical pulsar data analysis. I take a pragmatic approach, illustrating how our searches are performed in real life, and where possible directing the reader to codes or techniques that they can explore for themselves. The goal of this book is to provide theoretical background and practical recipes for data exploration that allow the reader to join in the exciting hunt for very low frequency gravitational waves.

I would not have been able to write this book without the continued collaboration with many excellent colleagues in NANOGrav and the International Pulsar Timing Array. Some have directly assisted in reviewing early chapter drafts, and I am particularly grateful to Dr. Joseph Romano, Dr. Xavier Siemens, and Dr. Michele Vallisneri for this reason. More practically, I would not have had the time to write this book without the overwhelming patience and support from my incredible wife, Erika, who tolerated many Saturday and Sunday afternoons of me shut away while I prepared this. Our very cute cat Olive also kept me company during many writing sessions. All my family back in Northern Ireland continue to be a source of strength and encouragement for me.   

\vspace{30pt}
\noindent Stephen R. Taylor

\noindent \textit{Nashville, Tennessee}

\noindent \textit{May 2021}

%% file: Author.tex
\chapter*{About the Author}

\begin{center}
\includegraphics[width=0.5\textwidth]{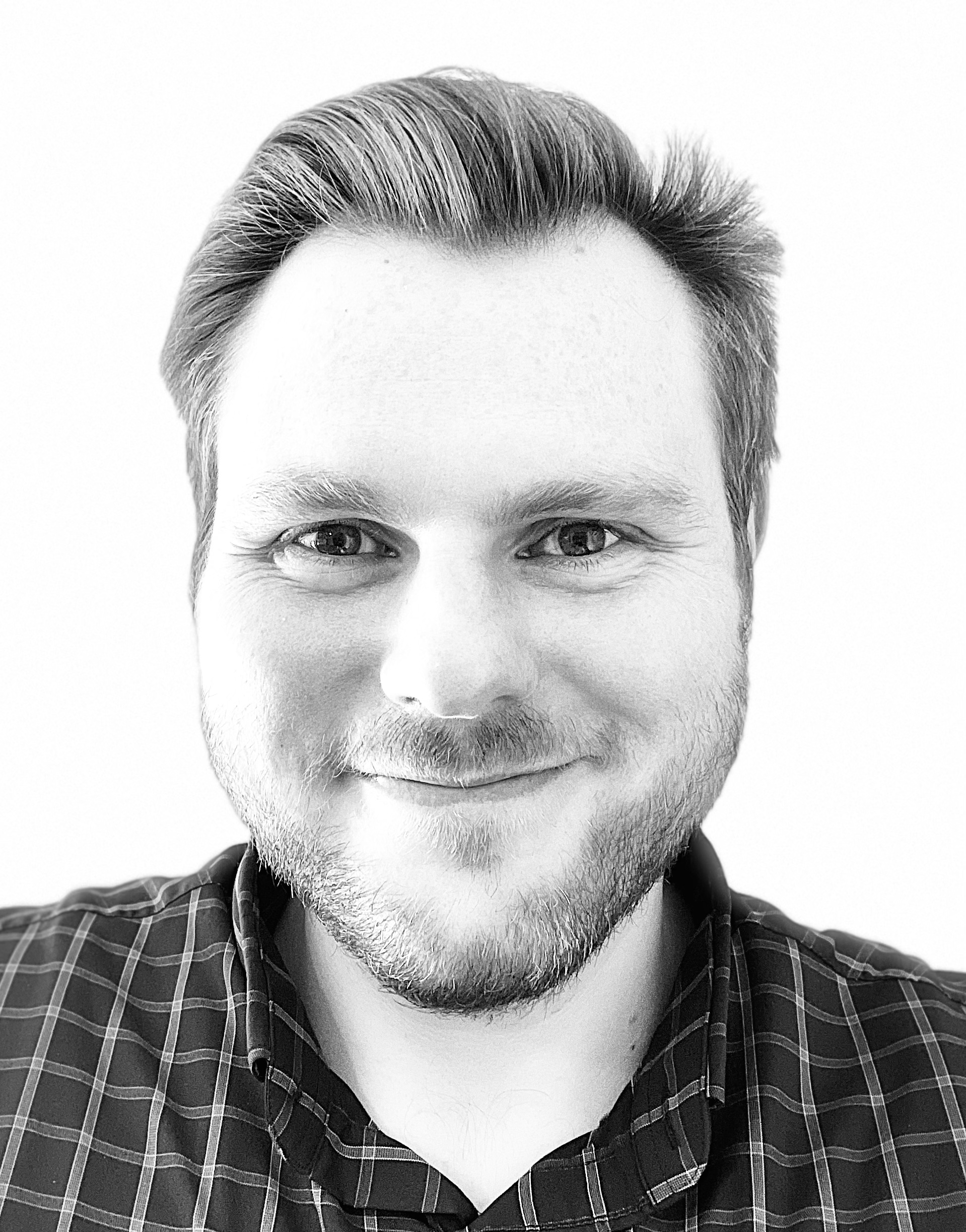}
\end{center}

\noindent Stephen R. Taylor is an Assistant Professor of Physics \& Astronomy at Vanderbilt University in Nashville, Tennessee. Born and raised in Lisburn, Northern Ireland, he went on to read Physics at Jesus College, Oxford from $2006-2010$, before switching to the ``other place'' for his PhD from the Institute of Astronomy at the University of Cambridge in $2014$. His positions have included a NASA Postdoctoral Fellowship at NASA's Jet Propulsion Laboratory, and a NANOGrav Senior Postdoctoral Fellowship at the California Institute of Technology in Pasadena, California. He currently lives in Nashville with his wife Erika and cat Olive.

%% file: 01.tex
\chapter{A Window Onto The Warped Universe}

\textit{Homo Sapiens}' split from \textit{Neanderthals} and other human species occurred roughly $500,000$ years ago, while the cognitive revolution that gifted us a fictive language took place $70,000$ years ago. It has been argued that this fictive language allowed us to develop abstract modes of thinking and planning that set us apart from other species. For most of the history of our species on this planet, the practice of \textit{``Astronomy''} has involved staring at the night sky to divine meaning. Indeed, when humanity first watched the twinkling fires in the sky, they envisioned a rich canvas on which their myths and collective stories took place. The heavens were the realm of the gods, and seeing patterns in the positions of stars may have been humanity's first attempt to make sense out of a complex Universe. The human eye was the most sensitive astronomical instrument for the vast majority of our time on Earth. Yet the history of Astronomy, much like the history of the human species, is one of widening panoramas. 

The last $500$ years has seen an explosion in creativity, knowledge, technology, art, and humanist values. Specifically to astronomy, the late 16th and early 17th centuries brought us the humble optical telescope, with which Galileo Galilei made precision observations of our Solar System that included the rings of Saturn and the four largest moons of Jupiter. His simple observation of those moons orbiting Jupiter caused a revolution, seeming to contradict Aristotelian cosmology. Yet reason and knowledge prevailed, eventually compelling society to change its worldview based on observations. 

Progress accelerates even more rapidly in the 19th century, when scientists learn that electromagnetic waves can have wavelengths greater and smaller than what the human eye can perceive. In fact, humans are innately blind to the vast majority of the electromagnetic spectrum! William Herschel, discoverer of Uranus, was the first to probe the spectrum beyond visible light in 1800. Through his investigation of the wavelength distribution of stellar spectra, Herschel discovered infrared radiation. Only one year after this, Johann Ritter discovered ultraviolet radiation, followed in 1886 by Heinrich Hertz's discovery of radio waves, then microwaves. In 1895, Wilhelm Roentgen revolutionizes diagnostic medicine with his discovery of X-rays that can travel through the human body. Finally, in 1900 Paul Villard finds a new type of radiation in the radioactive emission from radium, which is later ascribed to be a new, even shorter wavelength form of electromagnetic radiation called gamma-rays. See how the rate of progress accelerates, having taken us hundreds of thousands of years to find infrared radiation, yet less than a century to fill in the rest of the spectrum. 

These other parts of the electromagnetic-wave landscape were used throughout the 20th century to probe hitherto unexplored vistas of the cosmos. Infrared radiation has longer wavelengths than visible light and can pass through regions of dust and gas with less absorption, enabling the study of the origin of galaxies, stars, and planets.  
Ultraviolet radiation allows us to study the properties of young stars, since they emit most of their radiation at these wavelengths. Radio waves can easily propagate through the Earth's atmosphere, allowing observations even on a cloudy day. The impact of radio observations on astronomy is too great to chronicle here, but it has a particularly special relevance to this book, which will be explored later. Most people have microwave emitters in their homes to heat up food, yet the same type of radiation bathes the cosmos in its influence as a record of when the Universe was only $380,000$ years old-- this is the Cosmic Microwave Background that has revolutionized precision cosmology. X-ray astronomy has allowed us to peer into highly energetic environments like supernovae and the accretion flows that surround neutron stars or black holes. Finally, the class of objects known as gamma-ray bursts are the incredibly energetic and explosive result of the implosion of high mass stars and the collision of neutron stars. The latter is a literal treasure trove; double neutron star mergers are likely one of the environments that forged all the heavy elements (like gold, platinum, and silver) found on Earth. These collisions are the result of the inexorable inspiral of binary orbits due to the energy loss to gravitational wave emission. 

Just as electromagnetic waves are oscillations in the electromagnetic field caused by charge acceleration, gravitational waves are oscillating solutions to the space-time metric of Einstein's geometric theory of gravity in the presence of accelerating massive objects. They were predicted almost as soon as Einstein had developed his General Theory of Relativity (GR) in 1915, yet it was a full century before theory and technology could advance enough to detect them. In that century, Einstein would denounce them as coordinate artifacts, physicists would ignore them as mathematical curiosities that could not be practically measured, some pioneers would claim detections that turned out to be false flags, the first evidence of their existence would be seen in the orbital decay of binary pulsars, and a new millennium would arrive, all before the detection of two colliding black holes in 2015. 

But what are these specters of spacetime known as gravitational waves? We can not ``see'' them, nor perceive them in the slightest with any human sense. They travel at the speed of light, deforming spacetime as they propagate, and interacting very weakly with matter along the way. Their presence is exceptionally difficult to infer. In fact, the LIGO instrument that detected them first needed to be able to measure length deviations equivalent to the width of a human hair over the distance between Earth and Alpha Centauri, about $3$ lightyears. That's a length deviation of about 1 part in $10^{21}$; an almost unfathomably miniscule departure from perfection, and yet a departure that enables a fundamentally new form of astronomy.

Measuring gravitational waves (GWs) allows us to study systems from which no light is emitted at all. Pairs of black holes with no gas surrounding them will be completely invisible to conventional electromagnetic astronomy, yet will resound loudly in GWs. Black holes are extreme solutions to Einstein's theory, and are the crucible for successor theories to be tested that may link GR to a quantized description of gravity. Mining catalogs of GW signals will unearth clues about star formation, the progenitor environments of black holes and neutron stars, and even allow us to measure the rate of expansion of the Universe. As I write, Earth-based GW detectors like LIGO and Virgo have detected more than $50$ signals from inspiraling pairs of black holes and neutron stars. Such detectors are limited in sensitivity to GW frequencies of $\sim 10\,\,\mathrm{Hz}\leq f\leq 1\,\,\mathrm{kHz}$. But just as the electromagnetic landscape was ripe for discovery beyond the visible spectrum, so too is the GW landscape, where the most titanic black holes in the Universe lie in wait at the heart of massive galaxies for Galaxy-scale GW detectors to find them. Let us take a look at how precision timing of rapidly spinning neutron stars from all across the Galaxy will chart this ``undiscovered country'' of the low-frequency warped Universe.

%% file: 02.tex
\chapter[Gravity \& Gravitational Waves]{Gravity \& Gravitational Waves}\label{chap:gravity_and_gws}
\epigraph{\textit{``There's that word again. Heavy. Why are things so heavy in the future? Is there a problem with the Earth's gravitational pull?''}}{Dr. Emmett Brown, ``Back To The Future''}

\section{Gravity Before And After Einstein}

Gravity is the dominant dynamical influence throughout the Universe. Yet pick a ball off the ground, and the modest mechanical pull of your muscles has beaten the combined gravity of an entire planet.  

\subsection{Standing On The Shoulders Of Giants}

The Renaissance saw humanity climbing slowly out of the darkness and ignorance of the middle ages, and shedding their worshipful attitude to the knowledge of their ancient forebears. To question the cosmological wisdom of Aristole and Ptolemy was stupid and possibly even heretical! Nevertheless, in the 16th century the Polish astronomer Nicholaus Copernicus proposed a new heliocentric model of the Universe. This was later put to the test using the observations of Denmark's Tycho Brahe, who compiled high-accuracy stellar catalogues over more than 20 years of observations at his observatory, Uraniborg. Johannes Kepler, who had been Brahe's assistant, used these data to derive his famous laws of planetary motion. In his 1609 book {\it Astronomia nova}, Kepler presented two of these laws based on precision observations of the orbit of Mars, and in so doing ushered in the revolutionary new heliocentric model of the solar system with elliptical, rather than circular, planetary orbits.

Arguably the greatest scientific breakthrough of this age came in 1687, when Sir Isaac Newton published the {\it Principia} \cite{newton1850newton}. Ironically (given that the author of this book is writing it during the COVID-19 pandemic) Newton first glimpsed the concept of a universal law of gravitation $20$ years prior while under self-quarantine at his family's farm during a bubonic plague outbreak. Although it had been separately appreciated by contemporaries such as Hooke\footnote{In fact, the title of this sub-section is a phrase used by Newton in reference to Hooke, which some have interpreted as a thinly-veiled jab at the latter's diminutive height.}, Wren, and Halley, Newton's demonstrations of the universal inverse-square law of gravitation were remarkable not only for their ability to derive Kepler's laws, but most crucially for their predictive accuracy of planetary motion. Simply put, this law states that the mutual attraction between two {\it bodies} is proportional to the product of their masses, and the inverse square of their separation. When applied to the motions of solar system bodies it was extraordinarily successful, and remains so today-- it guided Apollo astronauts to the Moon and back. The mark of a successful theory is the ability to explain new observations where previous theories fall short; in 1821 the French astronomer Bouvard used Newton's theory to publish prediction tables of Uranus' position that were later found to deviate from observations. This discrepancy motivated Bouvard to predict a new eighth planet in the solar system that was perturbing Uranus' motion through gravitational interactions. Independent efforts were made by astronomers Le Verrier (of France) and Adams (of England) to use celestial mechanics to compute the properties of such an additional body. Le Verrier finished first, sending his predictions to Galle of Berlin Observatory, who, along with d'Arrest, discovered Neptune on September 23rd 1846.

Neptune's discovery was a supreme validation of Newtonian celestial mechanics. But the underlying theory of gravity suffered two fundamental flaws: it did not propose a mediator to transmit the gravitational influence, and it assumed the influence was instantaneous. Furthermore, Le Verrier, who had wrestled with complex celestial mechanics calculations for months, had by 1859 noticed a peculiar excess precession of Mercury's orbit around the Sun. Various mechanisms to explain this were proposed, amongst which was an idea drawn from the success of Neptune; perhaps there was a new innermost planet (named {\it Vulcan}) that acted as a perturber on Mercury's orbit. A concerted effort was made to find this alleged new planet, and there were many phony claims by amateur astronomers that were taken seriously by Le Verrier. But all attempts to verify sightings ultimately ended in failure. This problem would require a new way of thinking entirely, and a complete overhaul of the more than 200 years of Newtonian gravity.

\subsection{The Happiest Thought}

Newton's universal law of gravitation treated gravity as a force like any other within his laws of mechanics. Albert Einstein had already generalized Newtonian mechanics to arbitrary speeds $v\leq c$ in his 1905 breakthrough in special relativity, and went about doing the same for gravity. His eureka moment came, as it usually did for Einstein, in the form of a \textit{Gedankenexperiment} (a thought experiment). Imagine an unfortunate person falling from a roof; ignoring wind resistance and other influences, this person is temporarily weightless and can not \textit{feel} gravity. A similar restatement of this concept is that if we were to place a sleeping person inside a rocket without windows or other outside indicators, and accelerated the rocket upwards at $9.81\,\,\mathrm{ms}^{-2}$, then upon waking this person would feel a downwards pull so akin to Earth-surface gravity that they could not tell the difference. These thought experiments are consequences of the \textit{Equivalence Principle}, which states that experiments in a sufficiently small freely-falling laboratory, and carried out over a sufficiently short time, will give results that are identical to the same experiments carried out in empty space. 

In 1907, Hermann Minkowski developed an elegant geometric formalism for special relativity that recast the theory in a four-dimensional unification of space and time, unsurprisingly known as ``spacetime''. Even today we refer to spatially flat four-dimensional spacetime as \textit{Minkowski spacetime}. Accelerating particles are represented by curved paths through spacetime, and with the thought-experiment equivalence principle breakthrough that allowed Einstein to see acceleration as (locally) equivalent to gravitational influence, he began to build connections between geometric spacetime curvature and the manifestation of gravity. It was a long road with many blind alleys that required Einstein to learn the relevant mathematics from his friend and colleague Marcel Grossman. But finally, in late 1915 Einstein published his general theory of relativity (GR)\footnote{Interesting discussions in November/December 1915 between Einstein and the mathematician David Hilbert have raised some questions of priority, but Einstein undeniably developed the breakthrough vision of gravity as geometry.}. Formulated within Riemannian geometry, Einstein's new theory described how spacetime geoemtry could be represented by a metric tensor, and how energy and momentum leads to the deformation and curvature of spacetime, causing bodies to follow curved free-fall paths known as geodesics. In essence, gravity is not a ``pull'' caused by a force-field; it is the by-product of bodies travelling along free-fall paths in curved spacetime. The shortest possible path between two points remains a straight line (of sorts), but for the Earth or other planets, this straight-as-possible path within the curved geometry created by the Sun is one that carries it around in an orbit. In searching for a pithy summary, it's difficult to beat Wheeler: ``{\it Space-time tells matter how to move; matter tells space-time how to curve}''.

The phenomenal success of this new paradigm, connecting relativistic electrodynamics with gravitation, overturning notions of static space and time, and demonstrating gravity to emerge from the curvature of the fabric of spacetime itself, can not be understated. The first real demonstration of its power came through being able to predict the excess perihelion precession of Mercury's orbit that had been observed by Le Verrier and others; in GR this appears as a conservative effect due to the non-closure of orbits in the theory. After computing the correct result, Einstein was stunned and unable to work for days afterward. But this was a post-hoc correction of a known observational curiosity. The real test of any scientific theory is being able to make testable, falsifiable predictions. 

Einstein next set his sights on the deflection of starlight passing close to the Sun. Newtonian gravity does predict that such deflection will happen at some level, but precision observation of such deflection provided a perfect testbed for the new curved spacetime description of gravity. Before he had fully fleshed out his theory, Einstein made a prediction in 1913 that he encouraged astronomers to test. Before they could do so, World War I broke out, making the unrestricted travel of anyone to test a scientific theory a somewhat dangerous prospect. There is a serendipity to this though-- his calculation was wrong! It would only be using the final form of his field equations that Einstein made a testable prediction for the deflection of starlight by the Sun that corresponds to twice the Newtonian value. This was later confirmed in 1919 by the renowned astronomer Sir Arthur Eddington (already a proponent of Einstein's work) who made an expedition to the isle of Principe for the total solar eclipse on May 29th of that year. The eclipse made it possible to view stars whose light grazes close to the surface of the Sun, which would otherwise have been completely obscured. This result made the front page of the New York Times, catapulting Einstein from a renowned figure in academia to a celebrity scientist overnight.

Since those early days, GR has passed a huge number of precision tests \citep{will2006confrontation}, and sits in the pantheon of scientific theories as one of the twin pillars of 20th century physics, alongside quantum theory. In 1916, Einstein proposed three tests of his theory, now dubbed the ``classical tests''. Perihelion precession and light deflection were the first two of these, having been validated in very short order after the development of the theory. Gravitational redshift of light (the third and final classical test) would take several more decades, eventually being validated in 1960 through the \textit{Pound--Rebka Experiment} \cite{pound1960apparent}. The experiment confirmed that light should be blueshifted as it propagates toward the source of spacetime curvature (Earth), and by contrast redshifted when propagating away from it.

Beyond merely deflection, GR predicts that light will undergo a time delay as it propagates through the curved spacetime of a massive body. By performing Doppler tracking of the {\it Cassini} spacecraft en route to Saturn, this \textit{Shapiro delay} effect \citep{shapiro1964fourth} was experimentally tested, with the data agreeing to within $10^{-3}$ percent of the GR prediction \cite{bertotti2003test}. The 2004 launch of the {\it Gravity Probe B} satellite afforded independent verifications of the GR effects of \textit{geodetic} and \textit{Lense-Thirring precession} of a gyroscope's axis of rotation in the presence of the rotating Earth's curved space-time, exhibiting agreements to within $0.28\%$ and $19\%$, respectively \cite{everitt2011gravity}. Furthermore, there is a whiff of irony in the fact that Newtonian gravity was used to compute the trajectory of Apollo spacecraft to the Moon, yet one of the greatest legacies of the Apollo project is the positioning of retro-reflectors on the lunar surface that enable laser ranging tests of the \textit{Nordtvedt effect} (an effect which, if observed, would indicate violation of the strong equivalence principle), \textit{geodetic precession}, etc. \cite{williams2004progress}. A more exotic and extraordinary laboratory for testing the strong equivalence principle is through the pulsar hierarchical triple system J0337+1715 \citep{2014Natur.505..520R}, composed of an inner neutron star and white dwarf binary, and an outer white dwarf. The outer white dwarf's gravitational interaction with the inner binary causes the pulsar and its white dwarf companion to accelerate, but their response to this (despite their differences in compactness) differs by no more than $2.6\times10^{-6}$ \citep{2018Natur.559...73A}.  

Many more experiments have since validated GR. Those most relevant to the subject of this volume are through the indirect and direct detection of gravitational waves from systems of compact objects. These will be discussed in more detail in the next section. 

\section{Gravitational Waves}
\subsection{A Brief History of Doubt}

Gravitational waves (GWs) have had a colorful history. They were first mentioned by Henri Poincare in a $1905$ article that summarized his theory of relativity, and which proposed gravity being transmitted by an \textit{onde gravifique} (gravitational wave) \citep{poincare1906dynamique}. In Einstein's $1915$ general theory of relativity, an accelerating body can source ripples of gravitational influence via deformations in dynamic space-time \citep{einstein1916approximative,einstein1918gravitationswellen}. Whereas electromagnetic waves are sourced at leading order by second time derivatives of dipole moments in a charge distribution, mass has no ``negative charge'' and thus GWs are instead sourced at leading order by second time derivatives of the quadrupole moment of a mass distribution. 

Whether they even existed at all or were merely mathematical curiosities was the subject of much early speculation. By $1922$, Sir Arthur Eddington (who had conducted the famous solar eclipse experiment that tested Einstein's light deflection prediction), found that several of the wave solutions that Einstein had found could propagate at any speed, and were thus coordinate artifacts. This led to him quipping that GWs propagated ``\textit{at the speed of thought}'' \citep{eddington1922propagation}. But the most famous challenge to the theory of GWs came from Einstein himself, who, in collaboration with Nathan Rosen, wrote an article in $1936$ that concluded the non-existence of GWs. This chapter in the history of GWs is infamous, because when Einstein sent his article along to the journal \textit{Physical Review}, he became angry and indignant at receiving referee comments, stating in response to editor John Tate that ``\textit{we had sent you our manuscript for publication and had not authorized you to show
it to specialists before it is printed}'' \citep{einstein2005einstein}. Einstein withdrew his article. After Rosen departed to the Soviet Union for a position, Einstein's new assistant Leopold Infeld befriended the article's reviewer, Howard Robertson, and managed to convince Einstein of its erroneous conclusions arising from the chosen cylindrical coordinate system. The article was heavily re-written, the conclusions flipped, and eventually published in \textit{The Journal of the Franklin Institute} \citep{einstein1937gravitational}. 

The thorny issue of coordinate systems was solved two decades later by Felix Pirani in $1956$, who recast the problem in terms of the Riemann curvature tensor to show that a GW would move particles back and forth as it passed by \citep{pirani1956physical}. This brought clarity to the key question in gravity at the time: whether GWs could carry energy. At the first ``GR'' conference held at the University of North Carolina at Chapel Hill in $1957$, Richard Feynman developed a thought-experiment known as ``\textit{the sticky bead argument}''. This stated that if a GW passed a rod with beads on it, and moved the beads, then the motion of the beads on the rod would generate friction and heat. The passing GWs had thus performed work. Hermann Bondi, taking this argument and Pirani's work on the Riemann curvature tensor, fleshed out the ideas into a formal theoretical argument for the reality of GWs \citep{bondi1957plane}.  

\begin{figure}
    \centering
	\includegraphics[width=0.6\columnwidth]{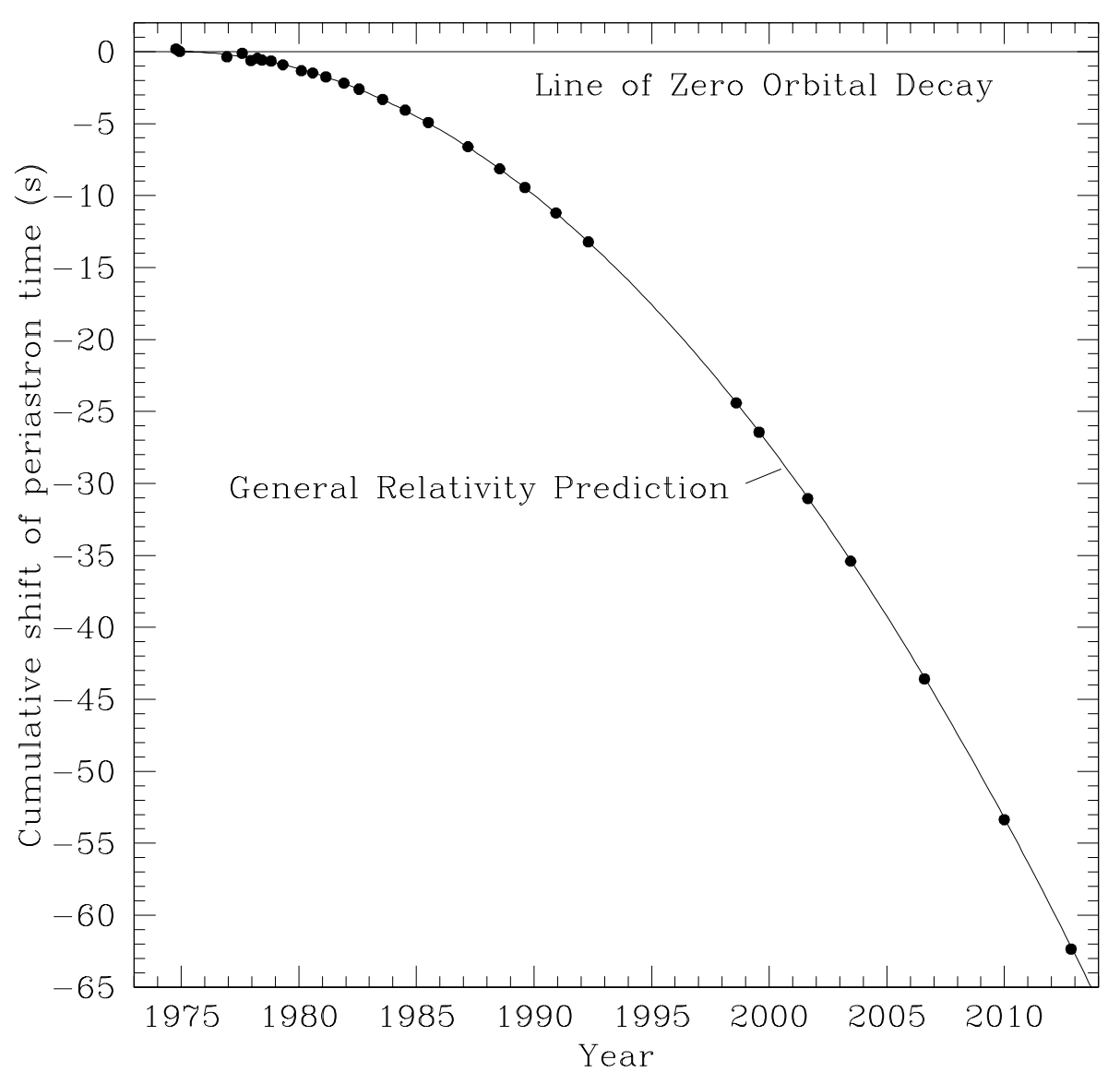}
    \caption{Data and GR prediction of the cumulative shift of periastron time of pulsar B$1913$$+$$16$, showing extraordinary agreement with orbital decay through GW emission. Figure reproduced with permission from Ref.~\citep{2016ApJ...829...55W}.}
    \label{fig:binary_pulsar}
\end{figure}

Bolstered by the theoretical existence of GWs, Joseph Weber established the field of experimental GW astronomy by developing resonant mass instruments (also known as \textit{Weber bars}) for their detection. By $1969/1970$ he made claims that signals were regularly being detected from the center of the Milky Way \citep{weber1969evidence,weber1970anisotropy}, however the frequency and amplitude of these alleged signals were problematic on theoretical grounds, and other groups failed to replicate his observations with their own Weber bars. By the late 1960s and early 1970s, astronomy was swept up in the excitement of the discovery of \textit{pulsars}. These rapidly rotating neutron stars acted like cosmic lighthouses as they swept beams of radiation around, allowing astronomers to record and predict radio-pulse arrival times with extraordinary precision. In $1974$, Russell Hulse and Joseph Taylor discovered B$1913$$+$$16$, the first ever pulsar in a binary system \citep{1975ApJ...195L..51H}. Over the next decade, the observed timing characteristics of this pulsar reflected its motion alongside the companion, allowing the binary system itself to be profiled. As shown in Fig.~\ref{fig:binary_pulsar}, the system exhibits a shift in the time to periastron corresponding to a decay in the orbital period, which matches to within $0.16\%$ the loss in energy and angular-momentum predicted by GR as a result of GW emission \citep{1982ApJ...253..908T}. The discovery of this pulsar binary system garnered the $1993$ Nobel Prize in Physics. Further precision binary-pulsar constraints on GW emission have been made possible by the double-pulsar binary system J$0737$$-$$3039$ (which constrains the GR prediction to within $0.05\%$ \citep{2008ARA&A..46..541K}), and the pulsar--white-dwarf system J$0348$$+$$0432$ \citep{2013Sci...340..448A}. 

The early 1970s also saw the origins of research into detecting GWs through laser interferometry, a concept that posed significant technological challenges. The foundations of GW interferometry were laid by Forward \citep{moss1971photon}, Weiss \citep{weiss1972electronically}, and others, which motivated the construction of several highly successful prototype instruments. These early detectors had arm lengths from $\sim10$m to $\sim100$m, and reached a strain sensitivity of $\sim 10^{-18}$ for millisecond burst signals, illustrating that the technology and techniques were mature enough to warrant the construction of much longer baseline instruments. Over decades of research, this concept bloomed into the Laser Interferometer Gravitational-wave Observatory (LIGO) in the USA, which became an officially-funded National Science Foundation project in 1994. LIGO is a project that includes two instruments: LIGO-Hanford (in Washington), and LIGO-Livingston (in Louisiana). Other instruments have been built since, including the Virgo detector in Italy, and now the KAGRA instrument in Japan. There are plans to continue to expand this global network of laser interferometers in order to drastically improve the sky localization of detected signals, and thus narrow search regions for electromagnetic follow-up. 

At this stage, history begins to impinge on the modern day, and I'll defer further discussions of LIGO until after introducing the theoretical framework of GWs. The last piece of history that I'll discuss requires us to jump forward to 2014, when the BICEP2 collaboration announced an incredibly exciting result \citep{2014PhRvL.112x1101B}. It appeared that a polarization pattern had been found in the cosmic microwave background that could be consistent with so-called B-modes (or curl modes). These modes could have been imprinted from quantum fluctuations in the gravitational field of the early Universe that had been inflated to cosmological scales. While the raw data was excellent, unfortunately such primordial GWs were not the only potential source of B-modes; galactic dust, if not properly modeled, could impart this polarization pattern as well. Subsequent analysis now suggests that dust is the main culprit, and primordial GW signatures are much smaller than initially thought \citep[e.g.,][]{2018PhRvL.121v1301B}. However, the hunt is still on for this elusive imprint from the dawn of time through later generations of BICEP and other detectors.

\subsection{Waves From Geometry} \label{sec:GWprimer}

Let's dig into some illuminating mathematics to grasp what these gravitational waves (GWs) really are. General Relativity is enshrined within the \textit{Einstein Field Equations},
\begin{equation}
    G_{\mu\nu}:= R_{\mu\nu} - \frac{1}{2}Rg_{\mu\nu} = 8\pi T_{\mu\nu},
\end{equation}
where $G_{\mu\nu}$ is the Einstein tensor, $R_{\mu\nu}:=R^\lambda_{\mu\lambda\nu}$ is the Ricci tensor derived from the Riemann curvature tensor, $R = g^{\mu\nu}R_{\mu\nu}$ is the Ricci scalar, $g_{\mu\nu}$ is the metric, $T_{\mu\nu}$ is the stress-energy tensor, and we assume natural units such that $G=c=1$. 

We consider the linearized treatment of GWs that Einstein originally studied \citep{einstein1916approximative,einstein1918gravitationswellen}. In the following, Greek indices refer to $4$-D spacetime coordinates, while Roman indices refer to $3$-D spatial coordinates. The derivation presented here closely follows the treatment in Maggiore~\citep{maggiore2008gravitational}. We start with spatially-flat $4$-D Minkowski spacetime, with $\eta_{\mu\nu} = \mathrm{diag}(-1,1,1,1)$, which upon adding a perturbation results in
\begin{equation}
    g_{\mu\nu} = \eta_{\mu\nu} + h_{\mu\nu},
\end{equation}
where $h_{\mu\nu}$ is a tensor metric perturbation quantity that we treat only in the weak field case, $|h_{\mu\nu}|\ll 1$. This metric perturbation is assumed to be so small that the usual index raising and lowering operations can be performed with the Minkowski metric. Expanding the Einstein tensor to linear order in $h_{\mu\nu}$ in this perturbed spacetime gives,
\begin{equation}
    G_{\mu\nu} = \frac{1}{2}\left(\partial_{\mu}\partial^{\alpha}h_{\alpha\nu} + \partial_{\nu}\partial^{\alpha}h_{\alpha\mu} - \partial_{\mu}\partial_{\nu}h - \Box h_{\mu\nu} + \eta_{\mu\nu}\Box h - \eta_{\mu\nu}\partial^{\alpha}\partial^{\beta}h_{\alpha\beta}\right),
\end{equation}
where $\partial_\mu\equiv \partial / \partial x^{\mu}$; $\partial^{\mu}\equiv \eta^{\mu\nu}\partial_{\nu}$; $h = \eta^{\mu\nu}h_{\mu\nu}$ is the trace of $h_{\mu\nu}$; and $\Box = \eta^{\mu\nu}\partial_{\mu}\partial_{\nu}$ is the flat space d'Alembertian operator. This is a bit cumbersome, so we change variables in an effort to tidy this up. We define the trace-reversed perturbation, $\bar{h}_{\mu\nu} = h_{\mu\nu} - n_{\mu\nu}h/2$, which also implies that $\bar{h} = \eta^{\mu\nu}\bar{h}_{\mu\nu}=h-2h=-h$, such that $h_{\mu\nu} = \bar{h}_{\mu\nu} - n_{\mu\nu}\bar{h}/2$. The Einstein tensor then becomes
\begin{equation}
    G_{\mu\nu} =     \frac{1}{2}\left(\partial_{\mu}\partial^{\alpha}\bar{h}_{\alpha\nu} + \partial_{\nu}\partial^{\alpha}\bar{h}_{\alpha\mu} - \Box\bar{h}_{\mu\nu} - \eta_{\mu\nu}\partial^{\alpha}\partial^{\beta}\bar{h}_{\alpha\beta}\right).
\end{equation}
Similar to electromagnetism, we can ditch spurious degrees of freedom from our equations by choosing an appropriate gauge. If we consider the coordinate transformation $x^{\alpha} \mapsto x^{\alpha} + \xi^{\alpha}$, the transformation of the metric perturbation to first order is $h_{\mu\nu}\mapsto h_{\mu\nu} - \left(\partial_{\mu}\xi_{\nu} + \partial_{\nu}\xi_{\mu}\right)$. Asserting $\partial_{\mu}\xi_{\nu}$ to be of the same order as $|h_{\mu\nu}|$, the transformed metric perturbation retains the condition $|h_{\alpha\beta}|\ll 1$. This symmetry allows us to choose the \textit{Lorenz gauge} (sometimes referred to as the \textit{De Donder gauge}, or \textit{Hilbert gauge}, or \textit{harmonic gauge}) where $\partial^{\nu}\bar{h}_{\mu\nu}=0$, such that the Einstein tensor reduces to the much more compact
\begin{equation}
    G_{\mu\nu} = -\frac{1}{2}\Box\bar{h}_{\mu\nu}.
\end{equation}
Note that in choosing the Lorenz gauge we have imposed $4$ conditions that reduce the number of degrees of freedom in the symmetric $4\times4$ $h_{\mu\nu}$ matrix from $10$ down to $6$. Finally, the field equations reduce to
\begin{equation} \label{eq:linearEinsteineq}
    \Box\bar{h}_{\mu\nu} = -16\pi T_{\mu\nu},
\end{equation}
which can be generally solved using the radiative Green's function:
\begin{equation}
    \bar{h}_{\mu\nu}(\vec{x},t) = 4\int d^4 x' \frac{T_{\mu\nu}(\vec{x}',t)}{|\vec{x}-\vec{x}'|}.
\end{equation}
Examining Eq.~\ref{eq:linearEinsteineq} far from the source gives $\Box\bar{h}_{\mu\nu} = 0$, whose solution is clearly wave-like with propagation speed $c$. 
We might think that all components of the metric perturbation are radiative. This is a gauge artefact \cite{eddington1922propagation}, where, in general, we can split the metric perturbation into (i) gauge degrees of freedom, (ii) physical, radiative degrees of freedom, and (iii) physical, non-radiative degrees of freedom. The Lorenz gauge is preserved under a coordinate transformation $x^{\mu} \mapsto x^{\mu} + \xi^{\mu}$, provided that $\Box\xi_\mu=0$, which also implies that $\Box\xi_{\mu\nu}\equiv\Box(\partial_\mu\xi_\nu + \partial_\nu\xi_\mu - \eta_{\mu\nu}\partial_\rho\xi_\rho)=0$. This means that the $6$ independent components of $\bar{h}_{\mu\nu}$ can have $\xi_{\mu\nu}$ subtracted, which depend on $4$ independent functions $\xi_\mu$, thereby distilling the GW information down to $2$ independent components of the metric perturbation. The conditions imposed on $\bar{h}_{\mu\nu}$ are such that $\bar{h}=0$ (therefore $h_{\mu\nu}=\bar{h}_{\mu\nu}$), and $h^{0i}=0$, which lead to the following properties of the metric perturbation that define the \textit{transverse-traceless} (TT) gauge:
\begin{equation}
    h^{0\mu}=0,\quad h^i_i=0,\quad \partial^jh_{ij}=0.
\end{equation}

Choosing a coordinate system such that we have a plane GW propagating in the $z$-direction in a vacuum, we can write the solution to $\Box \bar{h}_{\mu\nu}=0$ as
\begin{equation} \label{eq:hTT}
h^\mathrm{TT}_{\mu\nu}(t,z) =
 \begin{pmatrix}
  0 & 0 & 0 & 0 \\
  0 & h_{+} & h_{\times} & 0 \\
  0  & h_{\times}  & -h_{+} & 0  \\
  0 & 0 & 0 & 0
 \end{pmatrix}\cos\left[\omega(t-z)\right],
\end{equation}
where $h_{+}$ and $h_{\times}$ are the amplitudes of the two distinct polarizations of GWs permitted within general relativity, denoted as ``plus'' ($+$) and ``cross'' ($\times$) modes for how they tidally deform a circular ring of test masses in the plane perpendicular to the direction of propagation. 

\begin{figure}
    \centering
    \includegraphics[width=\columnwidth]{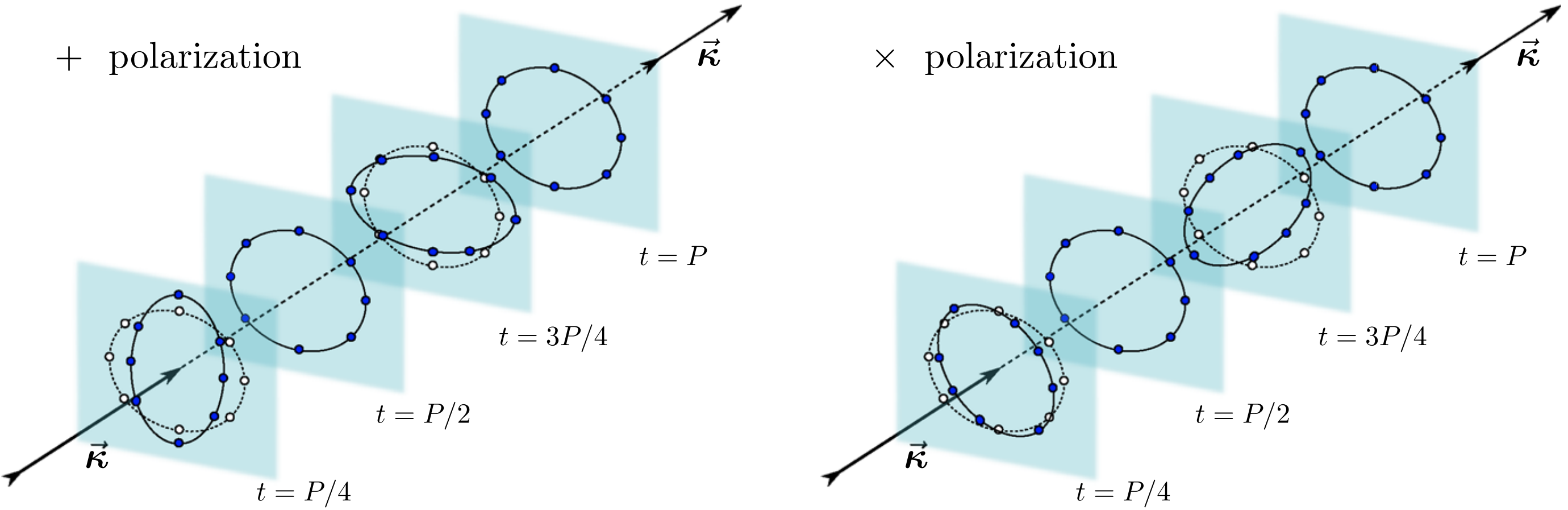}
    \caption{The periodic deformation of spacetime caused by a GW is illustrated for the two distinct polarizations permitted within GR. The influence of the GW is entirely in the plane perpendicular to the direction of propagation. Figure reproduced with permission from Ref.~\citep{2016LRR....19....2B}.}
    \label{fig:pluscross}
\end{figure}

The tidal deformation of spacetime caused by a GW as it propagates is at the heart of all ground-based (LIGO-Virgo-KAGRA), Galactic-scale (pulsar-timing arrays), and planned space-borne (LISA) detection efforts.  
Consider the \textit{coordinate} separation of two spatially-separated test masses as a GW sweeps by. The test masses fall along geodesics of the perturbed spacetime, such that in the weak-field regime, and to first order in amplitude of the wave, the coordinate separation of the test masses remains unchanged. However, the \textit{proper} separation between the test masses \textit{is} affected, and depends on the wave properties. The fractional change in the proper distance between two test masses separated by $\Delta x=L$ on the $x$-axis of a coordinate system due to the passage of a GW is given by $\delta L/L \simeq h_{xx}/2$, leading to a definition of the GW amplitude as the {\it strain}. For a periodic signal such that $h_{xx}(t,z=0) = h_+\cos\omega t$, we see that this proper distance separation oscillates according to $\delta\ddot{L} = -2\pi^2 f^2 L h_+\cos\omega t$. We can see from Eq.~\ref{eq:hTT} that the $+$-polarisation will lengthen distances along the $x$-axis while simultaneously contracting distances along the $y$-axis. The influence of the $h_{\times}$ mode is similar, but rotated by $\pi/4$ degrees counter-clockwise in the $xy$-plane. This tidal deformation in the plane perpendicular to the direction of propagation is illustrated for a full wave-cycle in Fig.~\ref{fig:pluscross}.

\subsubsection{The Quadrupole Formula}\label{chap:quadrupole_formula}

In the weak-field ($v\ll 1$) limit, far from a source, the leading order contribution to the solution of Eq.~\ref{eq:linearEinsteineq} is a function of the accelerating quadrupole moment of the source's mass distribution. Stress-energy conservation implies that the monopole makes no contribution due to conservation of the system's total energy, while the dipole moment makes no contribution due to conservation of momentum of the system's center-of-mass. Therefore the {\it quadrupole radiation formula} for the spatial components of the metric perturbation is \cite{einstein1918gravitationswellen},
\begin{equation} \label{eq:quadradform}
    h_{ij} = \frac{2}{r}\Lambda_{ij,kl}\frac{d^2Q_{kl}}{dt^2},
\end{equation}
where $r$ is the distance to the source, and $Q_{kl}$ is the {\it reduced} quadrupole moment of $\rho(t,\vec{x})$ (the source's mass density), which is defined as,
\begin{equation}
    Q_{ij} = \int d^3x\,\rho(t,\vec{x})\left(x_ix_j - \frac{1}{3}r^2\delta_{ij}\right),
\end{equation}
and $\Lambda_{ik,jl}$ is the {\it Lambda tensor} that projects the metric perturbation in the Lorenz gauge into the TT gauge (see Ref.~\citep{maggiore2008gravitational}). 

The most relevant source system for us is a binary composed of compact objects (COs), whether white dwarfs, neutron stars, or black holes.  
Let us consider that each CO (of mass $M$) orbits one another at a distance $R(t)$ from their common center of mass (assumed to be far enough apart that we can ignore any tidal disruption effects) with slowly varying angular velocity $\omega(t)$. The COs are assumed to be moving at non-relativistic speeds to simplify the calculation of the quadrupole moment tensor. Kepler's third law gives us $\omega^2 = M/4R^3$, such that the total energy of the system is $E = -M^2/4R$. The orbital geometry is defined such that it lies in the $x-y$ plane with the origin coinciding with the system's center of mass. Each CO's coordinates at $t=0$ are thus
\begin{equation}
    x_1 = -x_2 = R\cos\omega t,\quad y_1 = -y_2 = R\sin\omega t,\quad z_1 = z_2 = 0.
\end{equation}

Evaluating the second time-derivative of the reduced quadrupole mass-moment, and plugging into Eq.~\ref{eq:quadradform}, the radiative components of the GWs from this system are
\begin{equation} \label{eq:TTstrain}
    h_{\mu\nu} = -\frac{8MR^2\omega^2}{r}\begin{pmatrix}
      0 & 0 & 0 & 0 \\
      0 & \cos2\omega t & \sin2\omega t & 0 \\
      0 & \sin2\omega t & -\cos2\omega t & 0  \\
      0 &0 & 0 & 0
     \end{pmatrix},
\end{equation}
where the GW frequency is \textit{twice} the binary orbital frequency due to the quadrupolar nature of the emission. Note that the distance to the source is directly encoded in the amplitude of this leading-order radiative term.

Just like electromagnetic waves, or even waves on a string, the energy density in GWs is $\propto \omega^2 h^2$, where $h$ is the amplitude of the wave. The amplitude is already proportional to a second time derivative of the reduced quadrupole moment, where the the factor of $\omega$ implies an additional time derivative. Therefore, based on simple scaling arguments, the energy density in GWs should be proportional to a quadratic combination of the third time-derivative of the quadrupole mass moment. In a background spacetime that is approximately flat far from the source, the Issacson stress-energy tensor of a GW is
\begin{equation} \label{eq:stressenergyGW}
    T_{\mu\nu} = \frac{1}{32\pi}\langle\partial_{\mu}\hat{h}_{\alpha\beta}\partial_{\nu}\hat{h}^{\alpha\beta}\rangle,
\end{equation}
where $\langle\cdot\rangle$ denotes an average over several wave cycles. Evaluating the energy-flux elements of Eq.~\ref{eq:stressenergyGW} in the quadrupole approximation, and integrating over the sphere, the GW luminosity of the source system
\begin{equation} \label{eq:GWluminosity}
    L_{\rm GW} \equiv \frac{dE_{\rm GW}}{dt} =  \frac{1}{5}\left\langle\frac{d^3Q_{ij}}{dt^3}\frac{d^3Q^{ij}}{dt^3}\right\rangle,
\end{equation}
which, for a CO binary gives
\begin{equation}
    L_{\rm GW} = \frac{128}{5}M^2R^4\omega^6 = \frac{128}{5}4^{1/3}\left(\frac{\pi M}{P}\right)^{10/3}
\end{equation}

By equating the loss in the binary's orbital energy with GW emission, we can get a qualitative understanding of the frequency and strain evolution as the binary inspirals toward an eventual merger. The rate of change of the binary's orbital period is $dP/dt\propto P^{-5/3}$, such that the GW frequency evolution is $df_{\rm GW}/dt\propto f_{\rm GW}^{11/3}$. The strain amplitude is $h\propto R^2 f_{\rm GW}^2\propto f_{\rm GW}^{2/3}$. Hence the orbital evolution, as driven by GW emission, causes the strain amplitude to increase in tune with the frequency, with both said to be ``chirping'' as the binary progresses toward merger. 

\section{Stochastic Gravitational Wave Backgrounds}

Imagine yourself at a crowded party. All the guests are mingling and chatting to create a background hum of the usual social small-talk. You perk up your ears to attempt to hear whether your friend is stuck in an awkward conversation across the room, but alas, all you can hear is that damned background hum. Nevertheless, if we brought the lens of statistical inference to this party banter, there would be interesting information-- the distribution of laughter could tell us whether this is a party worth sticking around at, and by localizing regions of high laughter one may be able to zero-in on the life and soul of this get-together. Occasionally, someone who may have had a bit too much to drink may forget to regulate their volume, creating a distinct sentence that can be heard above the fray. 

Now imagine translating this scenario to GW signals. A cosmological population of systems emitting GWs of a similar frequency and comparable amplitude may not be able to be individually resolved by a detector. In such a scenario, the signals sum incoherently to produce a stochastic background of GWs (SGWB).  ``Stochastic'' here formally means that we treat this background as a random process that is only studiable in terms of its statistical properties. As a function of time, the SGWB may look like random fluctuations without any discernible information, but if we dig deeper and look at its spectral information then it begins to reveal its secrets. It may have more power at lower frequencies than at higher frequencies, indicating that it varies on longer timescales. It may have a sharp uptick at a few frequencies, potentially revealing a few of those loud ``voices'' at the party. Let us now see how we describe a SGWB, and how it manifests as signals in our detectors. Much of the material in the following subsections has been adapted from the excellent treatment in Romano \& Cornish \citep{2017LRR....20....2R}.

\subsection{The Energy Density Of A SGWB}

The energy content of the Universe today is dominated by three main components: dark energy makes up $\sim68\%$, dark matter makes up $\sim 27\%$, while baryonic matter (everything we encounter in everyday life) and everything else makes up $\lesssim 5\%$. The fractional energy density is usually defined in terms of closure density, $\rho_c=3H_0^2/8\pi$, which is the density of the Universe today that is required for flat spatial geometry, and where $H_0$ is Hubble's constant. For a SGWB, it makes most sense to consider the fractional energy density as a spectrum, in order to see how this energy density is distributed over frequencies that may or may not be accessible to our detectors. Hence, the energy-density spectrum in GWs is defined such that,
\begin{equation}
    \Omega_{\rm SGWB}(f) \equiv \frac{1}{\rho_c}\frac{d\rho}{d\ln f},
\end{equation}
where $\rho$ is the energy density in GWs. The gauge invariant energy density is given by evaluating the $00$ component of the stress-energy tensor, such that
\begin{equation}
    T^{00} = \rho = \frac{1}{32\pi}\langle\dot{h}_{ab}\dot{h}^{ab}\rangle,
\end{equation}
where an over-dot denotes $\partial_t$, and Roman indices denote the spatial components of the metric perturbation. This metric perturbation for a SGWB can be written in terms of an expansion over plane waves in the TT gauge,
\begin{equation} \label{eq:metricTTgw}
    h_{ab}(t,\vec{x}) = \sum_{A=+,\times}\int_{-\infty}^{\infty}df\int_{S^2}d^2\Omega_{\hat{n}}\; h_A(f,\hat{n}) e^{2\pi
  if(t+\hat{n}\cdot\vec{x})}e^A_{ab}(\hat{n}),
\end{equation}
where $h_A(f,\hat{n})$ are complex random fields whose moments define the statistical properties of the stochastic GW background\footnote{Since $h_{ab}(t,\vec{x})$ is real, the Fourier amplitudes satisfy $h^*_a(f,\hat{n})=h_A(-f,\hat{n})$};
$\hat{n}$ is a unit vector pointing to the origin of the GW; and $e^A_{ab}(\hat{n})$ are the GW polarisation basis tensors, defined in terms of orthonormal basis vectors around $\hat{n}$:
\begin{equation}
    e_{ab}^{+}(\hat{n}) = \hat{l}_a\hat{l}_b - \hat{m}_a\hat{m}_b,\quad e_{ab}^{\times}(\hat{n}) = \hat{l}_a\hat{m}_b + \hat{m}_a\hat{l}_b,
\end{equation}
where 
\begin{align}
    \hat{n} &= \sin\theta\cos\phi\,\hat{x} + \sin\theta\sin\phi\,\hat{y} + \cos\theta\,\hat{z}\equiv \hat{r} \nonumber\\
    \hat{l} &= \cos\theta\cos\phi\,\hat{x} + \cos\theta\sin\phi\,\hat{y} - \sin\theta\,\hat{z}\equiv \hat{\theta} \nonumber\\
    \hat{m} &= -\sin\phi\,\hat{x} + \cos\phi\,\hat{y} \equiv \hat{\phi}.
\end{align}
The GW energy density is thus
\begin{align} \label{eq:gwdensity}
    \langle\dot{h}_{ab}\dot{h}^{ab}\rangle =&
    \sum_A\sum_{A'}\int_{-\infty}^{\infty}df\int_{-\infty}^{\infty}df'\int_{S^2}d^2\Omega_{\hat{n}}\int_{S^2}d^2\Omega_{\hat{n}'}\, \langle h_A(f,\hat{n})h^*_{A'}(f',\hat{n}')\rangle \nonumber\\
    & \qquad\times e^A_{ab}(\hat{n})e^{ab}_{A'}(\hat{n}') \times 4\pi^2 ff' \nonumber\\
    & \qquad\times \exp{[2\pi i(f-f')t+2\pi i(\hat{n}-\hat{n}')\cdot\vec{x}]}.
\end{align}
For a \textit{Gaussian-stationary}, \textit{unpolarized}, \textit{spatially homogeneous and isotropic} stochastic background the quadratic expectation value of the Fourier modes are
\begin{equation} \label{eq:fourier-variance}
    \langle h_A(f,\hat{n})h^*_{A'}(f',\hat{n}')\rangle = \frac{\delta_{AA'}}{2}\frac{\delta^{(2)}(\hat{n},\hat{n}')}{4\pi}\frac{\delta(f-f')}{2}S_h(f),
\end{equation}
where $S_h(f)$ is the one-sided power spectral density (PSD) of the \textit{Fourier modes} of the SGWB. With identities $\int d\hat\Omega=4\pi$ and $\sum_Ae^A_{ab}e^{ab}_A = 4$, converting the frequency integration bounds in Eq.~\ref{eq:gwdensity} to $[0,\infty]$ gives
\begin{equation}
    \langle\dot{h}_{ab}\dot{h}^{ab}\rangle = 8\pi^2\int_0^{\infty}d f\; f^2S_h(f).
\end{equation}
Hence
\begin{equation} \label{eq:frac-energy-density}
    \Omega_{\rm SGWB}(f) \equiv \frac{1}{\rho_c}\frac{d\rho}{d\ln f}
    = \frac{2\pi^2}{3H_0^2}f^3S_h(f).
\end{equation}

\subsection{Characteristic strain}

The fractional energy density is often referenced in the cosmological literature, or even amongst particle physicists. But GW scientists typically talk about the \textit{characteristic strain} of the SGWB, defined as
\begin{equation}
    h_c(f) \equiv \sqrt{f S_h(f)},
\end{equation}
The characteristic strain accounts for the number of wave cycles the signal spends in-band through the $\sqrt{f}$ dependence (see also Ref.~\citep{2015CQGra..32a5014M}). Hence we can write
\begin{equation}\label{eq:omega_hc}
    \Omega_{\rm SGWB}(f) = \frac{2\pi^2}{3H_0^2}f^2h^2_c(f).
\end{equation}
As is often the case in astronomy and astrophysics, several SGWB sources predict a power-law form for $h_c(f)$, defined as
\begin{equation}
    h_c(f) = A_{\alpha,{\rm ref}}\left(\frac{f}{f_{\rm ref}}\right)^\alpha,
\end{equation}
where $\alpha$ is a spectral index, $f_{\rm ref}$ is a reference frequency that is typical of the detector's band, and $A_{\alpha,{\rm ref}}$ is the characteristic strain amplitude at the reference frequency. The fractional energy density then scales as $\Omega_{\rm SGWB}(f)\propto f^{2\alpha+2}$. Primordial SGWBs resulting from quantum tensor fluctuations that are inflated to macroscopic scales usually assume a scale-invariant spectrum for which $\Omega_{\rm SGWB}(f)\propto\mathrm{constant}$, thereby implying $\alpha=-1$ for the characteristic strain spectrum. We'll see in Chapter \ref{chap:sources} how a population of circular inspiraling compact-binary systems creates a SGWB with $\alpha=-2/3$ such that $h_c(f)\propto f^{-2/3}$ and $\Omega_{\rm SGWB}(f)\propto f^{2/3}$. 

\subsection{Spectrum of the strain signal}

We don't directly measure a metric perturbation in our detector; we measure the response of our detector to the influence of a metric perturbation. The strain signal measured by a detector is related to the source strain through the detector's response. The data in a given detector, $s_i(t)$, is a combination of measured signal, $h_i(t)$, and noise, $n_i(t)$:
\begin{equation}
    s_i(t) = h_i(t) + n_i(t),
\end{equation}
where the measured strain signal is the projection of the GW metric perturbation onto the detector's response, such that
\begin{equation}
    h_i(t) = d^{ab}_i h_{ab}(\vec{x}_i,t),
\end{equation}
and $d^{ab}_i$ is the GW detector tensor of the $i^{\rm th}$ detector. The one-sided cross-power spectral density (PSD) of the measured strain signal $S_s(f)$ in detectors $i$ and $j$ is
\begin{equation} \label{eq:strain_signal_psd}
    \langle \tilde{h}_i(f)\tilde{h}_j(f')\rangle =
    \frac{1}{2}\delta(f-f')S_s(f)_{ij},
\end{equation}
where tilde denotes a Fourier transform with the following convention:
\begin{equation}
    \tilde{h}_i(f) = \int_{-\infty}^{\infty} e^{-2\pi ift}h_i(t)dt.
\end{equation}
The Fourier domain strain signal is thus
\begin{align}
    \tilde{h}_i(f) &= \sum_A
    \int_{S^2}d^2\Omega_{\hat{n}}\int_{-\infty}^{\infty}d
    f\int_{-\infty}^{\infty}d t\;e^{-2\pi
      ift}F^A_i(\hat{n})h_A(f,\hat{n})e^{2\pi
      if(t+\hat{n}\cdot\vec{x})} \nonumber\\
      &= \sum_A\int_{S^2}d^2\Omega_{\hat{n}}\;F^A_i(\hat{n})h_A(f,\hat{n})e^{2\pi
      if\hat{n}\cdot\vec{x}_i},
\end{align}
where $F^A_i(\hat{n})$ is the antenna response beam pattern (or antenna pattern, or antenna response function) of the detector to mode-$A$ of the GW, defined as
\begin{equation} \label{eq:antennaresponse}
    F^A_i(\hat{n})\equiv d^{ab}_i e_{ab}^A(\hat{n}).
\end{equation}
Hence the quadratic expectation of the Fourier-domain signals is given by
\begin{align}
    \langle \tilde{h}_i(f)\tilde{h}^*_j(f')\rangle =&
    \sum_A\sum_{A'}\int_{S^2}d^2\Omega_{\hat{n}}\int_{S^2}d^2\Omega_{\hat{n}'}\langle
    h_A(f,\hat{n})h_{A'}^*(f',\hat{n}')\rangle \nonumber\\
    &\qquad \times F^A_i(\hat{n})F^{A'}_j(\hat{n}')e^{2\pi
      i(f\hat{n}\cdot\vec{x}_i-f'\hat{n}'\cdot\vec{x}_j)},
\end{align}
which, upon using Eq.~\ref{eq:fourier-variance}, becomes 
\begin{equation} \label{eq:fourier_crosspower}
    \langle \tilde{h}_i(f)\tilde{h}^*_j(f')\rangle = \frac{1}{16\pi}S_h(f)\delta(f-f')\sum_A\int_{S^2}d^2\Omega_{\hat{n}} F^A_i(\hat{n})F^{A'}_j(\hat{n})e^{2\pi
      if\hat{n}\cdot(\vec{x}_i-\vec{x}_j)}.
\end{equation}

\subsection{Overlap Reduction Function} \label{sec:orf}

If we compare Eq.~\ref{eq:fourier_crosspower} to Eq.~\ref{eq:strain_signal_psd}, we see that
\begin{align} \label{eq:signal_mode_psd_compare}
    S_s(f)_{ij} &= S_h(f)\times\frac{1}{8\pi}\sum_A\int_{S^2}d^2\Omega_{\hat{n}} F^A_i(\hat{n})F^A_j(\hat{n})e^{2\pi
      if\hat{n}\cdot(\vec{x}_i-\vec{x}_j)} \nonumber\\
      &= S_h(f)\times\tilde\Gamma_{ij}(f,\vec{x}_i-\vec{x}_j),
\end{align}
which relates the cross-power spectral density of the \textit{measured strain signal} to that of the \textit{Fourier modes of the GW background}. The term $\tilde\Gamma_{ij}(f,\vec{x}_i-\vec{x}_j)$ is referred to as the un-normalized overlap reduction function, or sometimes just the overlap reduction function (ORF) since we normalize it later as standard in order to be unity for co-located co-oriented detectors. Essentially the ORF measures how much GWB power is shared between pairs of detectors. It is the most essential element of searching for a GWB with PTAs, and in the next chapter we will see what this PTA ORF is (spoiler: it's the ubiquitous \textit{Hellings \& Downs curve}! \citep{1983ApJ...265L..39H}). 

\section{The Gravitational Wave Spectrum}
\begin{figure}
    \centering
	\includegraphics[width=\columnwidth]{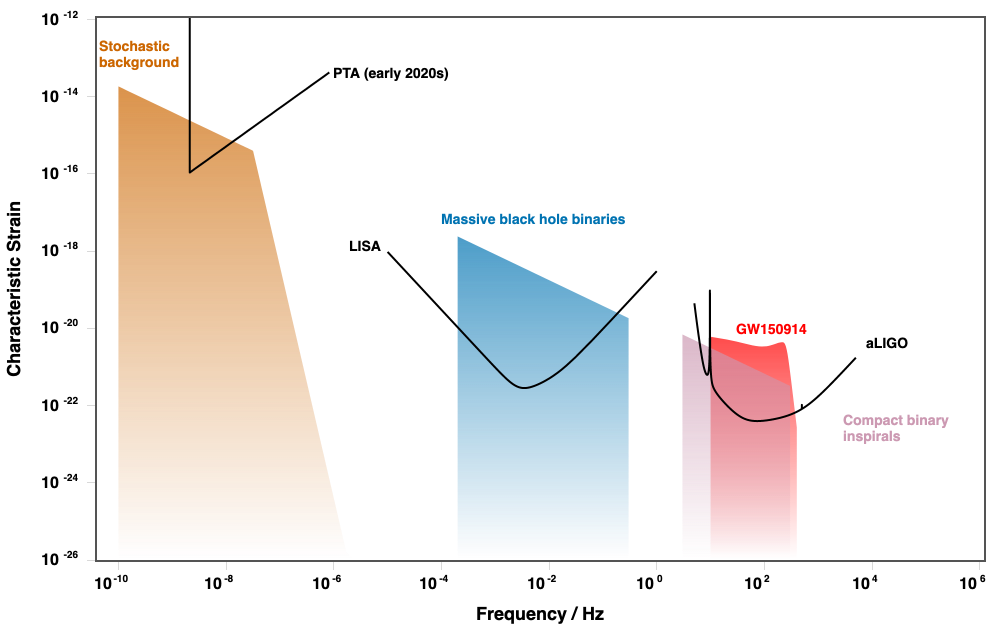}
    \caption{Overview of the detectors and sources covering the GW spectrum, spanning approximately twelve orders of magnitude in frequency. Figure created using \href{http://gwplotter.com}{http://gwplotter.com} \citep{2015CQGra..32a5014M}.}
    \label{fig:gwspectrum}
\end{figure}

Like electromagnetic radiation, GWs come in a spectrum of frequencies from many different sources, where (roughly speaking) the characteristic frequency of the emitting system scales inversely with its total mass. However, electromagnetic radiation from astronomical sources is usually an incoherent superposition of emission from regions much larger than the radiation wavelength. By contrast, the majority of the GW spectrum being targeted by current and forthcoming detectors is sub-kHz, and as such wavelengths are of comparable scale to their emitting systems. Thus GWs directly track the coherent bulk motions of relativistic compact objects. 

This volume is not intended to be a comprehensive overview of the enormous efforts being exerted towards detection across the full GW landscape. Nevertheless, there are three main detection schemes blanketing the spectrum from nanoHertz frequencies (with Pulsar Timing Arrays, the focus of this volume), milliHertz frequencies (with laser interferometry between drag-free space-borne satellites), to $10-100$~s Hz (ground-based laser interferometry). At higher frequencies there are efforts to detect MegaHertz GWs with smaller instruments \citep[e.g.,][]{2017PhRvD..95f3002C}, while even lower than Pulsar Timing Arrays at $\sim10^{-15}$~Hz there are ongoing efforts to detect GWs through the imprint of curl-mode patterns in the polarization signal of the Cosmic Microwave Background \citep{1997PhRvL..78.2058K,1997PhRvL..78.2054S,2018PhRvL.121v1301B}.

Figure \ref{fig:gwspectrum} illustrates the strain spectrum of GWs over many decades in frequency, with the sensitivity of current and planned detectors overlaid on the bands of compelling astrophysical targets. At the lowest frequencies we need precisely timed pulsars at $\sim$kiloparsec distances to probe nHz frequencies, where a stochastic background of merging supermassive binary black holes may be found. The $\sim1-10^4$~Hz band is the domain of terrestrial detectors, where kilometer-scale laser interferometers target the chirping signals of inspiraling stellar-mass compact binary systems, as well as core-collapse supernovae. The lowest frequency we can probe with terrestrial detectors ($\sim 1$~Hz) is restricted by local gravitational gradients and seismic noise isolation. The only way to overcome this ``seismic wall'' is by moving to space. Space-borne laser interferometers of $\sim 10^9$~m arm-length are planned which will dig into the $0.1-100$~mHz band, allowing for massive black hole mergers to be seen throughout the entire Universe, precision tests of gravity, and the entire Galactic white-dwarf binary population to be seen. 

Below is a brief review of ground-based and space-borne efforts. The Galactic-scale efforts of Pulsar Timing Arrays will be discussed in the remaining chapters of this volume.

\subsection{Ground-based Detectors}

Early terrestrial detectors were of the Weber bar variety \citep{weber1957reality,weber1960detection}, however it was soon realised that laser interferometry had the potential to greatly surpass the relatively narrow-band sensitivity of bars. The Weber bar was principally developed to infer the {\it transfer of energy} of GWs to the detection apparatus, while laser interferometery aims to measure the strain influence of a GW on the propagation of a precisely monitored laser beam. If we consider a GW impinging on a simple right-angled interferometer, with laser beams propagating along orthogonal directions, then the wave's influence is measured via the alternate stretching and compression of the proper-length of the arms, inducing a phase-shift in the recombined laser beams. As such, contemporary GW detectors aim to measure the {\it strain amplitude} of a passing wave.

GW interferometers aim to achieve a strain sensitivity of $\sim 10^{-21}$ or lower. The basic mode of operation is that of a Michelson interferometer, where laser light is injected into the interferometer and subsequently split into two beams to propagate along orthogonal arms. Each laser beam reflects off end test-mass optics and recombine at a beamsplitter to interfere at a photodiode. There are many sources of noise \citep[see, e.g.,][and references therein]{2020CQGra..37e5002A}, including photon shot noise and thermal motion of the reflecting test masses. The limiting noise sources at the lowest frequencies are seismic and gravity-gradient noise. The former can be ameliorated by a combination of pendulum isolation, spring suspension, and anti-vibration actuators. However gravity-gradient noise is caused by seismic waves that create local mass density fluctuations, whose gravitational influence couples to the test-masses; this can be minimised by monitoring seismic activity to subtract its signal, or moving the detector underground, but below $\sim 1$ Hz the detector must be completely distanced from these surface-wave density fluctuations by moving to space.

LIGO (the Laser Interferometer Gravitational wave Observatory)\footnote{\href{https://www.ligo.caltech.edu}{https://www.ligo.caltech.edu}} is the first kilometre-scale detection instrument for GWs, and is operated in partnership between the California Institute of Technology and the Massachusetts Institute of Technology. There are two $4$~km arm-length instruments in total, all located in the USA, with one sited in Hanford, Washington,\footnote{In the pre-2015 Initial and Enhanced LIGO stages there was a second half-length Hanford instrument contained within the same vacuum envelope} and one in Livingston, Louisiana. In addition to LIGO, there is the French/Italian $3$~km Virgo interferometer\footnote{\href{https://www.virgo-gw.eu}{https://www.virgo-gw.eu}} located at Cascina, near Pisa, in Italy. Also in Europe is the $600$ m arm-length GEO-$600$ interferometer\footnote{\href{https://www.geo600.org}{https://www.geo600.org}}, located near Hannover, Germany. With a smaller baseline and lower laser power, GEO-$600$ cannot match the sensitivity of LIGO/Virgo, but it has been a useful testbed for advanced technologies and techniques. Finally, in Japan there is the $3$ km arm-length KAmioka GRAvitational wave telescope (KAGRA)\footnote{\href{https://gwcenter.icrr.u-tokyo.ac.jp/en/}{https://gwcenter.icrr.u-tokyo.ac.jp/en/}}, which operates in the Kamioka mine under cryogenic conditions. Placing KAGRA underground dramatically suppresses seismic disturbances and gravity gradient noise.

Beyond these second-generation detector plans, there are concepts for new third-generation detectors aiming to achieve a broadband order of magnitude improvement in strain sensitivity and to push operation down into the $\sim 1-10$ Hz range\footnote{\href{https://gwic.ligo.org/3Gsubcomm/documents/science-case.pdf}{https://gwic.ligo.org/3Gsubcomm/documents/science-case.pdf}}. The most notable of these are the Einstein Telescope in Europe \citep{2020JCAP...03..050M}\footnote{\href{http://www.et-gw.eu}{http://www.et-gw.eu}}, and Cosmic Explorer in the USA \citep{2019BAAS...51g..35R}\footnote{\href{https://cosmicexplorer.org}{https://cosmicexplorer.org}}.  

\textit{\textbf{Sources--}} In its most recent incarnation, LIGO became an officially funded NSF project in 1994 under the leadership of Barry Barish. Construction broke ground in Hanford, Washington in 1994, and Livingston, Louisiana in 1995, with initial observations beginning in 2002. For approximately eight years of these initial observations LIGO did not see a single shred of evidence for GWs. However, this was not totally unexpected-- Initial LIGO was a dress rehearsal that acted as a technology concept and a rallying site for the community, but a remote prospect for a first detection. Around 2010, LIGO was taken offline and subjected to a multi-year overhaul to create Advanced LIGO, which saw first operations commence in 2015. 

The rest is now popular lore. On September 14, 2015, before the first science run officially commenced, LIGO detected the collision of two $\sim30M_\odot$ black holes at a distance of $\sim400$~Mpc from Earth, resulting in the emission of $\sim3M_\odot$ of mass-energy as GWs, leaving a remnant black hole of $\sim57M_\odot$ \citep{2016PhRvL.116f1102A}. The signal lasted $\sim200$ milliseconds in the sensitivity band of the detector, providing a signal-to-noise ratio of $\sim24$. Far from being the threshold event that had been expected for the first signal, this arrived with thunderous certainty. GW150914 (labeled by its discovery date) was announced to the world on February 11, 2016, inaugurating the field of direct GW astronomy. Since then, many other BH-BH signals have been detected, including GW190521 that has the largest constituent masses to date of $\sim85M_\odot$ and $\sim66M_\odot$ \citep{2020PhRvL.125j1102A}. These BH-BH detections have an enormous amount to teach us about their possible origin pathways \citep{2016ApJ...818L..22A,2020arXiv201014533T,2019ApJ...882L..24A}, including as isolated binary stellar systems that undergo successive supernovae and a common stellar-envelope phase, or as systems formed through dynamical capture in a stellar cluster environment, or even scenarios we can't envision yet. What's more, on August 17, 2017, LIGO and Virgo observed a double neutron-star collision, yielding not only a GW signal registered at all three sites, but a plethora of electromagnetic signals observed across the spectrum \citep{2017PhRvL.119p1101A}. GW170817 provided extraordinary multi-messenger insight into neutron star astrophysics \citep{2017ApJ...848L..12A}, tests of General Relativity \citep{2019PhRvL.123a1102A}, and even an anchor in spacetime with which we can calibrate the expansion rate of the Universe \citep{2017Natur.551...85A}. 

It's not just compact-binary coalescences that ground-based detectors can find. Other searches include modelled and unmodeled bursts (from core-collapse supernovae) \citep[e.g.,][]{2016PhRvD..94j2001A,2019arXiv190503457T}, continuous waves (from $<10$-cm pulsar ``mountains'' that generate a quadrupole mass moment) \citep[e.g.,][]{2017PhRvD..96l2006A,2019ApJ...875..122A,2019PhRvD.100b4004A}, and an unresolved background of GWs \citep[e.g.,][]{2017PhRvL.118l1101A,2019PhRvD.100f1101A}. As of writing, LIGO is offline due to the COVID-19 pandemic.  

\subsection{Space-borne Detectors}

Access to the frequency range $\sim0.1-100$~mHz requires much longer interferometer arm lengths, and a complete suppression of the seismic and gravity-gradient noise that plagues the low-frequency operation of terrestrial detectors. To this end, detection at these frequencies necessitates space-borne laser interferometers.

The canonical design for a mission in this band is the Laser Interferometer Space Antenna (LISA)\footnote{\href{https://www.elisascience.org}{https://www.elisascience.org}}$^\mathrm{,}$\footnote{\href{https://lisa.nasa.gov}{https://lisa.nasa.gov}}, a project led by the European Space Agency in collaboration with the National Aeronautics and Space Administration. This mission calls for an arrangement of three identical satellites maintaining a triangular constellation, each separated by $2.5\times10^9$~m. The test masses within each satellite are expected to be $\sim46$~mm, $2$~kg, gold-coated cubes of gold/platinum. Each satellite would achieve zero drag, where each test mass essentially floats in free fall while the surrounding satellite absorbs local gravitational influences. Micro-thrusters reposition the satellite around the test mass to maintain this drag-free configuration. The LISA satellite constellation would trail the Earth's motion by $20^\circ$ as it orbits around the Sun. With two optical links in each arm of the triangle, a total of six optical links should allow for the interferometer to operate in Sagnac mode, constructing a data-stream that will be completely insensitive to laser, optical-bench, and clock noise \citep[e.g.,][]{2004PhRvD..69b2001S}. In 2013, the European Space Agency selected ``The Gravitational Universe'' science theme for its L3 mission slot\footnote{\href{http://www.esa.int/Science_Exploration/Space_Science/ESA_s_new_vision_to_study_the_invisible_Universe}{http://www.esa.int/Science$\_$Exploration/Space$\_$Science/ESA$\_$s$\_$new$\_$vision$\_$to$\_$study$\_$the$\_$invisible$\_$Universe}}, wherein it committed to launch a space-borne GW mission, due for launch in $2034$ at the earliest. In 2017, LISA was proposed and accepted as the candidate mission\footnote{\href{https://www.lisamission.org/?q=news/top-news/gravitational-wave-mission-selected-esas-l3-mission}{https://www.lisamission.org/?q=news/top-news/gravitational-wave-mission-selected-esas-l3-mission}}. 

In 2015 a mission known as LISA Pathfinder (LPF)\footnote{\href{https://sci.esa.int/web/lisa-pathfinder}{https://sci.esa.int/web/lisa-pathfinder}}  was launched to act as a technology demonstration for the LISA drag-free concept. Rather than three satellites, LPF consisted of just one satellite enclosing an optical system that corresponded to an arm-length of $38$-cm. It reached its designated position at Lagrange point L1 on January 22, 2016. Far from a mere technology demonstration, LPF exceeded its scientific mandate by achieving exceptional noise precision that is comparable to the LISA requirement \citep{armano2016sub}. 

Despite LISA's prospective launch date being (as of writing) more than a decade away, possible follow-up missions are already being studied and advocated for. These missions are designed to bridge either the $\sim1-100$~$\mu$Hz gap between PTAs and LISA \citep{2019arXiv190811391S}, or the $0.1 - 10$~Hz gap between LISA and ground-based detectors \citep{2020CQGra..37u5011A}. Some notable examples include $\mu$Ares \citep{2019arXiv190811391S}, a straw-person $\mu$Hz detector concept suggested in the European Space Agency's \textit{Voyage 2050} long-range planning call. This ambitious mission would place at least three satellites in an approximately Martian orbit with the Sun at its center. At the deciHertz level, the Advanced Laser Interferometer Antenna (ALIA) \citep{2013CQGra..30p5017B,2019BAAS...51g.243M}, and DECi-hertz Interferometer Gravitational wave Observatory (DECIGO) mission \citep{2017JPhCS.840a2010S} have been proposed to capitalize on the treasure trove of science related to stellar-mass binary black holes, intermediate mass-ratio inspirals, and binary neutron stars that would otherwise be missed. 

\textit{\textbf{Sources--}} This frequency regime is a rich astrophysical zoo of sources, including a collection of Galactic white-dwarf binary systems whose properties are well-known electromagnetically, and hence should be detectable within a few weeks or months of instrument operation \citep{2018MNRAS.480..302K}. As a sure bet for GW detection by LISA, these systems are often referred to as \textit{verification binaries}, in that if no signal is observed then something is very wrong with the instrument. Additionally, $\sim25000$ other ultra-compact white-dwarf binary systems may be individually resolvable in GWs \citep{2019arXiv190305583L,2017arXiv170200786A}, while the remaining several million will create a stochastic GW foreground signal \citep[e.g.,][]{2020A&A...638A.153K,2017CQGra..34x4002R}. 

Massive BH binary systems in the mass range $10^4 - 10^7 M_{\odot}$ are prime targets for LISA, which should detect the inspiral, merger, and ringdown signals associated with coalescence \citep{2019arXiv190306867C}. Detailed parameter estimation of individual systems will be possible, permitting inference of such effects like spin-orbit precession, higher waveform harmonics, and eccentricity \citep[e.g.,][]{2011PhRvD..83h3001K,2018PhRvD..98j4043K,2020PhRvD.101l4008C,2020arXiv200300357M}. The detection rate is highly variable, ranging from $\sim 0.5$ per year to $\sim100$ per year depending on the underlying ``seed'' formation scenario and growth factors \citep[e.g.,][]{2020MNRAS.491.2301K,2019BAAS...51c..73N}. Nevertheless a catalog of such systems will empower demographic and ``genealogical'' studies of the massive BH population, unveiling the factors that drive their growth over cosmic time, and the fingerprints of their initial formation at high redshift \citep[e.g.,][]{2011PhRvD..83d4036S,2011MNRAS.415..333P}. 

At lower masses, a tantalizing possibility for LISA is that it will capture the very early inspiral stage of stellar-mass ($\mathcal{M}\gtrsim25M_\odot$) and intermediate-mass ($10^2M_\odot<M<10^4M_\odot$) BH-BH systems that will eventually also register (after $\sim$months) at higher frequencies in future ground-based detectors. By detecting the same systems at low frequencies (wide separation) and high frequencies (close separation through merger), multi-band GW astrophysics will yield insights into the dynamics of stellar-mass BH-BH over long timescales \citep[see, e.g.,][and references therein]{2019BAAS...51c.109C}. Combining the information from both detection schemes will break important degeneracies in parameter inference, allowing the systems to be characterized better \citep{2016PhRvD..94f4020N,2016PhRvL.117e1102V}. Early work in this area emphasized how LISA detection of these systems would allow precise time and sky-location forecasts of the eventual merger in order for ground-based detectors to be on the watch. But the reverse scheme is also important: detection of these systems in a future ground-based detector will allow archival LISA detector to be mined for marginal or sub-threshold signals. 

From a fundamental physics standpoint, arguably the most exciting sources in the LISA band are the Extreme Mass-Ratio Inspirals, where stellar-mass compact remnants gradually spiral in towards a much larger BH, and in so doing map out the geometry of the massive BH's space-time \cite[see, e.g.,][and references therein]{2019BAAS...51c..42B}. Beyond these extreme mass-ratio systems, LISA has huge potential to probe how modifications to General Relativity affect GW generation, GW propagation, BH spacetimes, and BH dynamics \citep{2019BAAS...51c..32B}. Indeed, even if no hints of GR departures are found, LISA may detect more exotic signals than mere compact binaries; a cosmological GW background signal of primordial origin may be detectable at these frequencies \citep[e.g.,][]{2017JPhCS.840a2030R}, or even a similar background formed later as the Universe undergoes phase transitions \citep[e.g.,][]{2020JCAP...03..024C}. LISA could even potentially detect cosmic strings that form during these phase transitions; these strings can intersect one another to chop off small loops, which then emit GWs as they vibrate relativistically under extreme tension \citep{2020JCAP...04..034A}. 

As of writing, LISA's prospective launch date in 2034 is 13 years away. But this does not diminish the vigor and excitement with which the GW and astrophysics communities are currently exploring its enormous science potential. 

\bibliographystyle{unsrt_new}
\bibliography{refs}

%% file: 03.tex
\chapter{Pulsar Timing} \label{chap:pulsar_timing}
\epigraph{\textit{``Time is a companion who goes with us on the journey, and reminds us to cherish every moment - because they'll never come again.''}}{Captain Jean-Luc Picard, ``Star Trek: Generations''}

\section{Pulsars}

Pulsars are extraordinary. They are a special class of neutron star, which in themselves are mind-boggling objects corresponding to the collapsed $\sim$~$M_\odot$ cores of massive stars that have undergone supernovae, leaving only small $\sim$~$10-15$~km carcasses that are supported against total collapse by neutron degeneracy pressure. Since their discovery in 1967 by Jocelyn Bell Burnell, Antony Hewish and collaborators \citep{hewish1968observation}, pulsars have shed light on strong-field gravity, the equation of state of nuclear matter, evolutionary scenarios for massive binary systems, the structure of the ionized interstellar medium, the existence of exoplanets, and much more. It would be difficult to overstate the exquisite astrophysical laboratories presented to us in the form of isolated and relativistic-binary pulsars. For deeper reviews, see Refs.\ \cite{2004hpa..book.....L,2008LRR....11....8L,2003LRR.....6....5S}. Various other sources have inspired the content of this chapter, including Verbiest et al. (2021) \citep{2021arXiv210110081V}, and Burke-Spolaor (2015) \citep{2015arXiv151107869B}.

The ``lighthouse model'' provides our basic framework for understanding and modeling pulsars as rapidly rotating, highly magnetized neutron stars resulting from stellar collapse \cite{pacini1968rotating,gold1968rotating}. Due to conservation of angular momentum and magnetic flux, these pulsars are far more rapidly spinning and magnetized than their progenitor stars. Their magnetic field (whose axis may not necessarily align with its rotational axis) is such that the star acts as a rotating magnetic dipole, generating a local electric field along which charged particles within the co-rotating magnetospheric plasma are accelerated. It is expected that these particles excite beams of radio emission high in the pulsar atmosphere that we observe whenever the rotating beam intersects our line-of-sight \cite{goldreich1969pulsar,sturrock1971model}. The pulse period is then a measure of the rotation period of the pulsar itself.

\begin{figure}
    \centering
	\includegraphics[width=1.0\columnwidth]{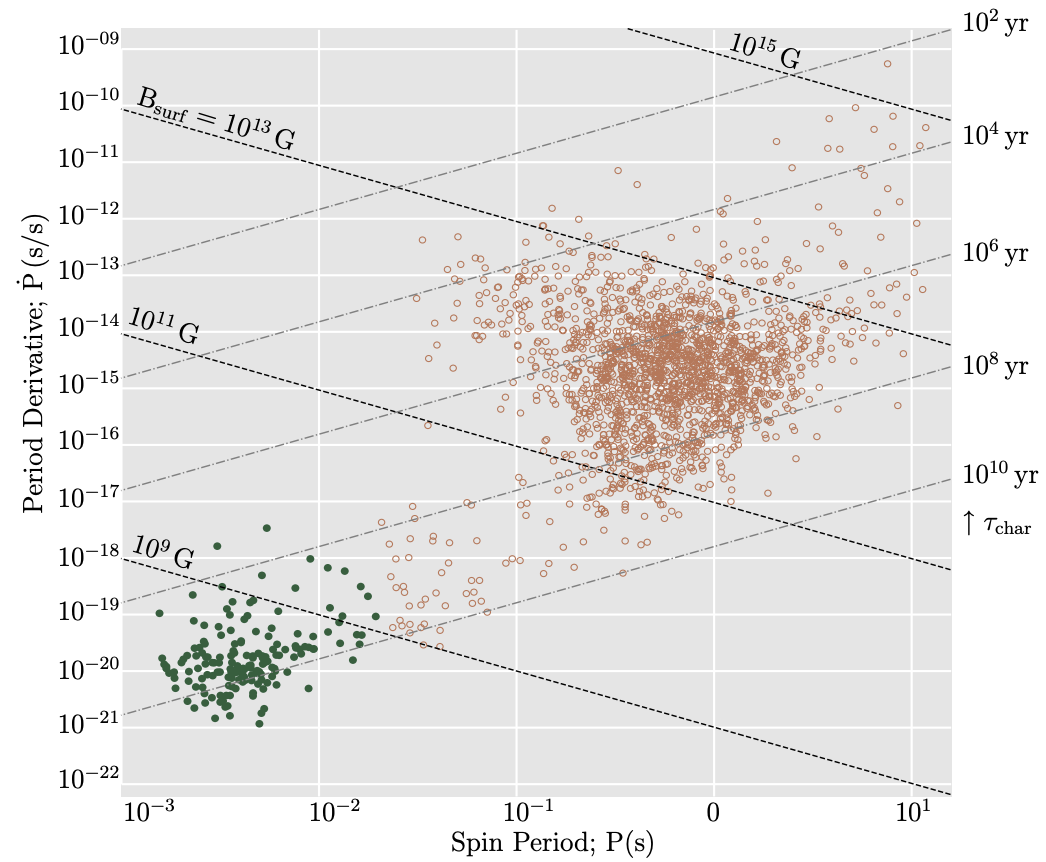}
    \caption{$P-\dot{P}$ diagram of known pulsars. MSPs are filled green circles, while canonical pulsars are open circles. Data were taken from the ATNF pulsar catalogue \citep{manchester2005australia} version 1.56 (see also www.atnf.csiro.au/research/pulsar/psrcat). The dashed black lines show estimated surface magnetic fields strengths ($B_\mathrm{surf}$), while the dot-dashed grey lines show the lines of characteristic age ($\tau_\mathrm{char}$). Figure reproduced with permission from Ref.~\citep{2018CQGra..35m3001V}.}
    \label{fig:ppdot_diagram}
\end{figure}

While the largest population of pulsars are in the class of initially discovered young $\sim1$-second rotators (so-called ``canonical pulsars''), they are not used in precision timing campaigns for GW searches. The broad reasons for this are that canonical pulsars can exhibit lower long-timescale rotational stability, and glitches, the latter of which are a technical term for a sudden spin-up in the pulsar that \textit{may} be related to reconnection of the internal neutron superfluid with the crustal lattice \citep[e.g.,][and references therein]{lyne2000statistical}. The 1982 discovery of the pulsar B1937+21, with its $1.5$~millisecond period, was the first of the new class of ``millisecond pulsars'' (MSPs) \cite{backer1982millisecond}. The demography of pulsars can be broadly split into these two varieties (see Fig.~\ref{fig:ppdot_diagram}); the canonical pulsars are ones that have formed relatively recently as a result of a supernova, while the millisecond pulsars are older, having spun down and been subsequently recycled back to millisecond periods via the accretion of material and angular momentum during mass transfer from a binary companion \citep{bhattacharya1991formation}.

\section{Precision pulsar timing} \label{sec:precision_pulsar_timing}

The key to using pulsars as astrophysical tools is that they can be used as excellent time-keepers\footnote{Full details of timing procedures can be found in Ref.\ \cite{2012hpa..book.....L}.}. We observe pulses of radio emission separated by the observational period of the pulsar. However, the shape of each pulse from one rotation to the next varies randomly, possibly associated with stochasticity in the emission region through which our line of sight is intersecting. But the pulse shape \textit{averaged} over rotations is remarkably stable and reproducible on timescale from minutes to decades \citep{helfand1975observations}\footnote{Investigations of the evolution of the standard pulse profile can yield rich information on the details of the emission region, and geodetic precession of the pulsar's spin axis in a binary system. See Ref.\ \citep{2003LRR.....6....5S}.}. It is this stability at a given radio frequency that permits precision timing; the pulse shape is unique to each pulsar and can be relied upon to mark the passage of rotations when receiving a train of radio pulses.

\begin{figure}
    \centering
	\includegraphics[width=1.0\columnwidth]{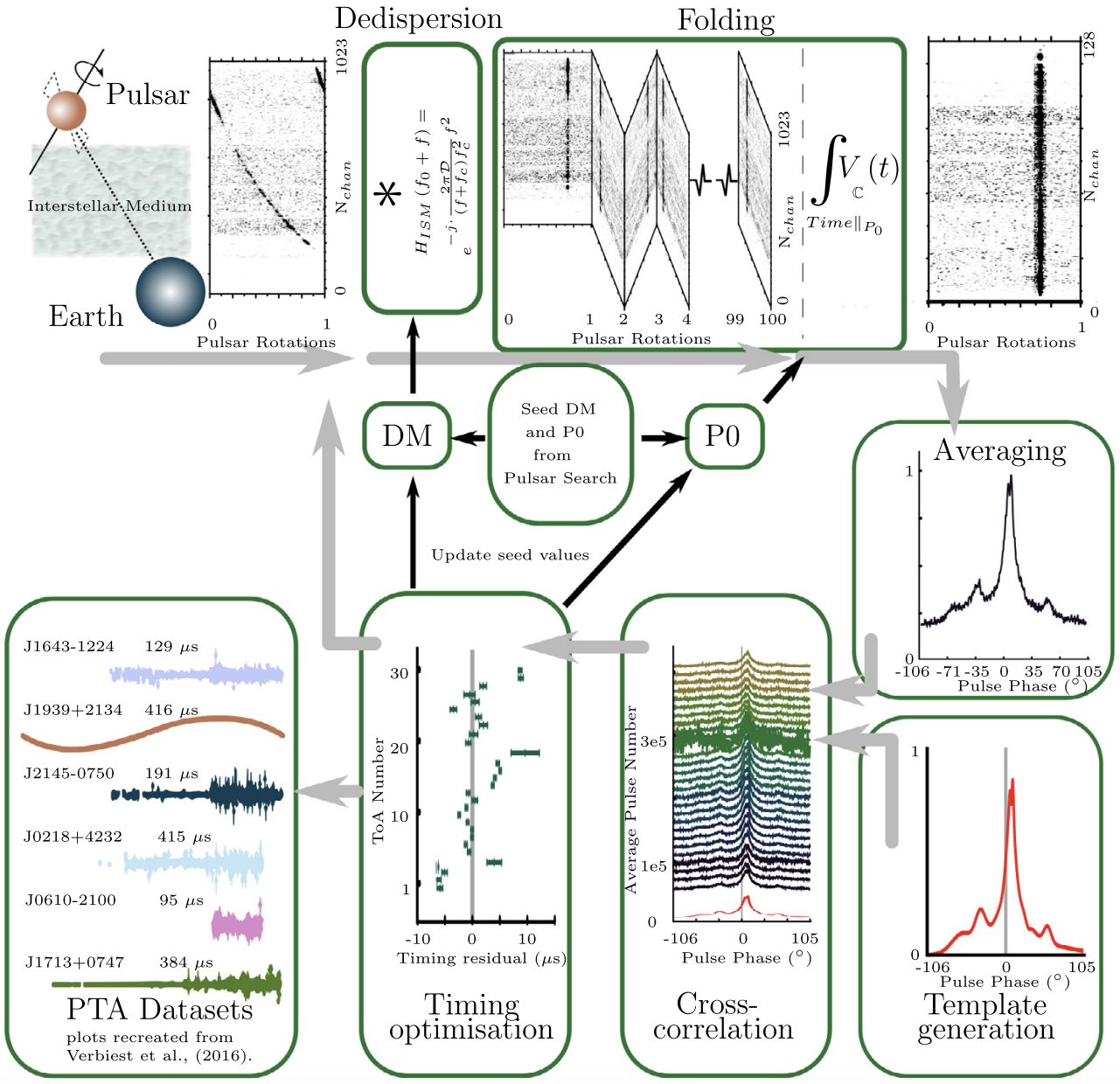}
    \caption{A representation of the main stages involved in the precision timing of pulsars. Figure reproduced with permission from Ref.~\citep{2018CQGra..35m3001V}.}
    \label{fig:pulsartiming_workflow}
\end{figure}

A schematic diagram of the main stages involved in pulsar timing is shown in Fig.~\ref{fig:pulsartiming_workflow}. Upon being accelerated in the pulsar's magnetosphere, high-energy charged particles excite beams of radiation with a steep, negative-slope radio spectrum. This radiation propagates through the ionized interstellar medium (ISM), suffering dispersion and other radio-frequency dependent delays. Dispersion arises from the frequency-dependent refractive index of the ISM, such that lower radio frequencies have a reduced group velocity, arriving at the telescope later than higher radio-frequency components of the radiation. The delay is determined by the distance travelled through the ISM, such that with an appropriate model of the line-of-sight electron-density distribution, the measured dispersion can be used to infer the pulsar's distance \citep[e.g.,][and references therein]{verbiest2012pulsar}. Dispersion can be overcome either by splitting the observed band into smaller sub-channels and delaying the higher-frequency components according to the dispersive $1/\nu^2$ relationship (incoherent dedispersion), or by convolving the raw observations with the inverse transfer function of the ISM (coherent dedispersion) \citep{hankins1975pulsar}. Further details of these effects, and how their residual influences are modeled alongside GWs, are given in Chapter \ref{chap:pta_likelihood}.

After the removal of dispersion, thousands of pulses are integrated over $\sim$minutes to an hour of observation, and folded by the current estimated rotational period to give a boost to the signal strength and stabilize the measured pulse profile. This measured pulse is then cross correlated with the {\it template profile} for the specific pulsar at the specific observing frequency. The phase offset between the measured pulse and template profile is added to the time-stamp of the observation, giving the pulse \textit{``Time Of Arrival''} (TOA). The template-fitting uncertainty of the phase offset, and thus the measurement uncertainty of the TOA, scales as
\begin{equation}
    \sigma_\mathrm{TOA}\propto \sqrt{n_P t_\mathrm{int}\Delta\nu}\times\frac{T_\mathrm{peak}}{T_\mathrm{sys}}\times\frac{\sqrt{W(P-W)}}{P},
\end{equation}
where $n_P$ is the number of combined polarizations, $t_\mathrm{int}$ is the integration time, $\Delta\nu$ is the radio bandwidth of observation, $T_\mathrm{peak}$ is the brightness temperature of the pulsar at the peak of its profile, $T_\mathrm{sys}$ is the brightness temperature (i.e. noise)
of the observing system, $W$ is the integrated pulse intensity divided by its peak intensity (the pulse's \textit{equivalent width}), and $P$ is the pulse period. The TOA measurement uncertainty is often referred to as "radiometer noise", corresponding to the limiting precision with which we can time a pulsar due to the radio telescope's sensitivity and the properties of the pulsar itself. The radiometer noise is not the end of the story for the per-TOA measurement uncertainty, as various modifying parameters are introduced to correct poorly-estimated $\sigma_\mathrm{TOA}$ values or account for additional sources of epoch noise; this is further described in Chapter \ref{chap:pta_likelihood}.

Over many repeated observations, a train of pulses are collected and the TOA computed for each as described above. The next stage is determining the \textit{timing model} for the pulsar (sometimes also referred to as the \textit{timing ephemeris}). This is a generative model that describes all deterministic influences that affect the arrival time of the pulses as they propagate from the pulsar system to the radio telescope on Earth. We write the pulse TOA as
\begin{equation}
    t_\mathrm{PSR} = t_\mathrm{TOA} - \Delta_\odot - \Delta_\mathrm{IISM} - \Delta_B, 
\end{equation}
where $t_\mathrm{PSR}$ is the pulse emission time at the pulsar and $t_\mathrm{TOA}$ is the pulse arrival time at the radio telescope. The term $\Delta_\odot$ accounts for timing corrections back to the quasi-inertial reference frame of the Solar System Barycenter (SSB) that include (a) Einstein delays due to time dilation and gravitational redshift in the presence of the Sun and other bodies in the Solar System; (b) Shapiro delays due to light propagating through the gravitational potential well of the Sun; (c) Roemer delays due to the classic light travel time across the Solar System from the Earth to the SSB; (d) Earth atmospheric propagation delays; (e) solar-wind induced radio-frequency dependent delays; and (f) clock corrections from the observatory standards to global timing standards. The term $\Delta_\mathrm{IISM}$ includes corrections for radio-frequency dependent propagation delays such as dispersion. While the pulse components are dispersion-corrected before the folding stage, the effects of interstellar dispersion have been shown to vary over long and short timescales, requiring additional modeling as described in detail in Chapter \ref{chap:pta_likelihood}. Finally, for pulsars in binary systems, a further transformation, $\Delta_B$, is needed to correct for time between the binary barycenter and the emitting pulsar itself. This includes Einstein, Shapiro, and Roemer delays within the pulsar binary orbit, in addition to a host of other higher order corrections \citep[see, e.g.,][and references therein]{edwards2006tempo2}.

After all of these corrections, the final model of the pulsar's phase evolution is remarkably simple. The lighthouse model describes a beam sweeping into our line-of-sight every pulsar rotation, where the rotational frequency of the pulsar is decreasing due to ``spindown'' that may be related to the electromagnetic outflow tapping the pulsar's rotational kinetic energy. Hence, for a pulsar with some rotational frequency $1/P$ measured at epoch $t_0$, the pulse phase is modelled as
\begin{equation}
 \phi(t_\mathrm{PSR}) = \phi_0 + 2\pi \frac{(t_\mathrm{PSR}-t_0)}{P} - \frac{1}{2}2\pi \left[\frac{(t_\mathrm{PSR}-t_0)}{P}\right]^2\dot{P} + \ldots,
 \end{equation}
where $\phi_0$ is the pulsar phase at $t_0$. With initial estimates of the dispersion measure, rotational period, period derivative, and location of the pulsar, we can perform a least-squares fit of the collection of measured TOAs with the field-standard software package \textsc{Tempo2} \citep{hobbs2006tempo2,edwards2006tempo2,hobbs2009tempo2} or the emerging heir-apparent \textsc{PINT} \citep{2020arXiv201200074L}. The differences between the measured TOAs and the predictions of the best-fit model are called the \textit{timing residuals}. By iterating and refining the timing model to remove systematic trends and miminize the residuals, we can construct extraordinarily precise predictions of the pulsar's phase. 

By definition, the residuals are generated by any phenomena that are not included in the timing model. Ideally this would be only radiometer noise (and GWs of course!), but there are many other sources of noise and uncertainty in pulsar-timing observations. For example, some pulsars are known to exhibit rotational irregularities. As mentioned earlier, discrete jumps in the rotational frequency of the pulsar (glitches) are thought to occur as a result of the sudden recoupling and angular momentum transfer between the neutron-superfluid and the crustal lattice, reducing the lag in their rotational frequencies which occurs due to the minimal friction between the two \citep{anderson1975pulsar}. This glitchy behaviour is suppressed in older and millisecond pulsars \cite{espinoza2011study,hobbs2010analysis}, so it is less of a concern for PTA GW searches. More relevant however is the fact that many pulsars exhibit timing noise with low-frequency structure ({\it red timing noise}). The origin of this may be due to the pulsar's magnetosphere rapidly and sporadically switching between stable configurations, leading to different pulse shapes and spindown rates \citep{lyne2010switched}. The variation in spindown rate causes the rotational frequency to wander over a period of years, contributing a source of red timing noise if unmodelled. While magnetospheric mode switching/nulling is not incorporated into the timing model, it can be accounted for as an extra red stochastic process \cite{2010ApJ...725.1607S}. Likewise, time-varying electron densities along the line-of-sight to each pulsar can result in time-dependent dispersion measure that can manifest as a radio-frequency dependent red stochastic process in the timing residuals \citep[e.g.,][and references therein]{2014MNRAS.441.2831L,2013MNRAS.429.2161K}. Further details of these and other noise sources, as well as the respective approaches we adopt to model them, are given in Chapter \ref{chap:pta_likelihood}. Ultimately, the target of our PTA searches is an effect that is not included in the pulsar timing model, specifically because this target phenomena influences all monitored pulsars in a correlated fashion. This target is of course the timing deviations induced by GWs.

The following sections describe the timing response of a pulsar to the influence of a GW signal, the inter-pulsar correlation pattern that we hunt for in searches for a stochastic background of GWs, and also additional sources of timing errors that could produce an apparent inter-pulsar correlated signal.

\section{Timing Response To Gravitational Waves} \label{sec:gw_timing}

We exploit the precision timing of millisecond pulsars to directly search for GWs, treating the pulsar and the SSB respectively as opposite ends of our laboratory setup. A passing GW perturbs the spacetime metric along the Earth-pulsar line of sight \citep{sazhin-1978,detweiler-1979,estabrook-1975,burke-1975}, deforming the proper separation, and thereby inducing irregularities in the perceived pulsar rotational frequency. We provide a derivation of the pulsar timing response due to a transiting GW below; this closely follows the treatment in Maggiore, Volume 2 \citep{maggiore2018gravitational}. We use the following line element for our GW spacetime:
\begin{equation}
    ds^2 = -dt^2 + [\delta_{ab} + h_{ab}^\mathrm{TT}(t, \vec{x})dx^a dx^b].
\end{equation}
where we follow the convention that Roman indices $a,b$ denote spatial components of the metric, while $i,j$ denote different pulsars. For a photon path travelling along the $x$-axis toward an observer at the origin, we have that $ds^2=0$ and thus
\begin{equation}
    dx \approx -\left\{ 1- \frac{1}{2}h_{xx}^\mathrm{TT}[t,\vec{x}(t)] \right\} dt.
\end{equation}
Integrating both sides for an Earth-pulsar coordinate separation of $L$, we have
\begin{equation}
    L = t_\mathrm{obs}-t_\mathrm{em} - \frac{1}{2}\int_{t_\mathrm{em}}^{t_\mathrm{obs}}dt'\,h_{xx}^\mathrm{TT}[t',\vec{x}(t')].
\end{equation}
Given that $h_{xx}^\mathrm{TT}$ is a small quantity, we are permitted to use $t_\mathrm{obs}\approx t_\mathrm{em}+L$ in the integral, with the photon path being approximately along its unperturbed trajectory $\vec{x}(t) = (t_\mathrm{obs}-t)\hat{p}$. We can also generalize to an arbitrary pulsar location by replacing $h_{xx}^\mathrm{TT}$ with $p^a p^b h_{ab}^\mathrm{TT}$. Thus
\begin{equation} \label{eq:gw_tobs}
    t_\mathrm{obs} = t_\mathrm{em} + L + \frac{p^a p^b}{2}\int_{t_\mathrm{em}}^{t_\mathrm{em}+L}dt'\,h_{ab}^\mathrm{TT}[t',(t_\mathrm{em}+L-t')\hat{p}].
\end{equation}
We now consider the arrival time of a subsequent pulse emitted after one rotational period of the pulsar, such that $t'_\mathrm{em} = t_\mathrm{em}+P$. The observed arrival time of this second pulse is simply given by substituting $t_\mathrm{em}\mapsto t_\mathrm{em}+L$ in Eq.~\ref{eq:gw_tobs}, and subtracting to get
\begin{equation}
    t'_\mathrm{obs} - t_\mathrm{obs} = P + \Delta P,
\end{equation}
where
\begin{equation}
    \Delta P = \frac{p^a p^b}{2}\int_{t_\mathrm{em}}^{t_\mathrm{em}+L}dt'\, \left\{ h_{ab}^\mathrm{TT}[t'+T,\vec{x}_0(t')] - h_{ab}^\mathrm{TT}[t',\vec{x}_0(t')] \right\},
\end{equation}
where $\vec{x}_0(t') = (t_\mathrm{em}+L-t')\hat{p}$. The observed arrival time difference between two subsequent pulses is thus equal to the spin period of the pulsar, plus an extra GW-induced term. The spin period of the pulsars are $\sim$~milliseconds, whereas the GW periods of interest span months to decades. Hence the first integrand term inside the curly brackets can be Taylor expanded to first order, leaving
\begin{equation} \label{eq:deltaP}
    \frac{\Delta P}{P} = \frac{p^a p^b}{2}\int_{t_\mathrm{em}}^{t_\mathrm{em}+L}dt'\, \left[ \frac{\partial}{\partial t'} h_{ab}^\mathrm{TT}(t',\vec{x})\right]_{\vec{x}=\vec{x}_0(t')}.
\end{equation}

Let us now consider a fiducial monochromatic wave solution propagating along the direction $\hat\Omega$:
\begin{equation}
    h_{ab}^\mathrm{TT}(t',\vec{x}) = \mathcal{A}_{ab}(\hat\Omega)\cos\left[\omega_\mathrm{GW}(t-\hat\Omega\cdot\vec{x}) \right].
\end{equation}
Upon substituting into Eq.~\ref{eq:deltaP} we get
\begin{align}
    \frac{\Delta P}{P} &= \frac{1}{2}\frac{p^a p^b \mathcal{A}_{ab}}{(1+\hat\Omega\cdot\hat{p})} \left\{ \cos\left[\omega_\mathrm{GW}t_\mathrm{obs}\right] - \cos\left[\omega_\mathrm{GW}t_\mathrm{em} - \omega_\mathrm{GW}(t_\mathrm{obs}-t_\mathrm{em})\hat\Omega\cdot\hat{p}\right] \right\} \nonumber\\
    &= \frac{1}{2}\frac{p^a p^b \mathcal{A}_{ab}}{(1+\hat\Omega\cdot\hat{p})} \left\{ \cos\left[\omega_\mathrm{GW}t_\mathrm{obs}\right] - \cos\left[\omega_\mathrm{GW}(t_\mathrm{em} - L\,\hat\Omega\cdot\hat{p})\right] \right\}.
\end{align}

We now define the GW-induced redshift of the pulse arrival rate as $z(t) \equiv (\nu_0 - \nu(t)) / \nu_0 = -(\Delta\nu/\nu) = \Delta P / P$. From the previous equation, and with a GW propagating in direction $\hat\Omega$, this can be written as
\begin{align} \label{eq:zredshift}
    z(t,\hat\Omega) &= \frac{1}{2}\frac{p^a p^b}{(1+\hat\Omega\cdot\hat{p})} \left[ h_{ab}(t,\vec{x}_\mathrm{earth}) - h_{ab}(t-L,\vec{x}_\mathrm{pulsar}) \right] \nonumber\\
    &= \frac{1}{2}\frac{p^a p^b}{(1+\hat\Omega\cdot\hat{p})} \Delta h_{ab}
\end{align}
where $t=t_\mathrm{obs}$, the position vector of the pulsar is $\vec{x}_\mathrm{pulsar}=L\hat{p}$, and the position vector of the Earth (or more precisely, the Solar System Barycenter) is $\vec{x}_\mathrm{earth}=0$. Thus the GW imparts two redshifting signatures on the pulse arrival times: an imprint of the metric perturbation as the wave washes over the Earth (the so-called \textit{Earth term}), and an imprint of the retarded metric perturbation from when the wave washed over the pulsar (the \textit{pulsar term}). The time difference between the pulsar term and Earth term is of order the light travel time of the Earth-pulsar distance, which for pulsars at $\sim$~kiloparsec distances can be thousands of years. This presents a very exciting possibility, since the imprinted GW signature carries a measure of the emitting source from thousands of years in the past. When compounded over many timed pulsars, this essentially creates a form of temporal aperture synthesis to allow the source evolution to be tracked over baselines much larger than our timing campaigns.

Given Eq.~\ref{eq:metricTTgw}, and assuming that the amplitude of the metric perturbation at the time of passing the pulsar and Earth is unchanged, we can write $\Delta h_{ab}$ as
\begin{equation} \label{eq:deltahij}
    \Delta h_{ab} = \int_{-\infty}^\infty df\, \left[e^{2\pi ift} \left(e^{-2\pi ifL(1+\hat\Omega\cdot\hat{p})}-1\right) \times\sum_A h_A(f,\hat\Omega)e^A_{ab}(\hat\Omega)\right],
\end{equation}
which has as its Fourier transform
\begin{equation}
    \Delta \tilde{h}_{ab}(f,\hat\Omega) = \left(e^{-2\pi ifL(1+\hat\Omega\cdot\hat{p})}-1\right) \sum_A h_A(f,\hat\Omega)e^A_{ab}(\hat\Omega).
\end{equation}
We will later be interested in the Fourier transform of $z(t)$ in Eq.~\ref{eq:zredshift}, which is thus
\begin{equation} \label{eq:ztilde}
    \tilde{z}(f,\hat\Omega) = \left(e^{-2\pi ifL(1+\hat\Omega\cdot\hat{p})}-1\right) \sum_A h_A(f,\hat\Omega)F^A(\hat\Omega),
\end{equation}
where $F^A(\hat\Omega)$ is the $A^\mathrm{th}$-mode GW antenna response pattern (see Eq.~\ref{eq:antennaresponse}) for an Earth-pulsar system, defined as
\begin{equation} \label{eq:pta_antenna}
    F^A(\hat\Omega) = \frac{1}{2}\frac{p^a p^b}{(1+\hat\Omega\cdot\hat{p})} e^A_{ab}(\hat\Omega). 
\end{equation}

The GW-induced timing perturbations measured with respect to a reference time $t=0$ are defined as
\begin{align} \label{eq:Rt}
    R(t) \equiv \int_0^t dt'\, z(t') &= \frac{1}{2}\frac{p^a p^b}{(1+\hat\Omega\cdot\hat{p})} \int_0^t dt'\, \left[ h_{ab}(t,\vec{x}_\mathrm{earth}) - h_{ab}(t-L,\vec{x}_\mathrm{pulsar}) \right] \nonumber\\
    &= R(t)_\mathrm{earth} - R(t)_\mathrm{pulsar}. 
\end{align}

\section{Overlap Reduction Function For A Background Of Gravitational Waves}

PTAs target a stochastic background of gravitational waves as the expected primary signal class. More details of why this is will be given in the next chapter, but suffice it to say that the GWs we are sensitive to are very low frequency, weakly evolving, and our detector's frequency resolution is also fairly limited. If the GW signal is composed of discrete systems, then their respective signals will likely pile up within each frequency resolution bin, defying our ability to full resolve them. Hence, in a practical search scenario, we need to understand how much this GWB will induce correlated power in the timing deviations between pulsars that are widely separated across the sky. We thus need to compute the PTA overlap reduction function (ORF) for a GWB. We draw from Section \ref{sec:gw_timing} and Section \ref{sec:orf} to compute this. The derivation has been given several times in the literature \citep[e.g.,][]{2009PhRvD..79h4030A,2013PhRvD..88f2005M,2014PhRvD..90h2001G,2015AmJPh..83..635J}, but we align most closely with that given in Maggiore, Volume 2 \citep{maggiore2018gravitational}. An important caveat is that this derivation is for an \textit{isotropic} GWB, with unifrom distribution of power across the sky. Further discussions of anisotropy and modifications for alternative GW polarizations are given towards the end of this section. 

We recall that Eq.~\ref{eq:ztilde} gives the GW-induced redshift of the pulse arrival rate in the Fourier domain. If we take the integral of this over the entire sky to account for GWs coming from any direction, and compute the expectation of the inter-pulsar correlation of these redshifts over many random SGWB realizations, we get
\begin{align}
    \langle \tilde{z}_i(f) \tilde{z}_j^*(f')\rangle =& \int_{S^2}\int_{S'^2}\,d^2\hat\Omega\, d^2\hat\Omega'\, \left[e^{-2\pi ifL_i(1+\hat\Omega\cdot\hat{p}_i)}-1\right]\left[e^{2\pi if'L_j(1+\hat\Omega'\cdot\hat{p}_j)}-1\right] \nonumber\\
    & \times\left\langle\sum_A h_A(f,\hat\Omega)F^A(\hat\Omega)\sum_{A'} h_{A'}^*(f',\hat\Omega')F^{A'}(\hat\Omega')\right\rangle. 
\end{align}
We can now invoke the assumption of a Gaussian, stationary, isotropic, and unpolarized SGWB using Eq.~\ref{eq:fourier-variance}, which reduces the above equation to
\begin{align}
    \langle \tilde{z}_i(f) \tilde{z}_j^*(f')\rangle &= \frac{1}{2}\delta(f-f')S_s(f)_{ij} \nonumber\\
    &= \frac{1}{2}\delta(f-f')S_h(f)\int_{S^2}\,\frac{d^2\hat\Omega}{8\pi}\, \kappa_{ij}(f,\hat\Omega)\sum_{A=+,\times} F^A_i(\hat\Omega)F^A_j(\hat\Omega),
\end{align}
where
\begin{equation}
    \kappa_{ij}(f,\hat\Omega) = \left[e^{-2\pi ifL_i(1+\hat\Omega\cdot\hat{p}_i)}-1\right]\left[e^{2\pi ifL_j(1+\hat\Omega\cdot\hat{p}_j)}-1\right].
\end{equation}
Thus the cross-power spectral density of the measured redshift to the pulse arrival rate is
\begin{equation} \label{eq:pta_crosspsd}
    S_s(f)_{ij} = \frac{1}{2}S_h(f)\int_{S^2}\,\frac{d^2\hat\Omega}{4\pi}\, \kappa_{ij}(f,\hat\Omega)\sum_{A=+,\times} F^A_i(\hat\Omega)F^A_j(\hat\Omega),
\end{equation}
with units of $[\mathrm{time}]$. It's worth discussing $\kappa_{ij}(f,\hat\Omega)$, which controls how rapidly the pulsar term of the measured signal spatially decorrelates. Even the closest known millisecond pulsars are $>100$~parsecs distant from us, while the minimum GW frequency that we can probe with current IPTA datasets is $\sim10^{-9}$~Hz, which still leaves $fL>10$. The complex exponentials are thus rapidly oscillating terms that contribute negligibly to the integral above, except in the case where the pulsars are \textit{identical} (i.e., same distance and sky location). Hence, $\kappa_{ij}(f,\hat\Omega) \rightarrow 2$ when $i=j$, and $\kappa_{ij}(f,\hat\Omega) \rightarrow 1$ otherwise. See Mingarelli \& Sidery (2014) \citep{2014PhRvD..90f2011M} for a complete discussion of this aspect. Bearing this in mind for later, we consider $i\neq j$ in the following.

\begin{figure} 
    \centering
	\includegraphics[width=1.0\columnwidth]{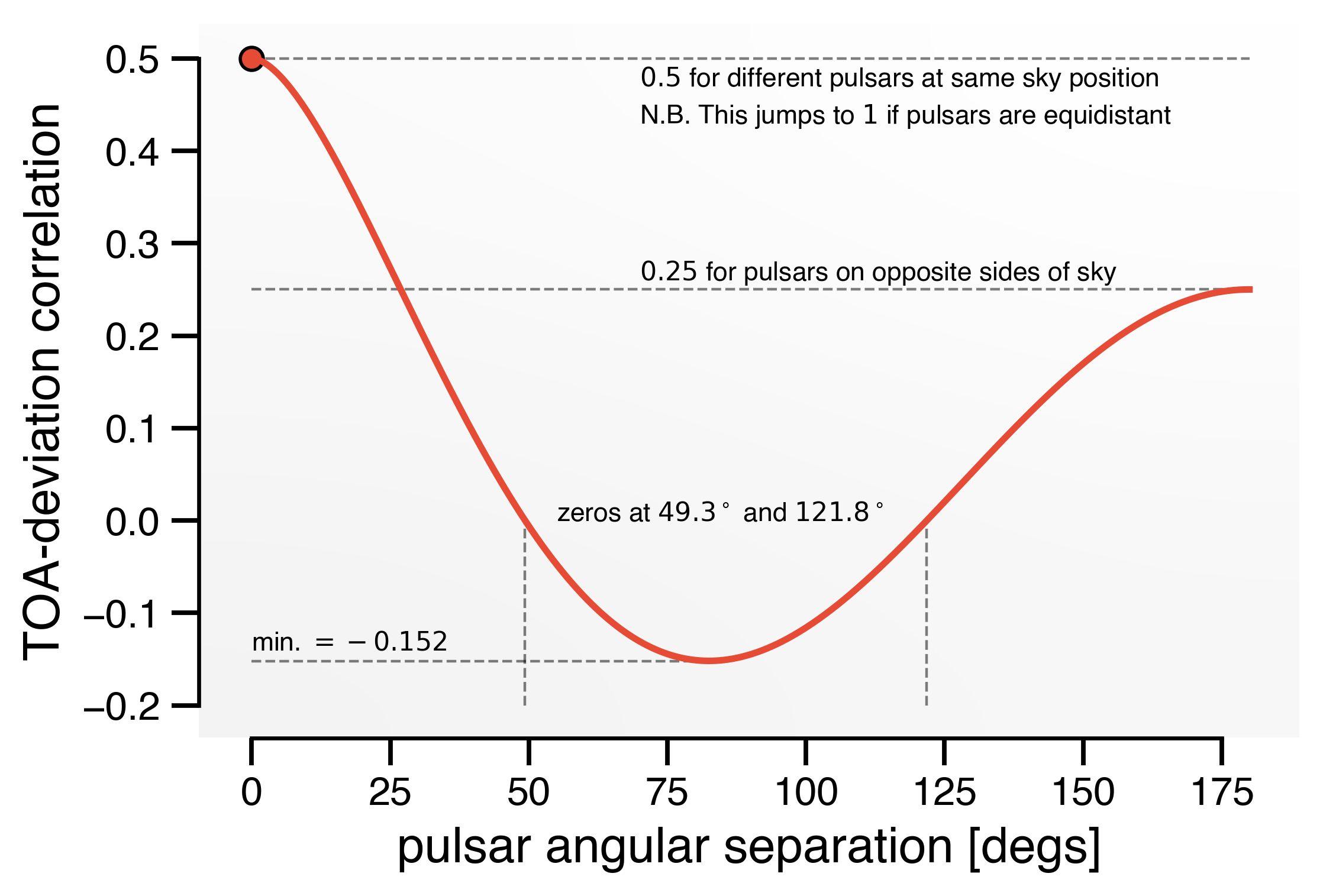}
    \caption{The normalized overlap reduction function for an isotropic SGWB in PTAs, more commonly referred to as \textit{The Hellings \& Downs Curve}, since it was first shown in Hellings \& Downs (1983) \citep{1983ApJ...265L..39H}. Some instructive features of the curve are labeled.}
    \label{fig:hellings_and_downs}
\end{figure}

\begin{figure} 
    \centering
	\includegraphics[width=1.0\columnwidth]{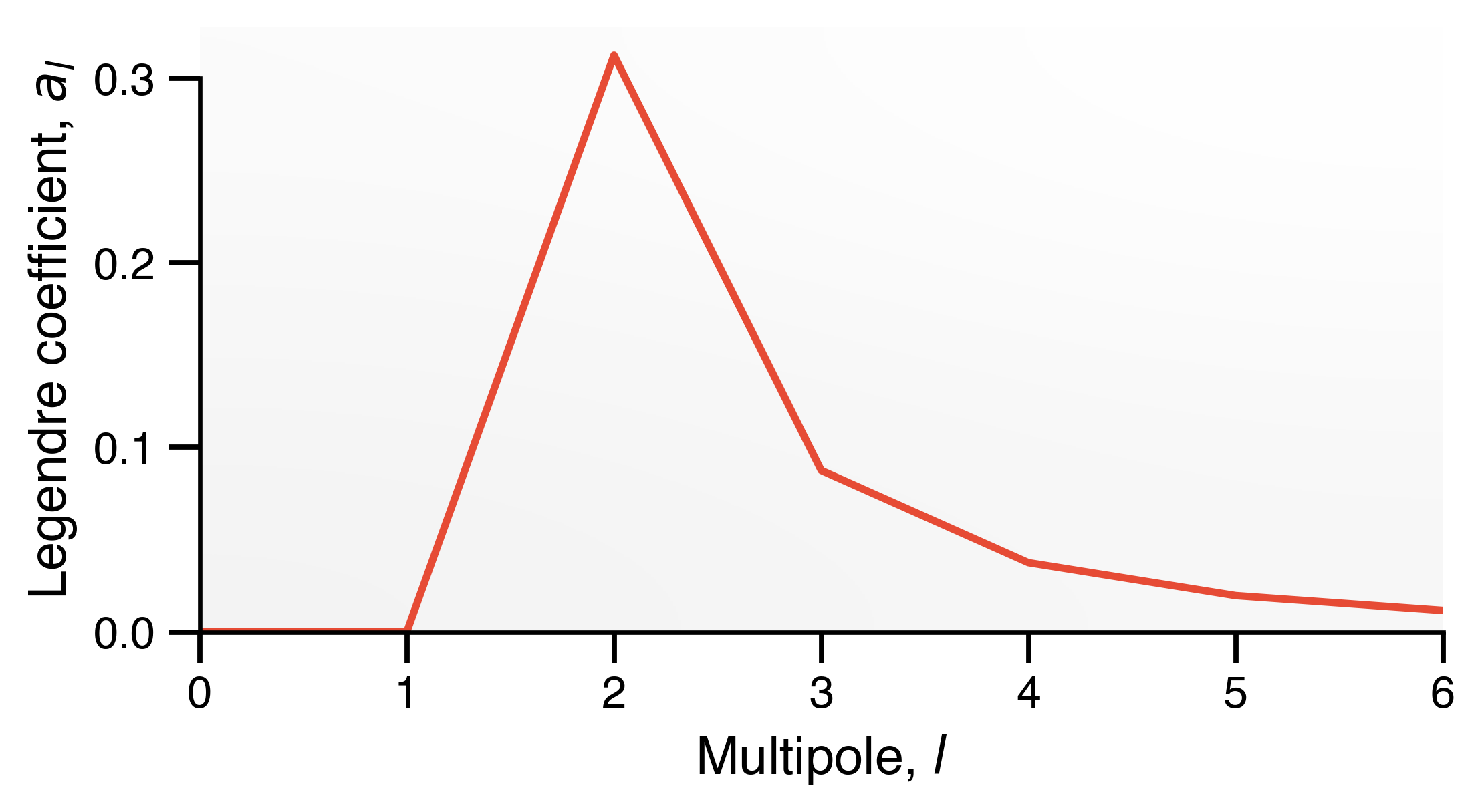}
    \caption{Legendre polynomial spectrum of the Hellings \& Downs curve \citep{2014PhRvD..90h2001G,2019ApJ...876...55R,2017ApJ...835...21R}. $63\%$ of the power is contained in $l=2$, while $80\%$ is contained within $l=2$ and $l=3$.}
    \label{fig:hd_legendre}
\end{figure}

Comparing Eq.~\ref{eq:pta_crosspsd} with Eq.~\ref{eq:signal_mode_psd_compare}, we see that our goal is to calculate the un-normalized ORF, $\tilde\Gamma_{ij}(f)$, for PTAs, which under our approximation for $\kappa_{ij}(f,\hat\Omega)$, is frequency independent. While elementary, any which way you do this integral is going to be tedious. I recommend following the detailed derivations cited in the initial paragraph above. The result is
\begin{align}
    \tilde\Gamma_{ij} &= \int_{S^2}\frac{d^2\Omega_{\hat{n}}}{4\pi} \sum_{A=+,\times} F^A_i(\hat{n})F^A_j(\hat{n}) \nonumber\\
    &= x_{ij}\ln(x_{ij}) - \frac{1}{6}x_{ij} + \frac{1}{3},
\end{align}
where $x_{ij} = (1-\cos(\theta_{ij}))/2$, and $\theta_{ij}$ is the angular separation between the position of pulsars on the sky. This was the initial calculation and normalization presented by Hellings \& Downs \citep{1983ApJ...265L..39H} (although it was initially presented in that paper without derivation steps). In the PTA literature, you'll more often see the general expression that accounts for $i=j$, and normalizes the ORF such that $\Gamma_{ij}=1$ for $i=j$. This expression is
\begin{equation} \label{eq:handd}
    \Gamma_{ij} = \frac{3}{2}x_{ij}\ln(x_{ij}) - \frac{1}{4}x_{ij} + \frac{1}{2} + \frac{1}{2}\delta_{ij},
\end{equation}
where $\delta_{ij}$ is the Kronecker delta function. This function is shown in Fig.~\ref{fig:hellings_and_downs} along with some notable features. We get maximal values in the pulsar autocorrelations. Even pulsars with very small angular separations will have a cross-correlation no greater than $0.5$, since the pulsar terms decorrelate rapidly once they are spatially separated by approximately more than a GW wavelength. The Hellings \& Downs curve exhibits a strong quadrupolar trend over angular separation as a result of the quadrupolar GW antenna response patterns of the Earth-pulsar systems. However it is not a pure quadrupole; the fact that the curve only returns to $0.25$ at $180^\circ$ instead of $0.5$ is evidence of that. This is because the denominator in the antenna response pattern (see Eq.~\ref{eq:pta_antenna}) introduces a preferred direction, where the response is largest to GWs propagating parallel to the radio pulses travelling from the pulsar to Earth. In fact, we can perform a decomposition of the Hellings \& Downs curve in terms of Legendre polynomials to inspect the power in different multipoles. The result is \citep{2014PhRvD..90h2001G,2019ApJ...876...55R,2017ApJ...835...21R}
\begin{equation}
    \Gamma_{ij} = \sum_{l=0}^\infty a_l P_l(\cos\theta_{ij}),
\end{equation}
where $i\neq j$, $a_0 = 0 = a_1$, and 
\begin{equation}
    a_l = \frac{3}{2}\frac{(l-2)!}{(l+2)!}(2l+1).
\end{equation}
Implicit in this decomposition is that the distribution of pulsars is isotropic across the sky, giving a distribution of pulsar angular separations that is  $\propto\sin\theta_{ij}$. The Legendre spectrum for the Hellings \& Downs curve is shown in Fig.~\ref{fig:hd_legendre}, with all power contained in $l\geq2$, meaning that the Hellings \& Downs curve is orthogonal to monopole and dipole inter-pulsar correlations in the limit of infinite precision data and uniform pulsar sky coverage. In fact, as expected, the quadrupole dominates such that $a_2 / \sum_{l=0}^\infty a_l = 0.63$, and the octupole adds another $0.17$; hence $80\%$ of the Hellings \& Downs curve can be described in terms of $l=2$ and $l=3$ Legendre polynomials.

The Hellings \& Downs curve is not the end of the story for cross-correlation SGWB searches with PTAs. As mentioned, it is only the ORF for an \textit{isotropic} SGWB, and also implicitly assumes GR as the correct theory of gravity through the transverse-tensor nature of the antenna response. Significant work has been done to generalize the ORF calculation to anisotropic SGWBs \citep{2013PhRvD..88f2005M,2013PhRvD..88h4001T,2015PhRvL.115d1101T,2014PhRvD..90h2001G}, allowing the angular structure to be probed using techniques similar to the Cosmic Microwave Background \citep{2014PhRvD..90h2001G,2019MNRAS.487..562C,2019PhRvD.100j3528H}, and for more bespoke methods to be developed that target discrete systems constituting the background signal \citep{2020PhRvD.102h4039T,2013CQGra..30v4005C}. An anisotropic ORF is no longer merely a function of the angular separation of pulsars on the sky, since the expected signal cross-correlation will also be dependent on the pulsar positions with respect to the distribution of angular GW power. Likewise, in a general metric theory of gravity there are a total of six allowed GW polarization states: two transverse-tensor (TT) GR states (TT$_+$, TT$_\times$), one scalar transverse state (ST), one scalar longitudinal state (SL), and two vector longitudinal (VL) states (VL$_x$, VL$_y$). The presence of these alternative polarization states modifies the ORF, and in the cases of the longitudinal modes make the ORF unavoidably dependent on the pulsar distances and GW frequency \citep{2015PhRvD..92j2003G,2018PhRvD..98j4025I,2018PhRvL.120r1101C,2012PhRvD..85h2001C,2008ApJ...685.1304L}. It is also possible to show that a massive graviton will distort the expected Hellings \& Downs curve through its effect on the graviton dispersion relation \citep{2020arXiv200711009Q,2010ApJ...722.1589L}.

Given Eq.~\ref{eq:handd}, the cross-power spectral density of the redshift to the pulse arrival rate can finally be written as
\begin{equation}
    S_s(f)_{ij} = \frac{1}{3}\Gamma_{ij}S_h(f).
\end{equation}

However, we don't measure shifted pulse arrival rates. We measure pulse arrival times, and their deviations away from modeled predictions. As shown in Eq.~\ref{eq:Rt}, the GW-induced timing perturbation is an accumulating shift over time. This will simply result in factors of $1/(2\pi if)$ (and the conjugate) when the $\Delta h_{ij}$ factor of Eq.~\ref{eq:deltahij} is integrated over time, and hence a factor of $1/4\pi^2f^2$ when these timing perturbations are correlated. Thus the cross-power spectral density of the GW-induced timing deviations is
\begin{equation}
    S_t(f)_{ij} = \frac{S_s(f)_{ij}}{4\pi^2f^2} = \Gamma_{ij}\frac{S_h(f)}{12\pi^2f^2} = \Gamma_{ij}\frac{h_c^2(f)}{12\pi^2f^3},
\end{equation}
which has units of $[\mathrm{time}]^3$, and where we have used the previously introduced relationship $S_h(f) = h_c^2(f)/f$. We can convert this process from the Fourier domain into the time domain using the \textit{Wiener-Khinchin Theorem}:
\begin{equation}
    C(\tau)_{ij} = \int_0^\infty df\, S_t(f)_{ij}\cos\left(2\pi f\tau \right),
\end{equation}
which gives the covariance of GW-induced timing deviations between pulsars $i$ and $j$ over a lag time between observations, $\tau$. This covariance has the expected units of $[\mathrm{time}]^2$.

As a final note, the SGWB is not the only process that can induce low-frequency inter-pulsar correlated timing deviations in PTA data. We will see in the next chapter that systematic errors in our model of the Solar System ephemeris could induce dipolar correlated timing deviations through incorrect TOA barycentering. Likewise, long timescale drifts in global timing standards would affect all pulsars equally, inducing monopole correlated timing deviations. Fortunately, since the Hellings \& Downs curve is orthogonal to these correlations, with sufficient pulsar sky coverage and data precision we can isolate our SGWB searches from these systematics.

\bibliographystyle{unsrt_new}
\bibliography{refs}

%% file: 04.tex
\chapter{Sources \& Signals}\label{chap:sources}
\epigraph{\textit{``Fortune and glory, kid. Fortune and glory\ldots''}}{Henry (Indiana) Jones Jr., ``Indiana Jones \& The Temple Of Doom''}

As in other portions of the GW spectrum, the overwhelming majority of signals in the PTA band are likely to be of compact-binary origin. Hence, I'll devote the bulk of this chapter to the dominant expected class of sources for PTAs: binary systems of supermassive black holes with masses $\sim10^8-10^{10} M_\odot$. While these systems should be plentiful, teasing them apart from one another will be challenging to the PTA detector response and frequency resolution, rendering our first target to be the statistical signal aggregation over the entire population in the form of a stochastic GW background. The properties of this background encode demographic information of the binary systems, as well as details of their dynamical evolution in the final parsec of their journey to coalescence. Nevertheless template-based searches for individual binaries are possible, and appropriate waveform models are given here. Beyond binaries, I will give a brief overview of other potential PTA sources, including primordial GWs, GWs from cosmic strings and phase transitions, and finally other non-GW sources of inter-pulsar correlated timing delays in the form of timing systematics and even scalar-field dark matter.

\section{Supermassive binary black holes}

There is now general agreement that massive black holes (MBHs) reside at the centers of most galaxies \citep{kormendy1995, Magorrian1998}, with several well-known scaling relations indicating a symbiosis with the properties of the host galaxy (e.g.\ $M$--$\sigma$, $M$--$M_\mathrm{bulge}$ \citep{2002ApJ...578...90F,2009ApJ...698..198G,Kormendy2013,2013ApJ...764..184M,2019ApJ...887..245S}). But conventional astronomical techniques are limited to studying them either locally or in quasar environments. The formation of MBH binaries should be a natural by-product of the hierarchical growth of galaxies through major and minor mergers (a process that also includes gas and dark matter accretion from cosmic web filaments) in $\Lambda$CDM cosmologies \citep{1978MNRAS.183..341W}. 

Indeed such MBH binaries (MBHBs) should be among the loudest GW sources in the Universe, and in the mass range $\sim 10^8-10^{10} M_\odot$ (referred to as \textit{super}-massive black hole binaries; SMBHBs) constitute a key target population for PTAs. This requires the SMBHs to reach very close separations (on the order of milliparsecs) such that their GW emission would fall within the sensitive frequency range of PTAs ($\sim 1-100$~nHz). The chain of interactions that govern the inward migration of these black holes is discussed later in this chapter. The existing observational electromagnetic evidence for such sub-parsec separation SMBHBs is tenuous. There are many known ``dual'' supermassive black hole systems separated by kiloparsecs to hundreds of parsecs that are observed across X-ray, optical, and radio \citep[see, e.g.,][and references therein for a comprehensive review]{2019NewAR..8601525D}. The closest known directly resolved binary is at a projected separation of $\sim 7$~parsecs \citep{rodriguez2006compact}. Below a parsec, the evidence becomes indirect since telescopes lack the requisite spatial resolution to separate the binary components. Binary candidates are inferred through either apparent periodic photometric variability of AGN lightcurves \citep[e.g.,]{2015MNRAS.453.1562G,2016MNRAS.463.2145C,2016ApJ...833....6L,2020MNRAS.499.2245C}, or large long-timescale velocity offsets in spectroscopically-derived AGN radial velocity curves \citep[e.g.,][]{2012ApJS..201...23E,2015ApJS..221....7R,2014ApJ...789..140L,2017MNRAS.468.1683R,2019MNRAS.482.3288G}. Both are interpreted as effects of binarity through the interplay of gas and accretion disks surrounding each black hole and the binary as a whole.

\subsection{Characteristic strain spectrum}

We saw in Chapter \ref{chap:gravity_and_gws} that the energy density in GWs is defined as
\begin{equation}
    \Omega_{\rm SGWB}(f) \equiv \frac{1}{\rho_c}\frac{d\rho}{d\ln f}.
\end{equation}
where $f$ is the GW frequency, $\rho$ is the GW energy density, and $\rho_c$ is the closure density that corresponds to flat cosmological geometry. I will first demonstrate a simple back-of-the-envelope scaling relationship for the GWB energy density in a population of circular inspiraling compact binary systems, which can then be trivially converted into a characteristic strain spectrum. While this derivation is specifically in the context of supermassive binary black holes (the major subject of this chapter), the derivation applies to any compact binary population. We write the term $d\rho/d\ln f$ as an integral over a continuous distribution of emitting sources:
\begin{equation}
    \frac{d\rho}{d\ln f} = \int_0^\infty dz \frac{dn}{dz}\frac{1}{(1+z)} \frac{dE}{d\ln f_r}\bigg|_{f_r = f(1+z)}
\end{equation}
where $dn/dz$ is the number density distribution of binaries over redshift, the factor of $1/(1+z)$ accounts for the cosmological redshifting of GW energy, and $dE/d\ln f_r$ is the source-frame GW energy emitted by the binary per logarithmic frequency interval. The interpretation of this is simple: within a given logarithmic frequency interval, we are adding together the energy from all binaries throughout the Universe whose emission results in source-frame frequencies redshifted into the relevant observed frequency range. By using scaling relations given in Chapter \ref{chap:quadrupole_formula}, we can write the GW energy spectrum as
\begin{equation}
    \frac{dE}{d\ln f_r} = f_r\frac{dE}{dt_r}\frac{dt_r}{df_r} \propto f_r \times f_r^{10/3} \times f_r^{-11/3} \propto f_r^{2/3}.
\end{equation}
The source-frame GW frequency is simply the redshift-corrected observed frequency, $f_r = f(1+z)$, and we have seen that the frequency of GW emission from a circular binary system is simply twice the orbital frequency. Plugging these scaling relationships into $d\rho/d\ln f$ and collecting frequency terms gives us $\Omega_{\rm SGWB}(f) \equiv (1/\rho_c)(d\rho/d\ln f) \propto f^{2/3}$. Finally, using Eq.~\ref{eq:omega_hc} we can write that the characteristic strain spectrum from a population of circular inspiraling compact binaries is \citep{1995ApJ...446..543R,p01,2003ApJ...590..691W,2003ApJ...583..616J}
\begin{equation}
    h_c(f) \propto f^{-2/3}.
\end{equation}
As mentioned, in PTA searches for a SGWB the relevant compact binary population consists of supermassive black holes. Given that we are probing a GW frequency range of $\sim 1-100$~nHz, historical convention opts for a reference frequency of $f_\mathrm{ref}=1\mathrm{yr}^{-1}$,\footnote{PTAs actually have terrible sensitivity at this frequency because of the need to fit for the pulsar's sky location in the timing model, but this is ultimately irrelevant since most of the signal significance derives from the lowest frequencies in the band regardless of what the model's reference frequency is.} such that
\begin{equation}
    h_c(f) = A_\mathrm{SGWB}\left(\frac{f}{1\mathrm{yr}^{-1}} \right)^\alpha
\end{equation}
where $\alpha=-2/3$ for our population of circular inspiraling supermassive binary black holes.

Let's now look more carefully at how we build up the characteristic strain spectrum from a population of SMBHBs. We will relax our previous assumption of circularity, and also retain all dependencies on the distribution of system properties. This discussion closely follows Ref.~\citep{2019A&ARv..27....5B} and references therein; we refer the reader there for further details. The squared characteristic strain spectrum can be written as
\begin{align} \label{eq:hc_spec}
    h_c^2(f) =& \int_0^\infty \int_0^\infty \int_0^1 dz dM_1 dq \frac{\mathrm{d}^4N}{\mathrm{d}z\mathrm{d}M_1\mathrm{d}q\mathrm{d}t_r}\times \nonumber\\
    & \quad\sum_{n=1}^\infty\left\{\frac{g[n,e(f_{K,r})]}{(n/2)^2} \frac{\mathrm{d}t_r}{\mathrm{d}\ln f_{K,r}} h^2(f_{K,r}) 
    \right\}, \\ 
\end{align}
where $M_1$ is the mass of the primary BH; $0 < q \leq 1$ is the binary mass ratio $(M_2/M_1)$; and $f_{K,r}$ is the source-frame Keplerian orbital frequency of the binary, where the GW emission from eccentric binaries will result in redshifted harmonics of this, $f_n = nf_{K,r}/(1+z)$. The other terms of the integrand split into population factors and individual system factors. The term $\mathrm{d}^4N / \mathrm{d}z\mathrm{d}M_1\mathrm{d}q\mathrm{d}t_r$ is the population weighting of each binary's contribution to the squared strain, corresponding to the comoving merger rate of SMBHBs per redshift, primary mass, and mass-ratio interval, where $t_r$ measures source-frame time. The remaining terms are all related to the contribution of each SMBHB to the squared strain. As noted above, the GW emission from an eccentric binary will be distributed over harmonics of the orbital frequency \citep{pm63}. We are thus summing up contributions of the GW emission from each binary into redshifted harmonics of their orbital frequency that coincide with the relevant observed GW frequency. The function $g(n,e)$ is an eccentricity-dependent function describing the distribution of GW emission into orbital-frequency harmonics \cite{pm63}. In the case of a circular binary with $e=0$, we have $g(n,e=0)=0$ for $n\neq 2$, resulting in our much simpler derivation above. The term $h(f_{K,r})$ is defined as
\begin{equation}
    h(f_{K,r}) = \sqrt{\frac{32}{5}}\frac{\mathcal{M}^{5/3}(2\pi f_{K,r})^{2/3}}{D_c},
\end{equation}
which corresponds to the orientation-averaged GW strain amplitude of a single SMBHB, where $\mathcal{M}:= (M_1 M_2)^{3/5} / (M_1+M_2)^{1/5}$ is known as the ``chirp mass'' (in the source frame), and $D_c$ is the radial comoving distance to the binary system.

Finally, the term $\mathrm{d}t_r / \mathrm{d}\ln f_{K,r}$ describes the amount of time that each SMBHB spends emitting in a particular logarithmic $f_{K,r}$ interval. If the binary is evolving purely through GW emission then this follows known relationships that account for enhancement in the rate of evolution due to eccentricity \citep{pm63,peters1964gravitational}. However, at wider separations the binaries may remain in contact with their ambient astrophysical environment, and their orbital evolution may be partially driven by these factors. Hence a more general description of this term sums over all processes that may be influencing the orbital evolution of a given SMBHB:
\begin{equation}
    \frac{dt_r}{d\ln f_{K,r}} = f_{K,r}\times\left[\sum_k \frac{df_{K,r}}{dt_r} \right]^{-1}.
\end{equation}

\subsection{Binary dynamical evolution}

\begin{figure}
    \centering
    \includegraphics[width=\columnwidth]{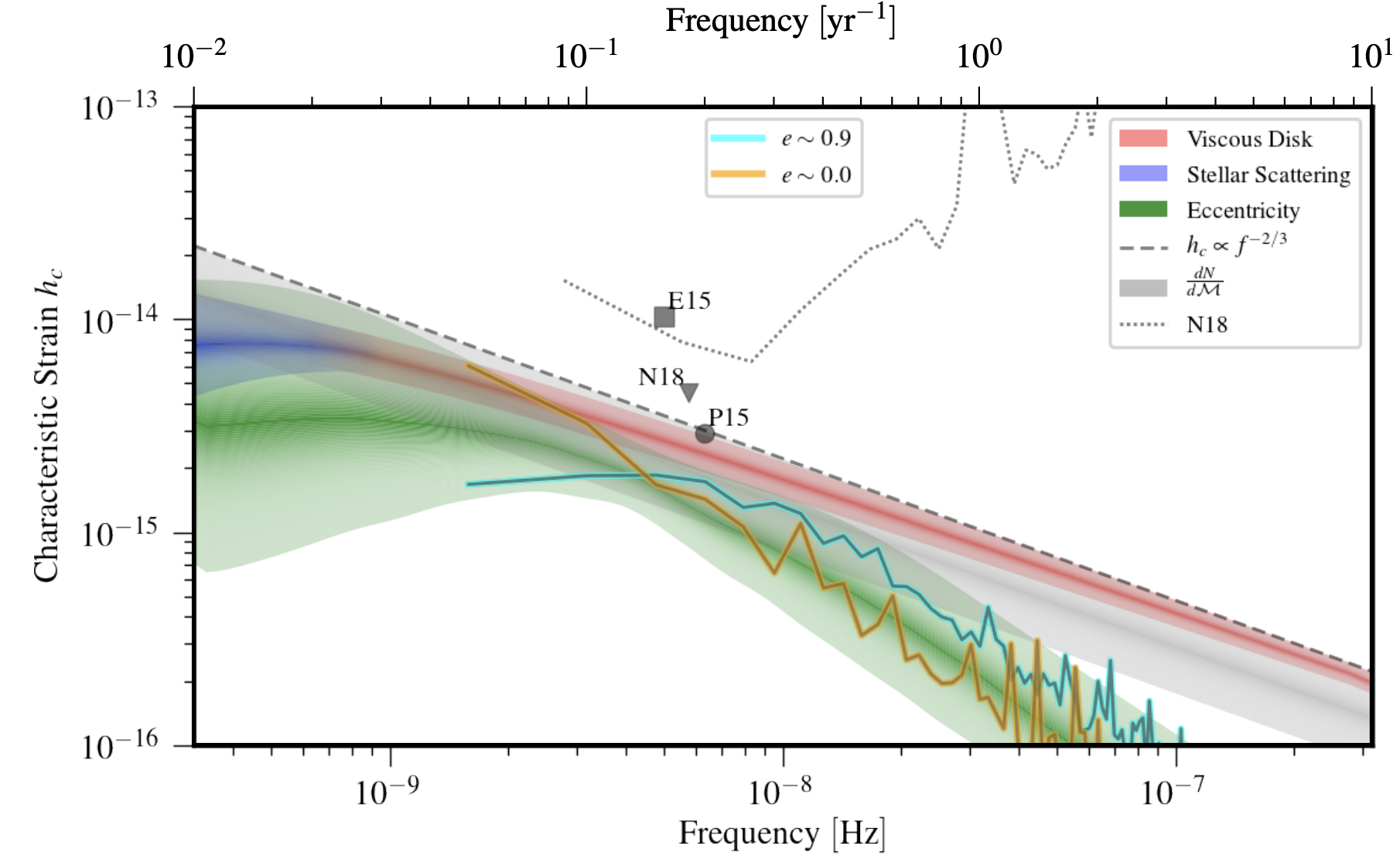}
    \caption{An example of the characteristic strain spectrum of a SGWB of SMBHB-origin, as a function of GW frequency in Hz (bottom axis) and inverse years (top axis). The grey dashed line shows the fiducial $f^{-2/3}$ spectrum, while the influences of dynamical interactions are shown as colored bands, for stellar loss-cone scattering (blue) and viscous disk coupling (red). The green band shows an ensemble of spectra from eccentric SMBHB populations, while the cyan and orange lines show individual realizations of high and low population-eccentricity spectra, respectively. Figure reproduced with permission from Ref.~\citep{2019A&ARv..27....5B} using populations based on Refs.~\citep{k+17,k+17a}.}
    \label{fig:all_gwb_shapes}
\end{figure}

The coalescence of two SMBHs occurs after a chain of dynamical interactions that begins with the merger of their host galaxies. Several of these interactions can imprint themselves on the SGWB characteristic strain spectrum through their influence on $dt_r / \ln f_{K,r}$, yielding an opportunity for PTAs to constrain the ensemble sub-parsec dynamical behavior of SMBHBs. The net result is that the strain spectrum can be attenuated at low frequencies and depart from the fiducial $f^{-2/3}$ behavior to give a \textit{turnover}, since these effects accelerate the evolution of the SMBHB orbit at wider separations. Ref.~\citep{Sampson2015} suggested a simple parametrization of the strain spectrum for a population of circular environmentally-influenced binaries:
\begin{equation} \label{eq:hc_bend}
    h_c(f) = A_\mathrm{SGWB} \frac{\left(f/1\mathrm{yr}^{-1}\right)^{-2/3}}{\left[1+\left(f_\mathrm{bend}/f\right)^\kappa\right]^{1/2}},
\end{equation}
where $f_\mathrm{bend}$ is the orbital frequency below which environmental interactions dominate the SMBHB evolution, and $\kappa$ is determined by the dominant dynamical interaction. We will see the values that $\kappa$ can take as each interaction is introduced below. As discussed later in Chapter~\ref{chap:pta_likelihood}, there are a variety of other astrophysically-driven spectral models of the SGWB in use \citep{2017MNRAS.468..404C,2017MNRAS.470.1738C,2019MNRAS.488..401C,2017PhRvL.118r1102T}. A range of possible SGWB shapes due to varying astrophysical conditions and dynamical interactions are shown in Fig.~\ref{fig:all_gwb_shapes}, along with some PTA strain upper limits for orientation.

\subsubsection{Dynamical friction} 

Dynamical friction is the term describing a viscous drag influence on SMBHs in the environment of a galactic merger. After this merger, the accumulated effect of many weak and long-range gravitational scattering events brings the ``dual'' system --- composed of the respective SMBHs and stellar cores --- from $\sim$kpc separations down to $\sim\!\!1$ parsec \citep{chandrasekhar1943,antonini2012,merritt2005,y02,k+17a,Dosopoulou2017}. For a SMBH spiraling toward the center of a spherical galaxy on a  circular orbit, the inspiral timescale is of the order of Gyrs \cite{BinneyTremaine,Dosopoulou2017}. Nevertheless, dynamical friction is an important initial step in reducing the separation of the two black holes from kiloparsecs. It does not yield a measurable influence on the shape of the SGWB strain spectrum.

\subsubsection{Stellar loss-cone scattering}

At parsec separations, the individual black holes are moving fast enough that dynamical friction no longer exerts an influential tightening drag. The dominant mechanism of binary hardening then results from individual scattering events between low angular-momentum stars in the galactic core and the SMBHB \cite{bbr80,fr76,mv92,q96}. The \textit{loss cone} (LC) \citep{fr76} describes the region of stellar-orbit phase space where orbits are centrophilic enough to scatter off the binary. The original definition of the \textit{final parsec problem} \cite{mm02,mm03} referred to the potential depletion of this phase-space reservoir, resulting in stalled SMBHBs. In practice the triaxiality and rotation of post-merger galaxies can ensure a healthy supply of such stars \citep{khb13,vm13,vam14,vam15}. The scattering and ejection of stars by the binary leads to hardening through the following semi-major axis and eccentricity evolution \cite{q96}:
\begin{equation} \label{eq:orbit-lc}
    \frac{{\rm d}a}{{\rm d}t} = -\frac{G\rho}{\sigma}H a^2, \quad \frac{{\rm d}e}{{\rm d}t} = \frac{G\rho}{\sigma} H K a,
\end{equation}
where $H\sim15$ is a dimensionless hardening rate, and $K\sim0.1$ is a dimensionless eccentricity \textit{growth} rate, i.e., stellar LC scattering can promote eccentricity growth. Both of these can be computed from numerical scattering experiments for specific mass ratio and eccentricity conditions \cite{shm06}. If we take stellar LC scattering to be the dominant term in $dt_r / \ln f_{K,r}$, then the resulting strain spectrum behavior is $h_c(f)\propto f$, which is clearly very different from the fiducial $\propto f^{-2/3}$ circular GW-driven behavior. This corresponds to $\kappa=10/3$ in \autoref{eq:hc_bend}. Eccentricity growth will further attenuate the strain spectrum at low frequencies \citep[e.g.][]{tss17,k+17}.

\subsubsection{Viscous circumbinary disk interaction}

At centiparsec to milliparsec separations, \textit{viscous dissipation of orbital energy to a gaseous circumbinary disk} could play a vital role in hardening the binary \citep{bbr80,ka11,ipp99,hbm09}. The binary torque will dominate over the viscous torque in the disk, leading to the formation of a cavity in the gas distribution and the accumulation of material at the outer edge of this cavity (i.e. Type II migration). The excitation of a spiral density wave in the disk torques the binary, and leads to hardening through the following semi-major axis evolution \cite{ipp99,hbm09}:
\begin{equation} \label{eq:orbit-disk}
\frac{{\rm d}a}{{\rm d}t} = -\frac{2\dot{M}_1}{\mu}(aa_0)^{1/2},
\end{equation}
where $\dot{M}_1$ is the mass accretion rate onto the primary BH, $\mu$ is the binary reduced mass, and $a_0$ is the semi-major axis at which the disk mass enclosed is equal to the mass of the secondary BH \citep{ipp99}.

As studied comprehensively in Ref.~\citep{ka11}, the strain spectrum under different disk--binary scenarios can vary from $h_c(f)\propto f^{-1/6}$ ($\kappa=1$ in \autoref{eq:hc_bend}), to $h_c(f)\propto f^{1/2}$ for the model in \autoref{eq:orbit-disk} ($\kappa=7/3$ in \autoref{eq:hc_bend}). Across all models, the characteristic strain spectrum can be flattened or even have positive slope due to disk coupling. There are many caveats to these simple spectral parametrizations, and the actual spectral shape will depend on the detailed dissipative physics of the disk, and the disk-binary dynamics. In fact, as of writing there is emerging evidence of viscous disk interaction actually \textit{widening} the binary separation (i.e., $da/dt$ has a positive sign) under certain conditions \citep[e.g.,][]{2019ApJ...875...66M,2021arXiv210309251D,2020A&A...641A..64H}.

\subsubsection{Gravitational-wave inspiral}

Once the binary decouples from its astrophysical environment at the smallest scales ($\lesssim$ milliparsec), the emission of gravitational radiation will dominate the orbital evolution. The dissipation of orbital energy then only depends on the binary component masses, the orbital semi-major axis, and the eccentricity. We have already seen that this leads to $h_c \propto f^{-2/3}$. However, this is an ensemble averaged behavior, and finiteness in a given realization of the emitting SMBHB population will lead to departures from this power-law spectrum at frequencies beyond $\sim 10^{-8}$ Hz, tilting it more steeply \cite{svc08}. 

But what happens if LC scattering and disk interactions don't bring the SMBHs close enough to actually coalesce within a Hubble time? What if all binaries stall before they can create a measurable signal in the relevant range of PTA-sensitive GW frequencies? A third mechanism may come to the rescue: \textit{triplet interactions}. Galaxies can undergo numerous merger events over cosmic time \citep[e.g.][]{r+15}, with the potential to form hierarchical MBH systems that undergo Kozai-Lidov oscillations \citep{k62,l62}, thereby driving up the eccentricity of the inner binary and accelerating coalescence due to GW emission \citep{me94,bls02,a+10,db17,Bonetti_2018,2018MNRAS.477.2599B,Ryu_2018}. Eccentricity itself can leave an imprint on the shape of the strain spectrum \citep{h+15,enn07,r+14,k+17,rm17b,tss17}, where its effect is to evolve each system more rapidly (thereby reducing the occupation fraction of frequency bins), and to distribute GW strain across higher orbital frequency harmonics. The result is that the spectrum can exhibit a flattening/turnover at low frequencies, a small enhancement at the turnover transition, and then a return to the usual $f^{-2/3}$ behavior at the highest frequencies (where the population will be mostly circularized).

\subsection{Signal from an individual binary}

\begin{figure}[t!]
    \centering
    \begin{subfigure}[t]{0.4\columnwidth}
        \centering
        \includegraphics[width=\columnwidth]{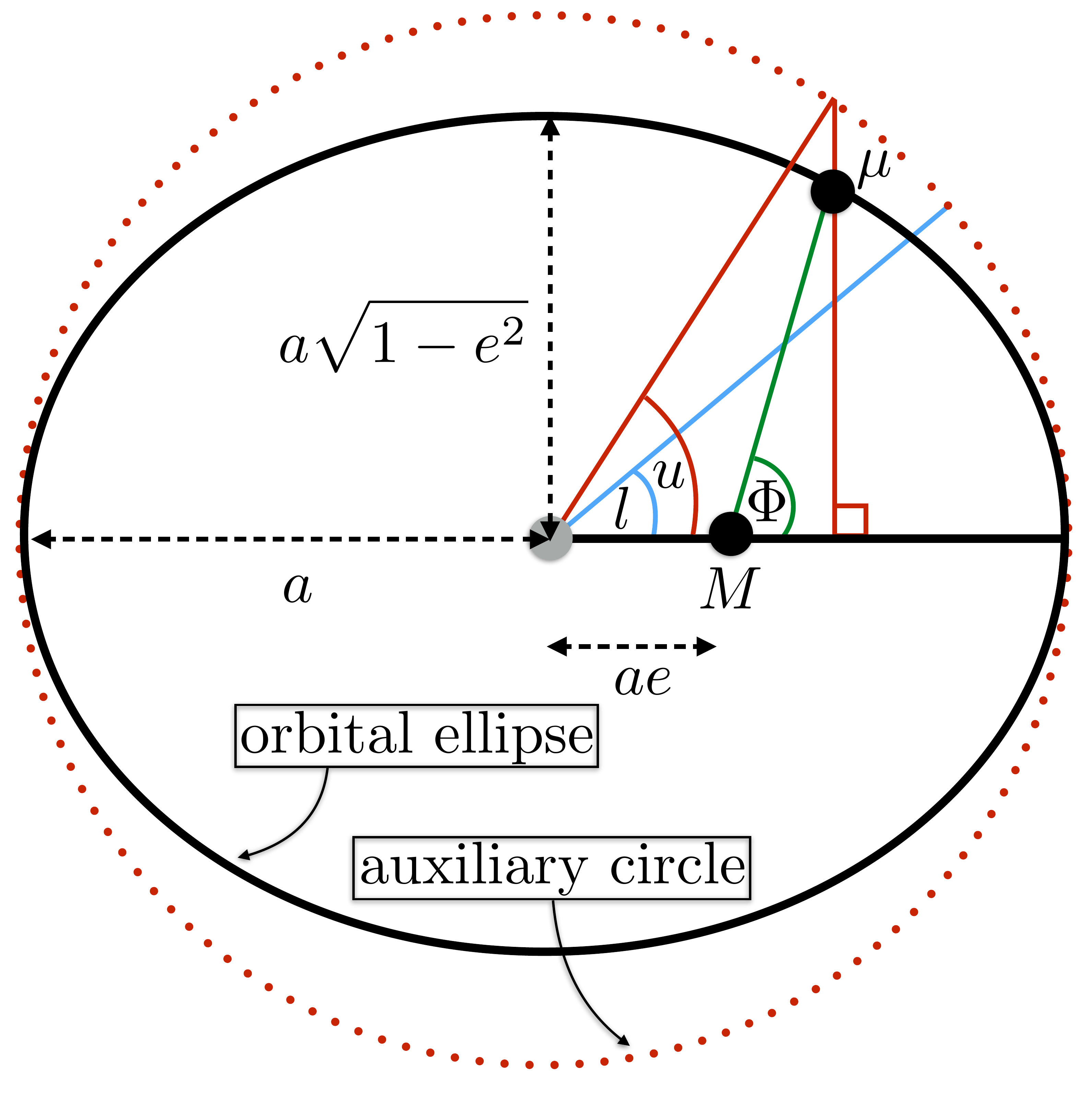}
    \end{subfigure}%
    ~ 
    \begin{subfigure}[t]{0.55\columnwidth}
        \centering
        \includegraphics[width=\columnwidth]{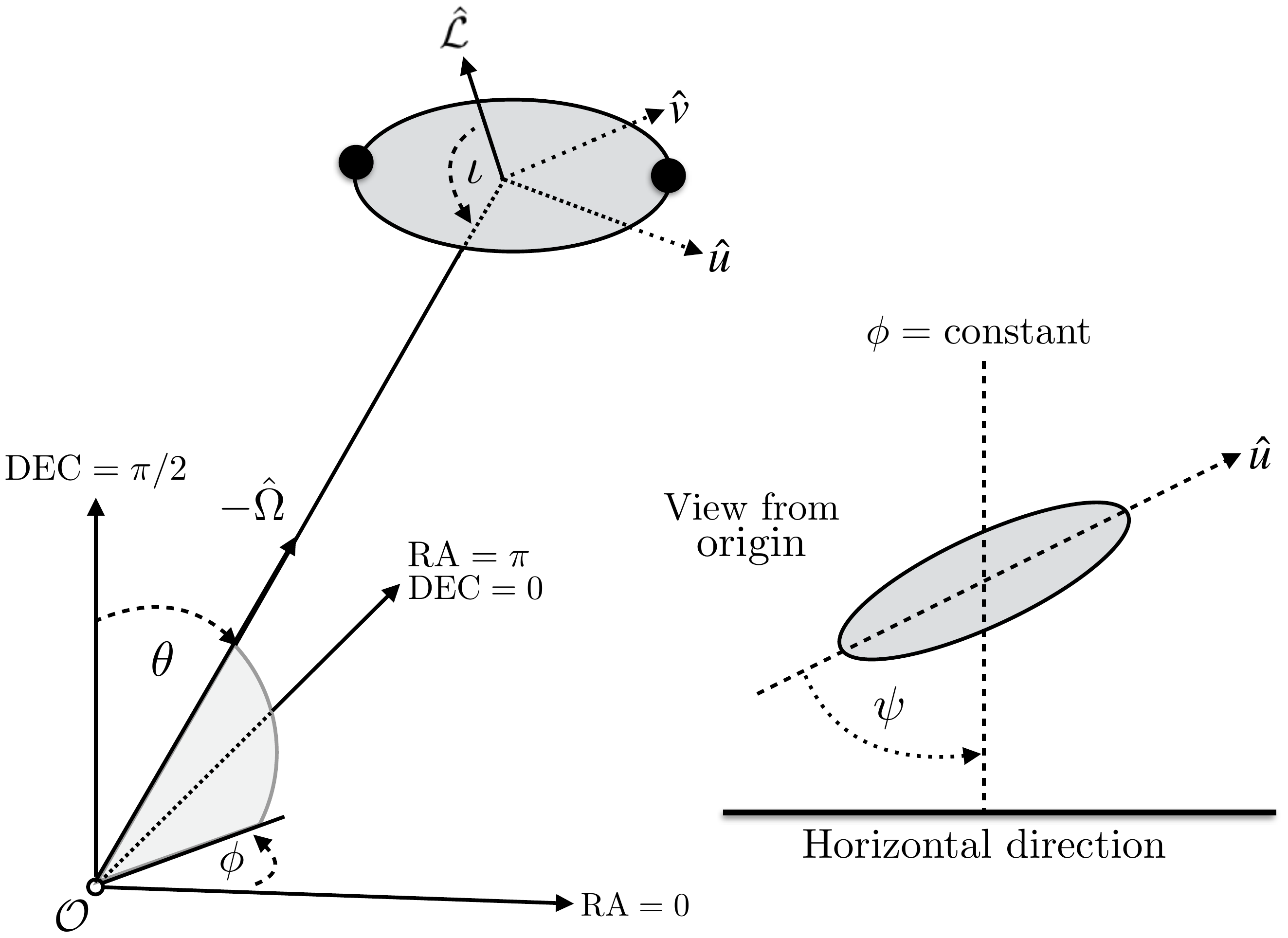}
    \end{subfigure}
    \caption{\textit{Left:} Orbital geometry of an eccentric binary. The motion is shown as the binary reduced mass, $M_1M_2/(M_1+M_2)$, orbiting the total mass, $(M_1+M_2)$, which is positioned at a focus of the ellipse. The semi-major axis is $a$, the eccentricity is $e$, the orbital phase is $\Phi$, the mean anomaly is $l$, and the eccentric anomaly is $u$. \textit{Right:} The orientation of the binary orbit with respect to the observer's coordinate system, where $\hat{n}=-\hat\Omega$, $\hat{u}$, and $\hat{v}$ form a basis triad with which to describe the GW polarization basis tensors, $\hat{\mathcal{L}}$ is a unit vector in the direction of the orbital angular momentum, $\psi$ is the GW polarization angle, and $\{\theta,\phi\}$ are the usual spherical-polar coordinates. Both figures reproduced from Ref.~\citep{2016ApJ...817...70T}.}
    \label{fig:eccentric_orbit}
\end{figure}

Equation \ref{eq:zredshift} showed the GW-induced fractional shift to the pulse arrival rate, which took the form
\begin{equation}
    z(t,\hat\Omega) = \frac{1}{2}\frac{p^a p^b}{(1+\hat\Omega\cdot\hat{p})}\Delta h_{ab},
\end{equation}
where $\Delta h_{ab}$ is the difference in the spatial components of the metric perturbation between the time at which the GW passes the Earth and when it passed the pulsar with positional unit vector $\hat{p}$. Previously we had only been interested in the statistical properties of this quantity for the purposes of describing a SGWB. However, for a single SMBBH we can write a determinstic waveform model to be deployed in searches for individual sources of GWs. In the TT gauge we write the metric perturbation for a GW propagating in direction $\hat\Omega$ at an arbitrary time as
\begin{equation}
    h_{ab}(t,\hat\Omega) = h_+(t) e^+_{ab}(\hat\Omega) + h_\times(t) e^\times_{ab}(\hat\Omega).
\end{equation}
The polarization basis tensors can be defined irrespective of the binary properties in terms of a basis triad:
\begin{align} \label{eq:polbasis}
    e^+_{ab} = \hat{u}_a\hat{u}_b - \hat{v}_a\hat{v}_b, \nonumber\\
    e^\times_{ab} = \hat{u}_a\hat{v}_b + \hat{v}_a\hat{u}_b,
\end{align}
\begin{align}
    \hat{n} \equiv &-\hat\Omega = \left(\sin\theta\cos\phi, \sin\theta\sin\phi, \cos\theta\right),\nonumber\\
    \hat{u} = &\left(\cos\psi\cos\theta\cos\phi - \sin\psi\sin\phi, \right.\nonumber\\
    &\left.\cos\psi\cos\theta\sin\phi + \sin\psi\cos\phi, -\cos\psi\sin\theta\right),\nonumber\\
    \hat{v} = &\left(\sin\psi\cos\theta\cos\phi + \cos\psi\sin\phi,\right. \nonumber\\
    &\left.\sin\psi\cos\theta\sin\phi - \cos\psi\cos\phi, -\sin\psi\sin\theta\right),
\end{align}
where $(\theta,\phi) = (\pi/2 - {\rm DEC}, {\rm RA})$ denotes the sky-location of the binary in spherical polar coordinates, and $\psi$ is the GW polarization angle that corresponds to the angle between $\hat{u}$ and the line of constant azimuth when the orbit is viewed from our coordinate system origin. See the right panel of Fig.~\ref{fig:eccentric_orbit}.

For the polarization amplitudes, I'll adopt the less conventional approach of tackling arbitrary binary eccentricity first, then showing how the model reduces down to its circular form. This discussion closely follows Ref.~\citep{2016ApJ...817...70T}, which employs leading-order Peters \& Matthews waveforms \citep{pm63} that are described in Ref.~\citep{2004PhRvD..69h2005B}. I'll be ignoring the influence of higher order post-Newtonian terms, although readily-usable waveforms inclusive of these do exist for PTA analysis, and are being actively implemented \citep{2020PhRvD.101d3022S}. This discussion also ignores BH spin, whose observable influences are likely to be challenging for PTAs to infer within the next decade \citep{2010PhRvD..81j4008S,2012PhRvL.109h1104M}. The relevant polarization amplitudes for the metric perturbation can be expressed analytically as 
\begin{align} \label{eq:hpluscross-sum}
    h_+(t) =& \sum_n -(1+\cos^2\iota)[a_n\cos(2\gamma)-b_n\sin(2\gamma)] +(1-\cos^2\iota)c_n, \nonumber\\
    h_{\times}(t) =& \sum_n 2\cos\iota[b_n\cos(2\gamma)+a_n\sin(2\gamma)],
\end{align}
where formally the summation is over integers $n=[1,\ldots,\infty]$, but in practice can be truncated \citep{2016ApJ...817...70T}, and
\begin{align} \label{eq:hpluscross-coeffs}
    a_n =&- n\zeta\omega^{2/3}\left[J_{n-2}(ne)-2eJ_{n-1}(ne)+(2/n)J_n(ne)\right. \nonumber\\
    &\left.\vphantom{J_{n-2}(ne)-2eJ_{n-1}(ne)+(2/n)J_n(ne)}+2eJ_{n+1}(ne)-J_{n+2}(ne)\right]\cos[nl(t)], \nonumber\\
    b_n =&- n\zeta\omega^{2/3}\sqrt{1-e^2}\left[J_{n-2}(ne)-2J_n(ne)+J_{n+2}(ne)\right]\sin[nl(t)], \nonumber\\
    c_n =&\;2\zeta\omega^{2/3}J_n(ne)\cos[nl(t)].
\end{align}
The parameters in these coefficients are defined as follows:

\begin{itemize}
    \setlength\itemsep{0pt}
    \item The amplitude is $\zeta=\mathcal{M}_z^{5/3}/D_L$, where $D_L=(1+z)D_c$ is the luminosity distance of the binary, and $\mathcal{M}_z = (1+z)\mathcal{M}$ is the \textit{redshifted} chirp mass.
    \item The quantity $l(t)$ is known as the \textit{mean anomaly}, defined by $l(t)-l_0 = 2\pi \int_{t_0}^t f_K(t') {\rm d}t'$, where $f_K=f_{K,r}/(1+z)$ is the observer-frame Keplerian orbital frequency, and $\omega=2\pi f_K$. The mean anomaly is essentially the orbital phase of the binary if we had been dealing with a circular system (see the left panel of Fig.~\ref{fig:eccentric_orbit}). 
    \item The angle $\gamma$ is an azimuthal angle measuring the direction of the system pericenter with respect to $\hat{x}\equiv (\hat\Omega+\hat{\mathcal{L}}\cos\iota)/\sqrt{1-\cos^2\iota}$, where $\hat{\mathcal{L}}$ is a unit vector pointing along the binary's orbital angular momentum.
    \item The binary orbital inclination angle, $\iota$, is defined by $\cos\iota = -\hat{\mathcal{L}}\cdot\hat\Omega$, i.e., $\iota=\pi/2$ corresponds to an edge-on binary. 
    \item The parameter $0<e<1$ is the eccentricity, describing the ellipticity of the system, where a perfectly circular orbit would have $e=0$.
    \item The functions $J_{(\cdot)}(\cdot)$ are Bessel functions.
\end{itemize}
This Fourier solution to Kepler's problem for a binary system makes explicit that the GW radiation from an eccentric binary is not monochromatic, and in fact occurs over a spectrum of harmonics of the orbital frequency. However, as expected, when $e=0$ we have Bessel terms $J_0(0)=1$ and $J_{n>0}(0)=0$, such that the only remaining GW emission occurs at $f = 2f_{K,r}/(1+z)$.

Finally, we can write our model for the pulsar timing delays that are induced by an individual SMBBH. As previously introduced, this is the integrated effect over the induced fractional shift to the pulse arrival rate:
\begin{equation}
    R(t) \equiv \int_0^t dt'\,z(t') = F^+(\hat\Omega)\Delta s_+(t) + F^\times(\hat\Omega)\Delta s_\times(t),
\end{equation}
where $\Delta s_{+,\times}$ are the corresponding differences in Earth-term and pulsar-term delays, and $F^{+,\times}$ are the previously introduced GW antenna response patterns for each polarization (see Eq.~\ref{eq:antennaresponse}). The time-dependent components of $s_{+,\times}$ can be computed analytically under the assumption that the system is non-evolving over the observational baseline of the pulsar (on average $\sim10-20$ years) at the time when the GW passes the Earth (for the Earth-term) and at the time at which the GW previously passed the relevant pulsar (for the pulsar-term). This means that quantities of time only appear linearly in the definition of the mean anomaly. In a modification to the description in Ref.~\citep{2016ApJ...817...70T}, we allow for evolution of the pericenter angle $\gamma$ over the observation time \citep{2004PhRvD..69h2005B}, for which we denote
\begin{equation}
    \dot\gamma\equiv \frac{d\gamma}{dt} = 3\omega \frac{(\omega M)^{2/3}}{(1-e^2)} \left[ 1 + \frac{(\omega M)^{2/3}}{4(1-e^2)}(26-15e^2) \right].
\end{equation}
Thus $\gamma(t) \approx \gamma_0 + \dot\gamma t$, and $l(t) \approx l_0 + \omega t$. Notice that the pericenter will advance even for a perfectly circular orbit. Therefore at a given time, either during the baseline of the Earth- or pulsar-term, the time dependence of the GW-induced delay from a single SMBBH is given by
\begin{align} \label{eq:splusscross-res}
    s_+(t) =& \sum_n -(1+\cos^2\iota)[\tilde{a}_n A_n(t)-\tilde{b}_n B_n(t)] + (1-\cos^2\iota)\tilde{c}_n, \nonumber\\
    s_{\times}(t) =& \sum_n 2\cos\iota[\tilde{a}_n C_n(t) - \tilde{b}_n D_n(t)],
\end{align}
where
\begin{align}
    A_n(t) &= \frac{1}{2}\left[\frac{\sin[nl(t)-2\gamma(t)]}{n\omega-2\dot\gamma} + \frac{\sin[nl(t)+2\gamma(t)]}{n\omega+2\dot\gamma}\right], \nonumber\\
    B_n(t) &= \frac{1}{2}\left[\frac{\sin[nl(t)-2\gamma(t)]}{n\omega-2\dot\gamma} - \frac{\sin[nl(t)+2\gamma(t)]}{n\omega+2\dot\gamma}\right], \nonumber\\
    C_n(t) &= \frac{1}{2}\left[\frac{\cos[nl(t)-2\gamma(t)]}{n\omega-2\dot\gamma} - \frac{\cos[nl(t)+2\gamma(t)]}{n\omega+2\dot\gamma}\right], \nonumber\\
    D_n(t) &= \frac{1}{2}\left[\frac{\cos[nl(t)-2\gamma(t)]}{n\omega-2\dot\gamma} + \frac{\cos[nl(t)+2\gamma(t)]}{n\omega+2\dot\gamma}\right],
\end{align}
and
\begin{align} \label{eq:splusscross-coeffs}
    \tilde{a}_n=&\; -n\zeta\omega^{2/3}\left[J_{n-2}(ne)-2eJ_{n-1}(ne)+(2/n)J_n(ne)\right. \nonumber\\
    &\left.\vphantom{J_{n-2}(ne)-2eJ_{n-1}(ne)+(2/n)J_n(ne)}+2eJ_{n+1}(ne)-J_{n+2}(ne)\right]\nonumber\\
    \tilde{b}_n =& -n\zeta\omega^{2/3}\sqrt{1-e^2}\!\left[J_{n-2}(ne)-2J_n(ne)+J_{n+2}(ne)\right] \nonumber\\
    \tilde{c}_n =&\,\, (2/n)\zeta\omega^{-1/3}J_n(ne)\sin[nl(t)].
\end{align}
This all seems quite complicated. But this is \textit{almost} as easy as it gets, since we have ignored binary evolution during the Earth- and pulsar-term baselines, as well as higher post-Newtonian terms and spin contributions. From a modeler's perspective, we are fortunate that PTAs are likely to detect binaries that are in the early adiabatic inspiral regime of their path to coalescence. I said this is \textit{almost} as easy as it gets, because of course all this simplifies significantly when $e=0$ and we ignore pericenter advance. The term $\tilde{c}_n = 0\,\,\forall n$, and only $\tilde{a}_2$ and $\tilde{b}_2$ are non-zero, taking values of $\tilde{a}_2 = \tilde{b}_2 = -2\zeta\omega^{2/3}$. Therefore the $s_{+,\times}$ terms reduce to
\begin{align} 
    s_+(t) &= \zeta\omega^{-1/3}(1+\cos^2\iota)\sin(2l_0 + 2\gamma_0 + 2\omega t) \nonumber\\
    &= \zeta\omega^{-1/3}(1+\cos^2\iota)\sin(2\Phi_0 + 2\omega t), \nonumber\\
    s_{\times}(t) &= 2\zeta\omega^{-1/3}\cos\iota\cos(2l_0 + 2\gamma_0 + 2\omega t) \nonumber\\
    &= 2\zeta\omega^{-1/3}\cos\iota\cos(2\Phi_0 + 2\omega t),
\end{align}
where typically the angular terms are packaged together as $\Phi_0\equiv l_0+\gamma_0$, rendering an even more compact expression.

\begin{figure}
    \centering
    \includegraphics[width=\columnwidth]{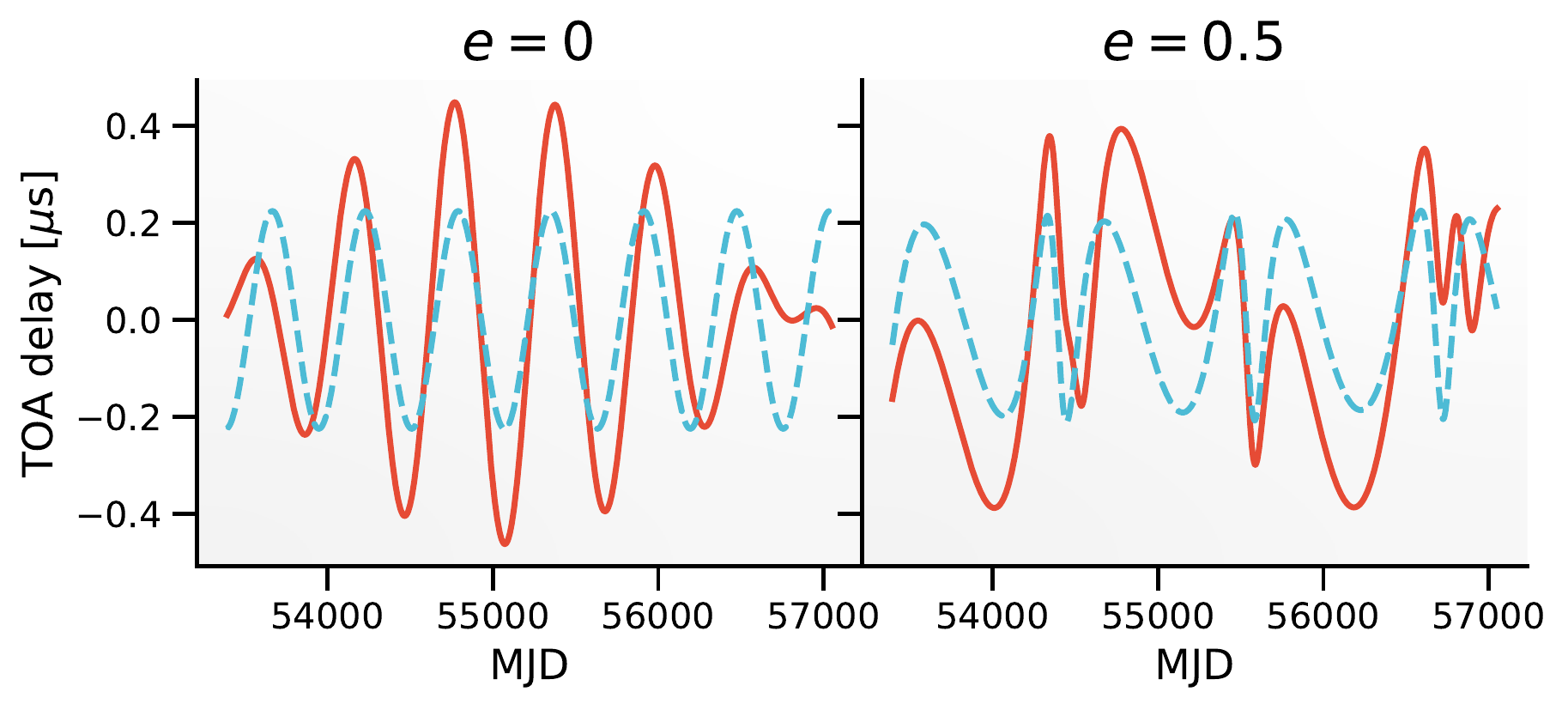}
    \caption{Example GW-induced TOA-delay time series for a circular binary ($e=0$) and an eccentric binary ($e=0.5$), over a baseline of $10$ years. The blue dashed line shows the Earth term, while the red solid line shows the entire signal that includes the pulsar term. The pulsar is assumed to be $1$~kpc distant, and in the sky location of PSR J$1713$$+$$0747$. The binary GW source has total mass of $3\times 10^9 M_\odot$, $q=1$, orbital frequency of $10$~nHz, and at a distance of $20$~Mpc.}
    \label{fig:eccentric_timeseries}
\end{figure}

Nevertheless even in the circular, static-pericenter scenario, there is some orbital evolution that we can not ignore; this is the evolution that occurs between the stage of the pulsar-term (earlier) and the Earth-term (always later). This time difference can be thousands of years. For an array of pulsars, the Earth-terms are at a common stage of source dynamics, whereas the pulsar terms are snapshots of the orbital dynamics that lag behind the Earth by $t_e-t_p = L_p(1+\hat\Omega\cdot\hat{p})$, i.e. they are pulsar specific. The orbital parameters of the source at the time of the pulsar-terms can be calculated by evolving the relevant Earth-term parameters backward in time using the evolution equations
\begin{align} 
    \frac{d\omega}{dt} &= \frac{96}{5\mathcal{M}^2}\left(\omega\mathcal{M}\right)^{11/3}\frac{1+\frac{73}{24}e^2+\frac{37}{96}e^4}{(1-e^2)^{7/2}},\nonumber\\
    \frac{de}{dt} &= -\frac{304}{15\mathcal{M}}(\omega\mathcal{M})^{8/3}e\frac{1+\frac{121}{304} e^2}{(1-e^2)^{5/2}},
\end{align}
where we see that GW emission causes a binary's orbital frequency to increase and its eccentricity to decrease. As the eccentricity decreases, the frequency of peak emitted GW power shifts to lower harmonics of the orbital frequency, eventually settling on $n=2$ at $e=0$. Eccentricity and orbital frequency co-evolve independently of the system mass \citep{pm63,p64}; a binary with an orbital frequency of $1$~nHz and eccentricity of $e=0.95$ will have partially circularized to $e\approx 0.3$ by the time that its orbital frequency has evolved to $100$ nHz. Calculating the backwards evolution of the Earth-term parameters can either be done numerically, or (for circular systems), analytically:
\begin{equation}
    \omega(t_p) = \omega_e\left( 1 - \frac{256}{5}\mathcal{M}^{5/3}\omega_e^{8/3}t_p \right)^{-3/8},
\end{equation}
where subscripts $e/p$ denote quantities at the time of Earth- or pulsar-terms. In fact, for weakly evolving systems we can simply perform a Taylor expansion around the Earth-term quantities such that
\begin{equation}
    \omega(t_p) = \omega_e + \frac{d\omega}{dt}\bigg|_e\times (t_p-t_e) = \omega_e - \frac{d\omega}{dt}\bigg|_e \times L_p(1+\hat\Omega\cdot\hat{p}).
\end{equation}
It is essential to include the pulsar-terms for accurate inference of SMBHB properties from PTA data. Ignoring them can, at the very least, lead to biased recovery of the source's sky location, rendering the difficult problem of localization even more challenging \citep{2010arXiv1008.1782C,2011MNRAS.414.3251L}. Indeed, for sources that do not evolve during the observational baseline of the PTA, it is not possible to disentangle the source's chirp mass and distance without leveraging the pulsar-term information. These different snapshots separated by thousands of years over-constrain the problem, and in principle allow amplitude terms to be split apart for mass and distance recovery. In practice, modeling the pulsar-terms in Bayesian single-source searches is incredibly challenging; the likelihood can be highly oscillatory in the pulsar distances (which must be searched over) \citep{2010arXiv1008.1782C,2011MNRAS.414.3251L,2013CQGra..30v4004E}, and current prior bounds on these distances are not where we need them to be for useful inference; ideally these would be tighter than a gravitational wavelength, which corresponds to $\sim 1$ parsec for a source with GW frequency of $10$~nHz. Some examples of TOA-delay time series caused by GWs from a circular and eccentric SMBHB are shown in Fig.~\ref{fig:eccentric_timeseries}. 

\subsection{Gravitational-wave memory burst}

Before moving away from SMBHBs entirely, it bears mentioning that there is another important signature that they can impart. PTAs are not sensitive to the final oscillatory gravitational waveform signature of the binary coalescence; this is too high in frequency, lying somewhere between $\sim 1-100$~$\mu$Hz (i.e., the gap between the high end of PTAs and the low end of LISA) for the relevant binary masses. However, they are sensitive to the associated broadband \textit{burst with memory} (BWM) \citep{1983PhRvD..28.1894P,1991PhRvL..67.1486C,1992PhRvD..46.4304B,1992PhRvD..45..520T}. There are two flavors of GW memory: (i) \textit{linear memory} that arises from non-oscillatory motion of the source system, particularly unbound masses (e.g., hyperbolic orbits, mass ejection), and (ii) \textit{nonlinear memory} that is a direct consequence of the nonlinear structure of General Relativity, where GWs themselves source other GWs. We are specifically interested in nonlinear GW memory bursts in binary systems \citep{2009ApJ...696L.159F,2009PhRvD..80b4002F}, where during the final coalescence the memory effect leads to a permanent offset in the baseline of the GW oscillations. This residual deformation in spacetime has a rise-time of $\sim 1$~day for a merger resulting in a $10^9 M_\odot$ BH \citep{2012ApJ...752...54C,2014ApJ...788..141M}. This is shorter in duration than the typical pulsar-timing observation cadence, meaning that for all intents and purposes we can model the GW memory burst as a step-function in strain (modulated by antenna response considerations), and thus a ramp-function in the induced timing delays. Since this kind of ramp feature is broadband, BWM signals provide an indirect probe of the final merger of SMBHs \citep{2010MNRAS.401.2372V,2012ApJ...752...54C,2014ApJ...788..141M}.

The signal model for a BWM in PTA searches is
\begin{equation}
    s_a(t) = h_\mathrm{mem}B_a(\theta,\phi,\psi) \times \left[ (t-t_0)\Theta(t-t_0) -  (t-t_a)\Theta(t-t_a) \right],
\end{equation}
where $B_a(\theta,\phi,\psi)$ describes the GW response of pulsar $a$ in terms of its sky location $(\theta,\phi)$ and GW polarization $\psi$; $t_0$ is the time at which the GW wavefront passes the Earth, $t_a$ is the retarded time at which the wavefront passed the pulsar, $t_a = t_0 - L_a(1+\hat\Omega\cdot\hat{p}_a)$, and $\Theta$ is the Heaviside step function. The memory strain is defined as
\begin{align}
    h_\mathrm{mem} &= \frac{1}{24}\frac{\eta M}{D_c} \sin^2\iota (17+\cos^2\iota)\left[\frac{\Delta E_\mathrm{rad}}{\eta M} \right], \nonumber\\
    \frac{\Delta E_\mathrm{rad}}{\eta M} &= 1-\frac{\sqrt{8}}{3} \sim 0.06,
\end{align}
where $M$ is the binary total mass, $\eta=M_1 M_2/M^2$ is the reduced mass ratio, $D_c$ is the radial comoving distance, $\iota$ is the inclination angle, and $\Delta E_\mathrm{rad}$ is the energy radiated from the system. NANOGrav, the European Pulsar Timing Array, and the Parkes Pulsar Timing Array have all searched for BWM signals, and in the absence of detection have constrained the strain amplitude as a function of sky location and burst epoch \citep{2010MNRAS.401.2372V,2015ApJ...810..150A,2015MNRAS.446.1657W,2020ApJ...889...38A}. 

\section{Exotic gravitational wave sources}

While SMBHBs form the likely initial source class for PTAs, there may be additional exotic sources of GWs lurking beneath the binary signals. These present a tantalizing probe of fundamental physics and the Universe at the earliest times. 

\subsection{Relic GWs}

Electromagnetic observations of the Universe can not peer through the veil of the last-scattering surface, rendering conventional astronomy limited to times later than $380,000$ years after the Big Bang. However, the fact that GWs couple so weakly to matter (which makes their detection technologically challenging) also means that they can propagate throughout the Universe essentially unhindered. 

If the Universe underwent an early period of exponential expansion (\textit{inflation}), microscopic phenomena should have been rapidly amplified to macroscopic scales. This applies to quantum fluctuations in early spacetime, where inflation caused superadiabatic amplification of zero-point quantum fluctuations in the early gravitational field, leading to the production of a broadband background of primordial GWs \citep{1976JETPL..23..293G,1980PhLB...91...99S,1982PhLB..108..389L}. These GWs are a major target for polarization studies of the CMB, wherein excess ``curl''-mode polarization could be the tell-tale signature \citep[][and references therein]{2016ARA&A..54..227K}.  

A simple model of the primordial/relic GW background can be expressed as a power-law in characteristic strain, where $h_c(f) = A(f/ 1 \mathrm{yr}^{-1})^\alpha$, and $\alpha = n_t/2 - 2/(3w+1)$ such that $n_t$ is the tensor index (which depends on the detailed dynamics of inflation), and $w=p/\rho$ is the equation of state in the immediate post-inflation (but pre-BBN) Universe \citep{2011PhRvD..83j4021Z}. For a scale-invariant primordial power spectrum ($n_t=0$), and a radiation-dominated equation of state ($w=1/3$), this predicts $\alpha=-1$ and $\gamma\equiv 3-2\alpha=5$ \citep{2005PhyU...48.1235G,2016PhRvX...6a1035L}.

\subsection{Cosmological phase transitions}

As the Universe cools and expands, the material within it may undergo a first-order phase transition if its temperature drops below the characteristic temperature for the transition \citep{2020arXiv200809136H}. Such phase transitions involve a discontinuity in first derivatives of the thermodynamic free energy, and are associated with the transitions we encounter in everyday life, e.g., solid/liquid/gas material transitions. Generally, they occur when the true minimum of a potential in the new phase is separated from a false minimum (i.e., the minimum from the previous phase) by a potential barrier through which a field must locally tunnel. In a cosmological scenario, bubbles of the new phase (``true vaccua'') nucleate within the old phase (``false vaccuum''), expanding rapidly until the bubble walls move relativistically. Collisions between these bubble walls should be copious sources of GWs \citep{2010PhRvD..82f3511C,2018CQGra..35p3001C,2020JCAP...03..024C}. Indeed, these bubble collisions precipitate phenomena that act as further sources of gravitational radiation, namely collisions of the sound waves that were generated within the false vaccuum by the original bubble wall collisions, and plasma turbulence generated by the expansion and collision of the sound waves. 

The spectrum of GWs generated by first-order phase transitions does not obey a power-law shape, but does encode dependencies on the underlying generating processes. For example, the peak frequency of the spectrum is linearly related to the temperature $T_*$, and the bubble nucleation rate $\beta_*$, at the time of the phase transition \citep{1997rggr.conf..373A,2021arXiv210413930A}:
\begin{equation}
    f^\mathrm{peak}_{0} \simeq 0.113\,\mathrm{nHz}\,\left(\frac{f^\mathrm{peak}_*}{\beta_*}\right)\left(\frac{\beta_*}{H_*}\right)\left(\frac{T_*}{\mathrm{MeV}}\right) \left(\frac{g_*}{10}\right)^{1/6},
\end{equation}
where $f^\mathrm{peak}_{0}$ is the peak frequency today, $f^\mathrm{peak}_*$ is the peak frequency at the time of emission, $H_*$ is the Hubble expansion rate at the time of emission, and $g_*$ is the number of relativistic degrees of freedom. The full shape of the GW spectrum will depend on the interplay of the three sourcing phenomena. While the typical frequencies probed by PTAs are far below the Standard Model electroweak scale ($\lesssim 100$~Gev), low-temperature phase transitions in \textit{hidden sectors} that are independent of Standard Model dynamics have been proposed \citep{2004JCAP...10..011C,2007PhLB..651..374S,2015PhRvL.115r1101S,2021arXiv210413930A}.

\subsection{Cosmic strings}

Cosmic strings are theorized linear (one-dimensional) topological spacetime defects formed as a result of symmetry-breaking cosmological phase transitions in the early Universe \citep{1976JPhA....9.1387K,1981PhLB..107...47V,1985PhR...121..263V,1995RPPh...58..477H,2000csot.book.....V}. Other defects that could form are monopoles and domain walls, although both are fairly confidently ruled out (in the monopole case by ``the monopole problem''). Thus cosmic strings are viewed as an exciting potentially-observable consequence of phase transitions in grand unified theories. Such a phase transition would lead to the formation of a \textit{network} of one-dimensional strings permeating the Hubble volume; these strings can ``intercommute'' when they meet one another (with probability, $p$), exchanging partners, while string self-interaction can lead to small closed loops being chopped off. The cosmic string network rapidly approaches an attractor scaling regime after formation, where the string correlation length, loop size, and other statistical properties, scales with cosmic time. 

The mechanism by which the network reaches the scaling regime is loop decay through GW emission. The cosmic string loops vibrate relativistically under tension (equal to their mass per unit length, $\mu$), leading to the emission of GW signals that causes them to eventually shrink and decay away. These GW signals can sum together incoherently to produce a stochastic GW background that is broadband, with the potential to be constrained by multiple experiments and GW detectors \citep{2013ApJ...764..108S,2012PhRvD..85l2003S,2015MNRAS.453.2576L,2018ApJ...859...47A}. While no models predict a purely power-law spectral shape across the entire GW spectrum, some approximations within the PTA band include $\alpha=-7/6$ and $\gamma=16/3$ \citep{2010PhRvD..81j4028O,2005PhRvD..71f3510D} (however this can vary quite a bit with modeling assumptions \citep{2013ApJ...764..108S}). One possible semi-analytic model that is valid within the PTA frequency range is \citep{2005PhRvD..71f3510D}
\begin{equation}
    h_c(f) = 1.74\times 10^{-14}\left(\frac{n_c}{p}\right)^{1/2} \epsilon_\mathrm{eff}^{-1/6} \left(\frac{h}{0.7} \right)^{7/6} \left(\frac{G\mu}{10^{-6}} \right)^{1/3} \left( \frac{f}{1\,\mathrm{yr}^{-1}}\right)^{-7/6}
\end{equation}
where $\mu$ is the cosmic string tension, $n_c$ is the average number of cusps per loop oscillation, $p$ is the intercommutation probability, $\epsilon_\mathrm{eff}$ is a loop length-scale factor, and $h$ is the Hubble constant in units of $100$~km$\,$s$^{-1}$Mpc$^{-1}$. There is also interest in searching for GW bursts from individual cosmic strings cusps and kinks \citep{2000PhRvL..85.3761D,2001PhRvD..64f4008D,2021MNRAS.501..701Y}. 

\section{Non-GW sources of correlated timing delays}

In an ideal world, the only sources of inter-pulsar correlations in pulsar timing delays would be GWs (modulo the usual statistical noise fluctuations present in any detection scenario). However, there are several sources of timing delays and inter-pulsar correlations that have origins on Earth, in the Solar System, and in the Galaxy, that we must consider.  

\subsection{Clock errors}

When first recorded, pulse TOAs are referenced to the local observatory time standard; however these local and national standards are not perfect. Countries distribute national atomic time-scales that are collected and combined by the Bureau International des Poids et Mesures (BIPM)\footnote{\href{https://www.bipm.org}{https://www.bipm.org}}, which then issues International Atomic Time (TAI). The latter exists as offsets from the original national atomic time-scales. TAI forms the basis for Coordinated Universal Time (UTC), and ultimately a realization of Terrestrial Time (TT). The relationship between the TAI realization of TT and TAI itself is $\mathrm{TT(TAI)} = \mathrm{TAI} + 32.184$~seconds. Once defined, TAI is never changed, instead being reviewed annually, and with departures from the SI second being corrected to steer TAI into a time-scale realization produced by BIPM \citep{2011Metro..48S.145A}. This realization is denoted TT(BIPM\textit{\{year\}}) with \textit{\{year\}} being the specific numerical year of creation, e.g., $2019$. The steering process can lead to a time-scale that is non-stationary, and whose long-timescale stability is difficult to judge. However, at least empirically we observe that recent TT(BIPM) versions differ from TT(TAI) by at least several microseconds since 1994 \citep{2012MNRAS.427.2780H,2020MNRAS.491.5951H}.

Any systematic errors in the realization of TT will be shared by \textit{all} pulsars, since they will all be referenced to the same time standard. This means that every pulsar includes exactly the same clock-error time series, or at least a window of it when baselines are different. The result is a \textit{monopolar} inter-pulsar correlation induced by systematic clock errors, where $\Gamma_{ab}=1$ for all pulsar pairs (a horizontal line on a correlation--angular-separation diagram) \citep{2012MNRAS.427.2780H,2016MNRAS.455.4339T,2020MNRAS.491.5951H}. This is often discussed along with \textit{spatially-correlated} processes, or shown on a correlation--angular-separation diagram with the Hellings \& Downs curve. While instructive, it is important to note that this is a systematic time-standard error that is common to all pulsars, not some extraterrestrial process that induces monopolar correlations.

\subsection{Solar-system ephemeris errors}

As described in Chapter~\ref{chap:pulsar_timing}, pulsar timing residuals are obtained by subtracting the observed TOAs from a best-fit model. The observed TOAs must be referenced back to the notional emission time at the pulsar, requiring a chain of timing corrections, amongst which are the aforementioned clock corrections. But another important link in this chain is the referencing of observatory-measured TOAs to the equivalent arrival time at the quasi-inertial reference frame of the Solar System Barycenter (SSB). The dominant term in this step is the calculation of the Roemer delay \citep{roemer121676}, corresponding to the classical light-travel time between Earth and the Sun ($\sim 500$~seconds):
\begin{equation} \label{eq:roemer}
    \Delta_{R\odot} = t^\mathrm{obs} - t^\mathrm{SSB} = -\left[ \vec{r}^\mathrm{\,obs}(t^\mathrm{obs}) - \vec{r}^\mathrm{\,SSB}(t^\mathrm{obs}) \right]\cdot \hat{p}
\end{equation}
where $\vec{r}^\mathrm{\,obs/SSB}$ are the coordinate vectors of the observatory and SSB at the observed pulse arrival time, respectively, and $\hat{p}$ is a unit vector pointing in the direction of the pulsar. 

Any imperfections in our knowledge of the observatory's or SSB's coordinate vector will induce systematic errors in the barycentering process of Eq.~\ref{eq:roemer}. Radio observatory positions with respect to the Earth's barycenter are known to within sub-nanosecond precision \citep{2006MNRAS.372.1549E}, leaving uncertainties in the SSB position as a possible error source. The resulting systematic timing error can be written as
\begin{equation}
    \delta t_{R\odot} = -\delta\vec{x}^{\,(3)}(t^\mathrm{obs})\cdot\hat{p}
\end{equation}
where $\vec{x}^{\,(3)}$ is the Earth barycenter's position in a coordinate frame that has the SSB as the origin. Thus $\delta\vec{x}^{\,(3)}$ is the time-dependent systematic error in our Solar System Ephemeris' model of $\vec{x}^{\,(3)}$. The SSB itself is not an observable position; it is the center of mass of the entire Solar System, whose computation requires accurate masses and orbits of all important dynamical objects, and with quantifiable precision on these values. For pulsar timing analysis, these positions and the SSB calculation rely on published ephemerides, the most prominent of which are the \textit{Development Ephemeris (DE)} series produced by NASA JPL \citep[e.g.,][]{folkner2009planetary,folkner2014planetary,park2021jpl}, and the \textit{Int\'egration
Num\'erique Plan\'etaire de l'Observatoire de Paris (INPOP)} series produced by IMCCE-Observatoire de Paris \citep[e.g.,][]{2013arXiv1301.1510F,2014arXiv1405.0484F,2020jsrs.conf..293F}. These ephemeris models require painstaking regression over heterogeneous datasets of different Solar System bodies that stretch back decades, many of which have uncertainties and data quality that is difficult to assess and synthesize with more recent data. Thus formal ephemeris uncertainties and covariances are not considered reliable, and are not published with the best-fit orbits; the latter are used as point estimates in the Roemer delay calculations in pulsar timing.

For reasonable astrophysical models of SMBHBs, the level of induced timing delays is $\sim 100$~ns, requiring the position of the SSB to be known to within $\sim \mathcal{O}(100$~m). Until $\sim2016-2017$, these kinds of Solar System ephemeris errors were thought to constitute a sub-dominant source of systematic errors for PTA GW searches. However several groups found that there were large systematic differences between timing residuals computed under different JPL DE models ranging from DE418 to DE436, whose publication dates differed by more than a decade. This led to an exhaustive analysis by the NANOGrav Collaboration, who found that limits on the SGWB varied significantly when considering ephemerides DE421 to DE436, and the Bayes factor for a common-spectrum process varied by over an order of magnitude \citep{2018ApJ...859...47A}. This led to the development of a perturbative Bayesian modeling scheme, \textsc{BayesEphem} \citep{2018ApJ...859...47A,2020ApJ...893..112V}, (which updates the earlier approach of Champion et al.~\citep{2010ApJ...720L.201C}) to model uncertainties on gas giant masses, Jupiter and Saturn's orbital elements, and coordinate frame rotations. This model updates the Roemer delay calculation ``in real time'' during an MCMC search over the joint SGWB and intrinsic pulsar noise parameter space, contributing additional ephemeris parameters to this global analysis. In the analysis, a baseline DE or INPOP ephemeris is chosen, and the exploration of \textsc{BayesEphem} parameters allows these different ephemerides to ``bridge'' the systematic differences between them. Therefore current PTA GW searches attempt to constrain the Solar System ephemeris simultaneously with a search for GWs of extragalactic origin. Additional modeling approaches by other PTA groups and the IPTA are in development \citep{2018MNRAS.481.5501C,2019MNRAS.489.5573G}. 

The systematic timing error from Solar System ephemeris uncertainties has a cosine dependence on the angle between the SSB error vector and the pulsar's position vector. Thus pulsars on opposite sides of the sky will experience the same magnitude timing error but with an opposite sign. Provided that the SSB error vector is uncorrelated with the pulsar position vectors, Solar System ephemeris errors will induce \textit{dipolar} inter-pulsar correlations, where $\Gamma_{ab}=\cos\theta_{ab}$ \citep{2016MNRAS.455.4339T,2019ApJ...876...55R}. This is only an approximation, because of course SSB errors can be modeled deterministically through planetary mass and orbit perturbations (this is how \textsc{BayesEphem} works). Nevertheless, in the regime where there are large number of error sources, it may be effective to treat Solar System ephemeris uncertainties as a dipolar-correlated stochastic process.

\subsection{Dark matter}

\subsubsection{Cold dark matter substructure}

Despite being one of the major components of our current concordance cosmology, dark matter is poorly constrained by observations on sub-galactic scales \citep[e.g.,][]{2000ApJ...542..281A,2007A&A...469..387T,2011MNRAS.416.2949W,2011PhRvL.107w1101G,2017PhRvD..95h3006C}. This presents problems since well-motivated models of dark matter predict characteristic structure on such small scales, amongst which is the concordance $\Lambda$CDM model (Cosmological Constant + Cold Dark Matter) that has a mass function of dark matter haloes that extends down to the free-streaming scale, corresponding to $\sim 10^{-6}M_\odot$ for Weakly Interacting Massive Particle (WIMP) dark matter \citep{2005JCAP...08..003G}. Constraining dark matter substructure would provide an important validation of CDM and inform limits on the mass of the candidate particle. 

High precision pulsar timing could provide a powerful probe of sub-galactic dark matter structure, through Doppler and Shapiro effects \citep[e.g.,][]{2007MNRAS.382..879S,2007ApJ...659L..33S,2011PhRvD..84d3511B,2012MNRAS.426.1369K}. The Doppler effect occurs when dark matter substructure (e.g., a clump, primordial BH, etc.) passes by the Earth or a pulsar, pulling on either and inducing accelerations that change the measured arrival rate of radio pulses. If the substructure passes the Earth then the timing delay will be dipolar-correlated amongst all observed pulsars in an array (similar to the previously discussed Solar System ephemeris systematics), whereas substructure passing close to a pulsar(s) will create uncorrelated timing delays. The Shapiro effect is an integrated timing delay that accumulates as photons propagate through the gravitational potential of substructure along the line of sight between Earth and a given pulsar. As such, it is akin to chromatic timing delays induced by the ionized interstellar medium, in that it builds over the entire path and is uncorrelated between different lines of sight. 

The fractional shift to the pulse arrival rate from each effect is given by \citep{2019PhRvD.100b3003D,2021arXiv210405717L} 
\begin{align}
    \left( \frac{\delta \nu}{\nu}\right)_D &= \hat{p}\cdot \int \nabla\Phi(\vec{r},M)dl, \nonumber\\
    \left( \frac{\delta \nu}{\nu}\right)_S &= -2\int \vec{v}\cdot \nabla\Phi(\vec{r},M)dl,
\end{align}
where $\hat{p}$ is a unit vector in the direction of a pulsar, $\Phi$ is the dark matter gravitational potential, $M$ and $\vec{v}$ is the mass and velocity of the dark matter, and $l$ parametrizes the photon propagation path from the pulsar to the Earth.

\subsubsection{Fuzzy dark matter}

Fuzzy Dark Matter (FDM) refers to a proposed ultralight axion (essentially a generalization of the QCD axion) as a candidate for dark matter \citep{2000PhRvL..85.1158H,2017PhRvD..95d3541H}. Such an ultralight axion propagating at velocity $v$ would have a macroscopic de Broglie wavelength,
\begin{equation}
    \frac{\lambda_\mathrm{dB}}{2\pi} = \frac{\hbar}{mv}\approx 60\,\mathrm{pc}\left(\frac{10^{-22}\,\mathrm{eV}}{m} \right) \left(\frac{10^{-3}}{v} \right),
\end{equation}
whose wave-like property would effectively suppress power on small scales, thereby resolving some of the problems that cold dark matter models have with predicting more small-scale structure than is observed \citep{2017arXiv170704591B}. It is worth noting that FDM can refer to a broader class of ultralight bosons as dark matter candidates.  

Tantalisingly, FDM is predicted to produce a measurable influence on pulsar timing observations \citep{2014JCAP...02..019K,2014PhRvD..90f2008P,2017PhRvL.119v1103D}. FDM should couple gravitationally to regular matter in the Milky Way, inducing periodic oscillations in the gravitational potential at frequencies related to twice the mass of the ultralight scalar field:
\begin{equation}
    f = \frac{2m}{h} \approx 4.8\times 10^{-8}\,\mathrm{Hz}\left(\frac{m}{10^{-22}\,\mathrm{eV}} \right).
\end{equation}
The propagation of photons through this oscillating gravitational potential will lead to shifts in the arrival rate of radio pulses from pulsars. These shifts are not GW-induced; they are entirely due to the time-dependent potential through which the photons propagate. The periodicity in FDM-induced timing delays matches the oscillation of the potential, which happens to lie within the nanohertz PTA sensitivity band for the masses of reasonable FDM candidates. The rms timing delay induced by FDM follows
\begin{equation}
    \delta t_\mathrm{rms} \approx 0.02\,\mathrm{ns} \left(\frac{m}{10^{-22}\,\mathrm{eV}}\right)^{-3} \left(\frac{\rho_\mathrm{SF}}{0.4\,\mathrm{GeV}\,\mathrm{cm}^{-3}}\right),
\end{equation}
where $\rho_\mathrm{SF}$ is the local scalar-field dark matter density, and $0.4\,\mathrm{GeV}\,\mathrm{cm}^{-3}$ is the measured local dark matter density \citep[e.g.,][]{2012ApJ...756...89B}. A full signal model is given in Ref.~\citep{2018PhRvD..98j2002P}. As the particle mass decreases, the oscillation frequency decreases while the amplitude of induced timing delays increases, thus presenting an exciting opportunity as PTA baselines grow longer \citep{2018PhRvD..98j2002P}.

\bibliographystyle{unsrt_new}
\bibliography{refs}

%% file: 05.tex
\chapter{Data Analysis}
\epigraph{\textit{``Never tell me the odds!''}}{Han Solo, ``Star Wars: The Empire Strikes Back''}

\section{Statistical inference}

Statistical inference is the science of extracting meaning from observation. An astrophysicist can not test a new model without thinking shrewdly about what the observable ramifications will be. Perhaps this new model posits that black holes at the centers of galaxies have a strong influence on the growth and evolution of the galaxy itself. Well...great...but what shall we look at? How do we compute the masses of black holes at the center of galaxies, and what characteristics of the galaxies are measurable in order to test a correlation? In this specific case, my example is a real empirical relationship that exists, showing that there is a strong correlation between central black hole masses and the masses of host-galaxy stellar bulges, bulge luminosity, and bulge velocity dispersion. But even deriving these quantities involves a chain of processes that start at a telescope somewhere, passes through some pipeline to reduce the raw observations to some more digestible form, and involves fitting a model to the reduced observations in order to quote a value with associated precision. Along the way, we must be fastidious in keeping track of sources of systematic and statistical uncertainties so that our final value is a proper reflection of our confidence. 

In our case, we are interested in using pulsar-timing data in the form of pulse TOAs (already a reduced form of the initial voltage readings in a receiver) to answer questions like ``is there evidence for a stochastic background of gravitational waves?'', ``what is the most massive binary that could be present in a certain galaxy center as constrained by our data?'', and ``what kind of radio-frequency dependent noise are the pulses from a certain pulsar experiencing?''. There are two main approaches to statistical inference: \textit{frequentist} (or classical) inference, and \textit{Bayesian} inference. The former is predominant in most of the early literature on gravitational waves and even in PTA searches for them, but Bayesian inference now has a foothold as the flagship approach (as in much of astrophysics research). It has grown steadily in popularity for two key reasons: computing power has made it possible to numerically explore high-dimensional probability distributions, and by incorporating prior model information one can extract meaningful results even from sparse datasets. 

Frequentist inference considers the statistical spread of experimental outcomes over a number of trials; hence the use of ``frequentist`` to denote how probabilities are phrased in this approach. All physical phenomena have true values in the absence of an observation, and all sources of uncertainty and variation are in the data itself. Therefore the natural question for a frequentist statistician to ask is ``how many times should I expect to have measured this data given an assumed model parameter?'' So we form a probability distribution for data realizations given a set of model parameters, $p(d|\theta, \mathcal{H})$, where $d$ is data, and $\theta$ are parameters of the hypothesis model $\mathcal{H}$. This probability is more often referred to as the \textit{likelihood} function, but note that we talk about the likelihood of parameters given some data, not the likelihood of data-- the reader will excuse this dip into pedantry, but there is precedent in the literature \cite{jaynes_2003}. Hence, $L(\theta|d)\equiv p(d|\theta)$. By contrast, in Bayesian statistics there is no concept of a true underlying model parameter-- there is only the measured data with which we infer some probabilistic distribution of the model parameter. The shift in thinking is subtle, but significant. We are no longer thinking about repeated experimental trials providing a distribution of measured values of a parameter; instead we consider a single measured dataset from which we deduce a probability distribution of a parameter. This Bayesian probability distribution is not a distribution over experimental trials! It is a formal probability distribution that encodes the spread in our belief of what the model parameter is. The terminology for this is the posterior probability $p(\theta|d)$, i.e., the probability of the model parameter given the data. The probabilities at the core of frequentist and Bayesian inference are linked through \textit{Bayes' Theorem}:
\begin{equation} \label{eq:bayes_theorem}
    p(\theta|d) = \frac{p(d|\theta)p(\theta)}{p(d)},
\end{equation}
where $p(\theta|d)$ is the posterior probability distribution of $\theta$, $p(d|\theta)$ is the likelihood of $\theta$, $p(\theta)$ is the prior probability of $\theta$, and $p(d)$ is a normalization constant that is unimportant in parameter estimation but very important to model selection. It is sometimes called the fully-marginalized likelihood, or evidence, and we will meet it again soon. 

In the following I will outline key points in the usage of frequentist and Bayesian statistics. However, the pact of confidence between writer and reader compels me to disclose that I am a card-carrying Bayesian, and this discussion will skew toward that proclivity. Much of the discussion in the sub-section on frequentist inference is inspired by the excellent treatments in Allen \& Romano (1999)~\citep{1999PhRvD..59j2001A} and Romano \& Cornish (2017)~\citep{2017LRR....20....2R}, while the details of Bayesian inference can be read about in more detail in Jaynes (2003)~\citep{jaynes_2003}, Gelman \textit{et al.}~(2013)~\citep{gelman2013bayesian}, and Gregory (2010)~\citep{2010blda.book.....G}. 

\section{Frequentist inference}

Classical hypothesis testing assumes that we have repeatable experiments (or trials) with which we can define probabilities as being the fraction of such identical trials in which certain outcomes are realized. A statistic or optimization function is constructed from the data to describe the fitness of a particular model, and although this is often some form of the likelihood function, it need not be. The flexibility in defining the form of statistic is because we are ultimately more interested in its sampling distribution, i.e., the distribution of the statistic over many trials under different hypotheses. These hypothesis distributions for the statistic will determine its detection efficacy. The objectively best statistic is one that meets the \textit{Neyman-Pearson criterion} for maximizing the \textit{detection probability} at a given fixed \textit{false alarm probability}. We will learn exactly what these different terms mean very soon. 

\subsection{Significance}

\begin{figure}
\centering
	\includegraphics[width=0.5\columnwidth]{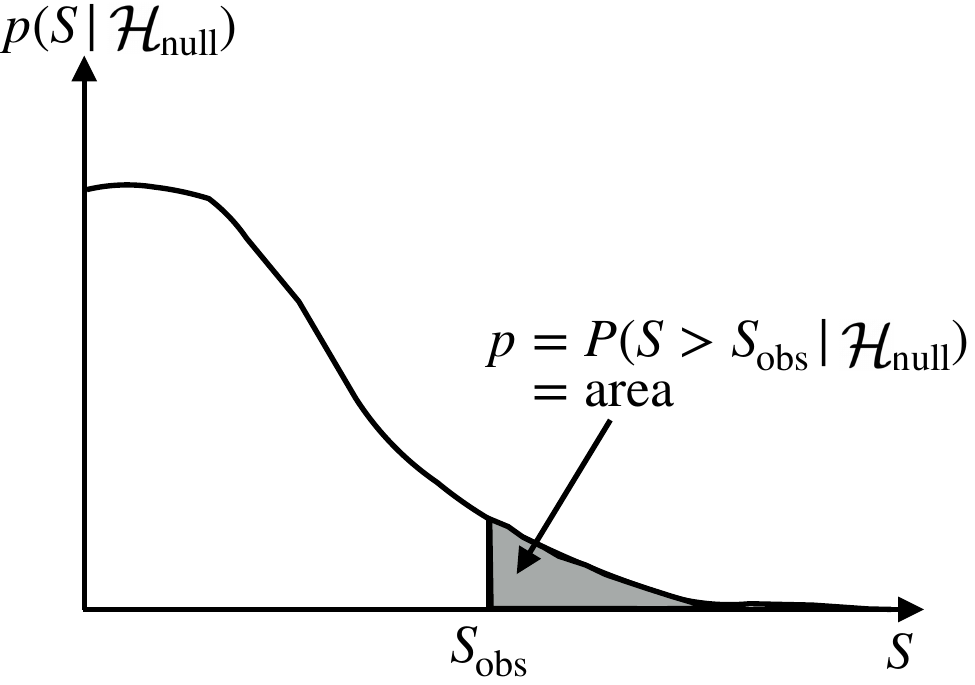}
    \caption{The $p$-value is the probability of a detection statistic exceeding the value measured from some data under the noise-only hypothesis. It reports how often spurious noise fluctuations could give statistic values more extreme than observed. Adapted from Ref.~\citep{2017LRR....20....2R}.}
    \label{fig:pvalue}
\end{figure}

Imagine data being collected from an experiment, with which we compute a detection statistic, $S(d) = S_\mathrm{obs}$. As a measure of the fitness of a signal model to the data, we would like to know how often noise fluctuations alone could spuriously produce a statistic as loud as that observed. Thus we need to know the \textit{null hypothesis distribution}, $p(S|\mathcal{H}_\mathrm{null})$, i.e. the statistical distribution under the noise-only hypothesis. This distribution could be computed either analytically or through numerical Monte Carlo trials. Either way, it allows us to make a statement about the significance of the measured statistic, denoted by its \textit{$p$-value} (see also Fig.~\ref{fig:pvalue} for a graphical description.):
\begin{equation}
    p = P(S>S_\mathrm{obs}|\mathcal{H}_\mathrm{null}) = \int_{S_\mathrm{obs}}^\infty p(S|\mathcal{H}_\mathrm{null}) dS.
\end{equation}

There are many abuses of $p$-values, the most egregious of which is interpreting them in terms of support for the signal hypothesis, rather than their true interpretation as arguing against the specific, assumed null hypothesis. Supporting the signal hypothesis requires the statistical distribution under the signal hypothesis, which we next consider.

\subsection{Type I \& II errors}

A pre-defined $p$-value can be assumed in order to set a decision threshold for the statistic. For example, we can adopt a very small $p$-value that maps to a value of the statistic $S_\mathrm{threshold}$, above which there is very little chance for noise alone to generate more extreme data. But the chance is not zero! Consider the case where noise does produce a more extreme measured statistic; this would pass our decision threshold, causing us to reject the null hypothesis. This is called a Type I (or more commonly a \textit{false alarm}) error, where $S_\mathrm{obs}>S_\mathrm{threshold}$ under $\mathcal{H}_\mathrm{null}$. The converse can also occur, where the statistical distribution of the signal hypothesis allows for the possibility that $S_\mathrm{obs}<S_\mathrm{threshold}$ under $\mathcal{H}_\mathrm{signal}$, causing us to reject the signal hypothesis in a Type II (or \textit{false dismissal}) error. We define the false alarm and false dismissal probabilities as
\begin{align}
    \alpha = P(S>S_\mathrm{threshold}|\mathcal{H}_\mathrm{null}), \nonumber\\
    \beta = P(S<S_\mathrm{threshold}|\mathcal{H}_\mathrm{signal}),
\end{align}
where in general the false dismissal probability will depend on the strength of the signal, since this directly influences the spread of $p(S|\mathcal{H}_\mathrm{signal})$.

Related to the false dismissal probability is the \textit{detection probability}, $\mathrm{DP} = 1-\beta$, which measures the fraction of trials in which the signal is correctly identified when truly present (assuming a fixed $\alpha$ that sets the decision threshold). The DP is a quantity that is often tracked in order to determine at what strength we would expect to confidently detect a signal. The strength need not refer simply to the amplitude of the signal; in fact, for PTA GW searches the total observation time and number of pulsars in the array are at least as important factors. Hence we can forecast the detection probability along a number of axes, determining the conditions under which it would surpass a pre-determined threshold of, say, $95\%$ at a fixed false alarm probability of $\alpha$.

Note that on occasion, people will merely quote the conditions under which the expectation value of the detection statistic $S$ passes the threshold value $S_\mathrm{threshold}$. To see why this would be a very weak claim of detection, imagine the statistical distribution of $S$ under the signal hypothesis to be centered on $S_\mathrm{threshold}$. Trivially, one can see that the observed value of $S=S_\mathrm{obs}$ (which is drawn from $p(S|\mathcal{H}_\mathrm{signal})$) only surpasses $S_\mathrm{threshold}$ in $50\%$ of the trials. So under these conditions the detection probability is a paltry $50\%$, no better than a coin flip.

\subsection{Upper limits}

Without a detection, we can still place constraints that are referenced to upper confidence limits on the amplitude of any signal present. In the frequentist framework, this amounts to saying that if the signal were any louder than this limit, then we would have detected a more extreme value of the statistic than that observed, and with even greater frequency. Thus the upper limit is referenced to a confidence level, and requires knowledge of the statistical distribution of $S$ under the signal hypothesis. 

\begin{figure}
\centering
	\includegraphics[width=0.5\columnwidth]{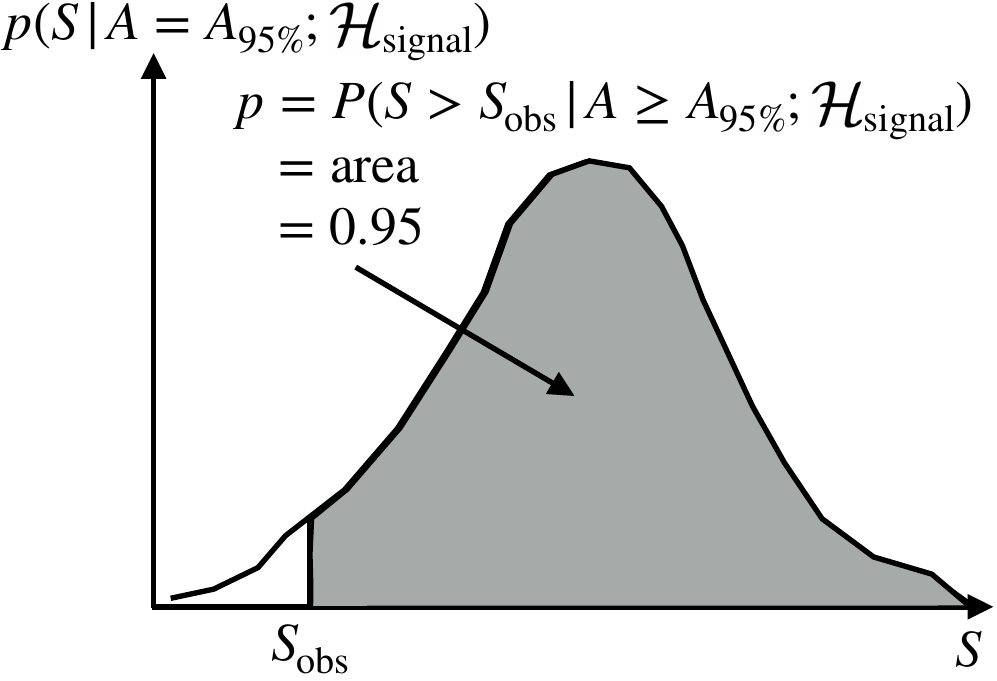}
    \caption{The frequentist definition of an upper confidence limit. The distribution of the detection statistic, $S$ is shown under the signal hypothesis when $A=A_{95\%}$. $95\%$ of this distribution lies above the measured value of the statistic, hence we quote $A_{95\%}$ as the frequentist $95\%$ upper limit. Adapted from Ref.~\citep{2017LRR....20....2R}.}
    \label{fig:freq_ul}
\end{figure}

We imagine that in our signal model there is an amplitude parameter, $A$ that we can place upper constraints on. The frequentist $95\%$ upper limit on $A$ is defined as
\begin{equation}
    P(S\geq S_\mathrm{obs}|A\geq A_{95\%};\mathcal{H}_\mathrm{signal}) = 0.95,
\end{equation}
and shown graphically in Fig.~\ref{fig:freq_ul}.

\section{Bayesian inference}

Equation \ref{eq:bayes_theorem} states Bayes' Theorem in a form that makes clear the importance of prior knowledge. Indeed, despite the fact that you may hear some people say ``but I have not assumed a prior'', it is impossible to perform Bayesian inference without assuming a prior, even if it is unconscious. The aforementioned quote usually means that this person has allowed the prior on a parameter to be uniform over an unbounded range, which is often referred to as an \textit{improper} prior because it is not normalizable. We usually attempt to ensure that the posterior probability distribution is data-dominated by employing weakly informative priors, e.g., a wide Gaussian distribution (which is in fact the maximum entropy distribution amongst all real-valued distributions with a specified variance), or a bounded uniform prior in log-space (if the parameter can vary over a large dynamic range). With prior functions that bound the parameters of our model, and a likelihood function that assesses the fitness of some model parameters to the measured data, we can use Bayes' Theorem to deduce the posterior probability distribution, $p(\theta|d)$. I will gloss over exactly how that distribution is deduced until later, but it usually involves numerical sampling techniques for which we can ignore the evidence as an unimportant normalization factor. The posterior probability distribution can have arbitrary forms depending on the model and the data; it need not be Gaussian or something else simple. Even if the likelihood is Gaussian, it is only Gaussian in the \textit{data}, not the \textit{parameters}, so one should not expect a simple functional form for the posterior in arbitrary cases. 

\subsection{Parameter estimation}

Consider that our model has more than one parameter, such that $\theta = \{x,y,z\}$. The posterior probability distribution will be multi-dimensional, and we may not even care about all of the parameters being represented. Bayesian inference gives us the power to treat those as \textit{nuisance parameters} that we can \textit{marginalize over}. Marginalization equals integration, so that we collapse our posterior distribution down to a smaller dimensional representation that contains parameters of interest:
\begin{equation}
    p(x|d) = \int p(x,y,z|d)\,\, dydz.
\end{equation}
However, we have not simply fixed the nuisance parameters to some arbitrary values; we have integrated over their entire probabilistic support, and in so doing have propagated all of the uncertainties in those parameters down through to our inference of the parameters of interest.

\begin{figure}
\centering
	\includegraphics[width=0.5\columnwidth]{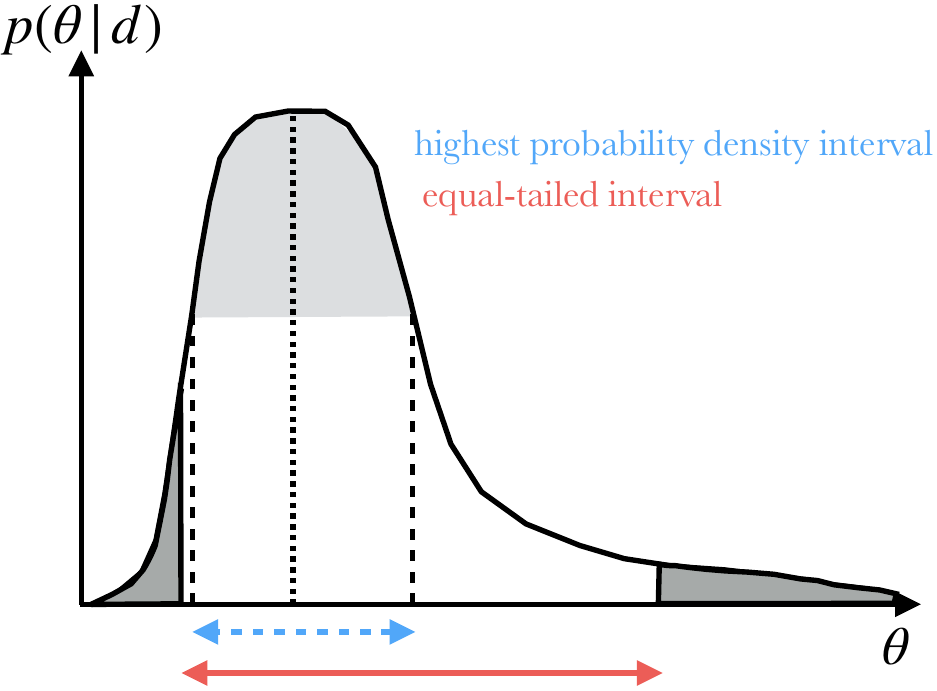}
    \caption{Bayesian credible regions encode how well we believe a parameter is constrained within a certain range. However, they are not unique. One can derive an $X\%$ credible interval by integrating the distribution downwards from the maximum-a-posteriori point (this is the \textit{highest probability density interval}), or by integrating inwards from each tail by $(X/2)\%$ (this is the \textit{equal-tailed interval}). Adapted from Ref.~\citep{2017LRR....20....2R}.}
    \label{fig:bayes_credregion}
\end{figure}

Parameter estimation in Bayesian inference usually involves quoting quantiles of the one-dimensional marginalized posterior probability distribution of each parameter. Point estimates can be quoted as the mean, median ($50\%$ quantile), maximum likelihood, or \textit{maximum-a-posteriori} (MAP)\footnote{\textit{A posteriori} is a Latin phrase for ``from what comes after''. You may see it in other textbooks or articles.} value. Uncertainties are often quoted by defining \textit{credible regions} (named for how Bayesian posterior probabilities encode the spread in our belief of parameters) that are analagous to Gaussian $\sigma$'s. For example, we may quote the parameter ranges that enclose $68\%$ and $95\%$ of the posterior distribution, which are analagous to Gaussian $1$- and $2$-$\sigma$ uncertainties. Credible regions are not unique. There are two ways we can compute an $X\%$ credible region; $(i)$ we integrate (upwards)downwards from $(-)\infty$ until we enclose $X/2\%$ to get the bounding values-- this provides an \textit{equal-tailed interval}; or $(ii)$ we imagine lowering a horizontal line from the posterior maximum downwards until we have enclosed $X\%$ of the posterior, where the credible boundaries are where this level intersects the posterior distribution-- this is sometimes referred to as the \textit{highest probability density interval}. See Fig.~\ref{fig:bayes_credregion} to understand how these credible regions are computed in operationally different ways.

\begin{figure}
\centering
	\includegraphics[width=0.6\columnwidth]{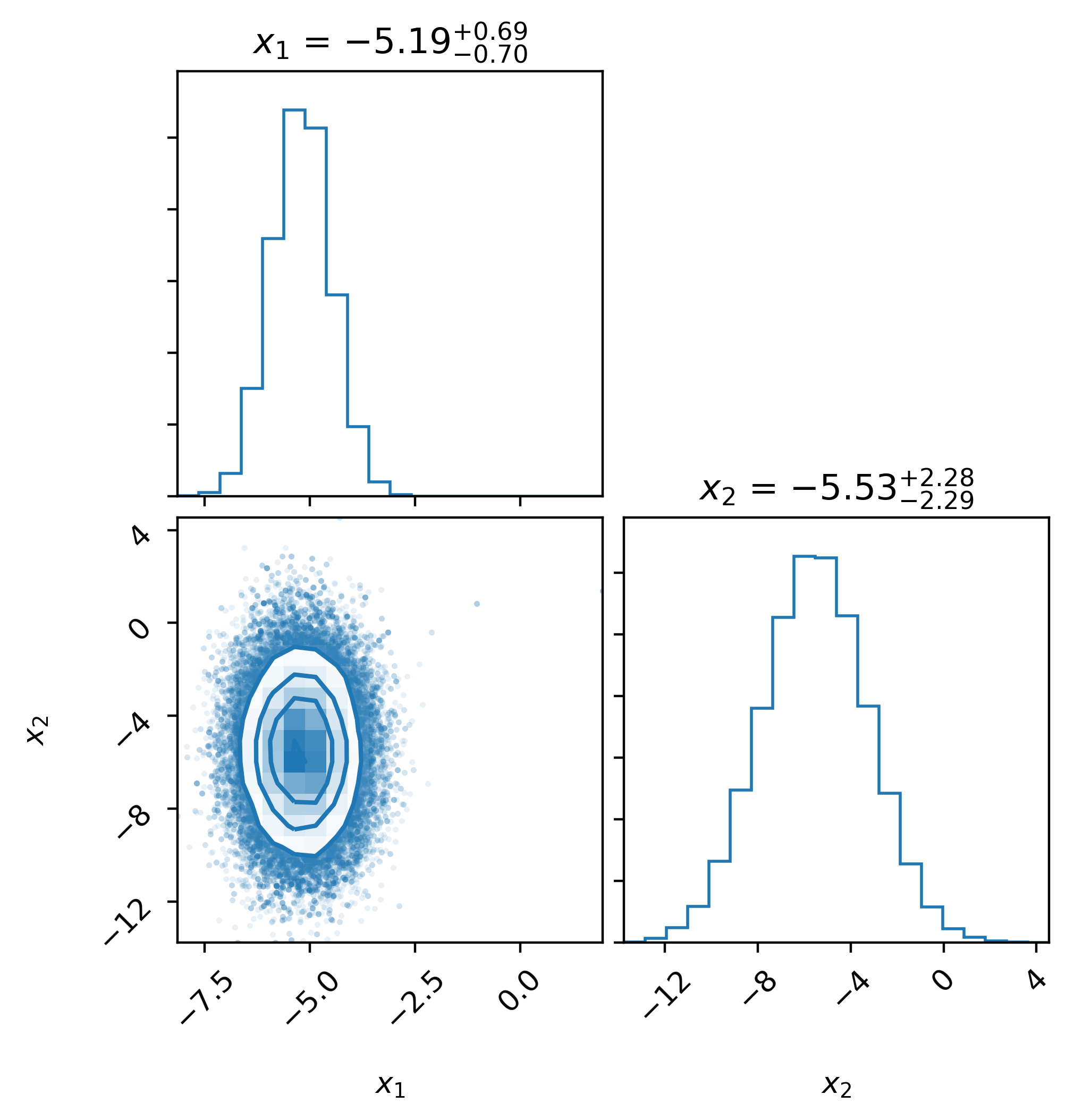}
    \caption{An example ``corner''/``pairwise''/``triangle'' plot, showing marginalized $1$-dimensional posterior distributions along the diagonal, and marginalized $2$-dimensional distributions of all unique parameter pairs in the lower left corner. Adapted from \href{https://jellis18.github.io/post/2018-01-02-mcmc-part1/}{https://jellis18.github.io/post/2018-01-02-mcmc-part1}, and using Ref.~\citep{corner}.}
    \label{fig:corner}
\end{figure}

It is often useful to investigate covariances amongst parameters in our posterior distribution. To that end we often plot $2$-dimensional marginalized posterior distributions. In this case, it makes most sense to compute point estimates and credible regions through the second ``level filling'' approach mentioned in the previous paragraph. We can even extend our visualization of such covariances by making a ``corner'' plot (e.g., Fig.~\ref{fig:corner}, also known as a ``pairwise'' or ``triangle'' plot) with the $2$-dimensional distributions of various pairs of parameters occupying positions in a [parameter $\times$ parameter] grid, with the relevant $1$-dimensional distributions along the diagonal.

\subsection{Upper limits}

Unlike in frequentist inference, Bayesian inference casts upper limits as a purely parameter estimation issue, by computing one-sided credible regions. To obtain the $95\%$ upper limit on some parameter $\theta$, one simply integrates upward from $-\infty$ (or the relevant prior lower boundary) until $95\%$ of the posterior distribution is enclosed, i.e.,
\begin{equation}
    \int_{\theta_\mathrm{low}}^{\theta_{95\%}} p(\theta|d)\,\,d\theta = 0.95,
\end{equation}
where $\theta_{95\%}$ is the $95\%$ upper limit value of $\theta$. In Bayesian inference, an upper limit is a boundary above which there is low credibility for a parameter to have a value.

\begin{figure}
\centering
	\includegraphics[width=0.5\columnwidth]{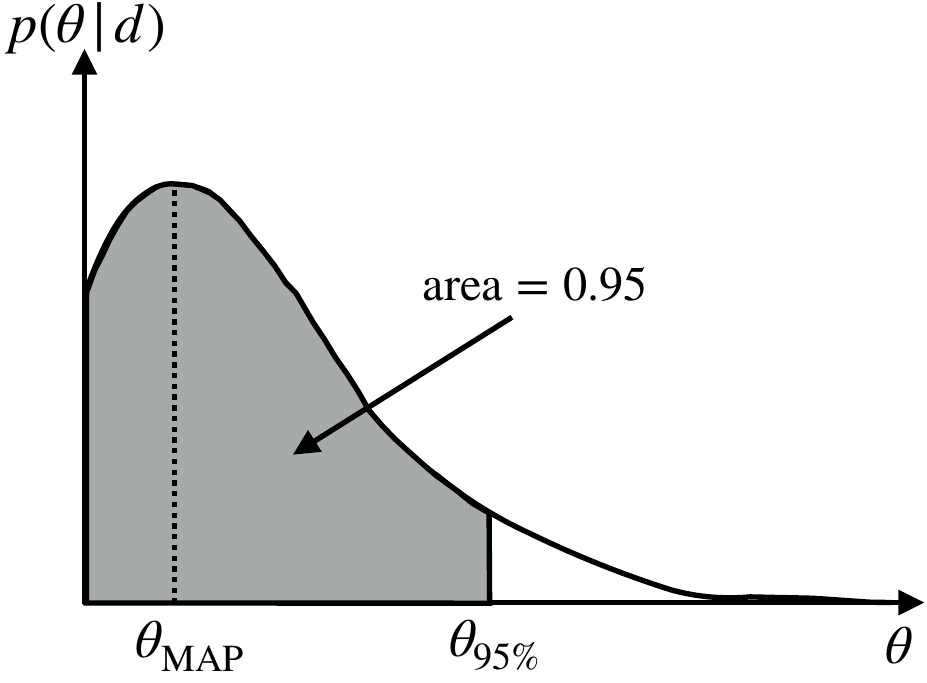}
    \caption{Bayesian parameter upper limits are computed by integrating the marginalized parameter posterior distribution upwards until $X\%$ of the posterior mass is enclosed; it is thus treated as a parameter estimation problem. This is markedly different from frequentist upper limits that are based on detection statistics. Adapted from Ref.~\citep{2017LRR....20....2R}.}
    \label{fig:bayes_ul}
\end{figure}

\subsection{Model Selection}

Parameters belong to a specified hypothesis model. If there are competing parametrized hypotheses to explain a physical phenomenon then we may wish to arbitrate which one the data supports best. There are two main ways that this has so far been performed in PTA searches: $(i)$ \textit{in-sample} techniques that summarize a model's fitness in one number; and $(ii)$ \textit{out-of-sample} techniques that test a model's predictive power.  

\subsubsection{In-sample model selection} 

In Bayesian inference this is performed by computing evidences for each model. As alluded to earlier, the evidence is the fully-marginalized likelihood that appears as a normalization factor in parameter estimation (because in that case we are exploring a fixed model), but must be accounted for in model selection. Basically, it is the average of the prior-weighted likelihood over the entire model parameter space:
\begin{equation}
    p(d) := \mathcal{Z} = \int p(d|\theta)p(\theta)d^n\theta.
\end{equation}

This is a challenging integral to evaluate, and I will explain some of the numerical techniques in the next chapter. However with an evidence for each model, one can compute evidence ratios that are more commonly known as \textit{Bayes factors}, which themselves may be weighted by prior odds ratios to compute posterior odds ratios:
\begin{equation}
    \mathcal{O}_{12} = \frac{P_1}{P_2}\times \frac{\mathcal{Z}_1}{\mathcal{Z}_2},
\end{equation}
where $\mathcal{O}_{12}$ is the posterior odds of model 1 versus model 2, $P_1/P_2$ is the prior odds of model 1 versus model 2, and $\mathcal{B}_{12}=\mathcal{Z}_1/\mathcal{Z}_2$ is the Bayes factor of model 1 versus model 2. Prior odds between models are most often set to be equal, giving equivalence between Bayes factor and posterior odds. 

But what does this posterior odds ratio actually mean? Well, it's your betting odds, telling you the strength of belief in one model over another as conditioned on the measured data. The interpretation of Bayes factors and odds ratios are problem specific, but there are some very loose rules of thumb as listed in Table \ref{jeffreys-scale-bayes}. However, one must always remember the old acronym \textit{GIGO} from the dawn of the information age, which stands for \textit{``Garbage In, Garbage Out''}. Odds ratios can only rank user-defined models, and if none of them encapsulate a physical model closest to reality then we are just selecting between some bad visions of the phenomenon.

Bayesian evidences ostensibly incorporate the spirit of Occam's razor by penalizing models with excess parameters that may lead to more spread-out likelihood support over the parameter space. However, note that the evidence only penalizes models with parameters that are constrained by the data-- no penalty is incurred if a model has many nuisance parameters that are unconstrained, as these will simply integrate against their prior to return one. Priors have a huge effect on Bayesian evidences; naively choosing a prior range that spans orders of magnitude beyond the likelihood support of a dataset can dilute the evidence value, making it more challenging to arbitrate between models. 

\begin{table} 
\caption{Loose rule-of-thumb for interpreting a Bayes factor, $\mathcal{B}_{12}~=~\mathcal{Z}_1/\mathcal{Z}_2$ between models $\mathcal{H}_1$ and $\mathcal{H}_2$ \citep{jeffreys1983}. \vspace{5pt}}\label{jeffreys-scale-bayes}
\centering
\begin{tabular}{c c l}
\hline
Bayes factor, $\mathcal{B}$ & $\ln(\mathcal{B})$ & Strength of evidence\\
\hline
    $< 1:1$ & $< 0$ & Negative (supports $\mathcal{H}_2$)\\
    $1:1$ to $3:1$ & $0 -1.1$ & Barely worth mentioning\\
    $3:1$ to $10:1$ & $1.1 - 2.3$ & Substantial\\
    $10:1$ to $30:1$ & $2.3 - 3.4$ & Strong\\
    $30:1$ to $100:1$ & $3.4 - 4.6$ & Very strong\\
    $> 100:1$ & $> 4.6$ & Decisive\\
\hline
\end{tabular}
\end{table}

\subsubsection{Out-of-sample model selection} 

This is performed via posterior predictive checks, wherein we test the efficacy of a data-trained model in predicting new data. The posterior predictive density for new data $\tilde{d}$ conditioned on a model $\mathcal{H}$ that has been trained on data, $d$ is
\begin{equation} \label{eq:posterior_predictive}
    p(\tilde{d}|d,\mathcal{H}) = \int p(\tilde{d}|\theta, \mathcal{H})p(\theta|d,\mathcal{H})\,\,d^n\theta,
\end{equation}
where we see that \textit{training} on $d$ in this context refers to recovering the posterior probability distribution of model parameters $\theta$, $p(\theta|d,\mathcal{H})$. The other term in the integrand is the probability of new data $\tilde{d}$ given parameters $\theta$, $p(\tilde{d}|\theta)$. One can see the strong analogy with the Bayesian evidence here, except that in this case the prior has been replaced with the posterior of $\theta$ as trained on some data $d$. Given how similar this is in form to the evidence, why would it benefit us to use this posterior predictive density instead? The most important reason is that, despite our best intentions, prior choices can sometimes be somewhat arbitrary in functional choice, and especially in choosing boundaries. As mentioned above, poor prior choices can have a huge impact on evidence values. This is tempered upon using the posterior predictive density, where arbitrary prior choices are replaced by the compact support of a data-trained posterior distribution. 

Rather than wait for new data to roll in, we can simply partition the data that we have into training and holdout samples. This is referred to as \textit{cross-validation}, where a common implementation is through $k$-fold partitioning. The dataset is split into $k$ exclusive subsets, where for each $k$ we obtain the posterior distribution of model parameters as conditioned on the data \textit{not in} $k$. The posterior predictive density can then be computed for each $k$ subset, and averaged over all subsets. This procedure is repeated for each model under consideration, with differences in the log predictive densities corresponding to ratios between models. 

\bibliographystyle{unsrt_new}
\bibliography{refs}

%% file: 06.tex
\chapter{Numerical Bayesian techniques} \label{chap:nbayes}
\epigraph{\textit{``I know Kung Fu''}}{Neo, ``The Matrix''}

All of the techniques mentioned in the previous chapter sound great, but how do we actually implement them? The posterior distributions under study will in general not have a simple functional form, and will have more than just a handful of parameter dimensions. This is important to stress, because there is nothing about the principles of Bayesian inference that demands numerical sampling techniques (like Markov chain Monte Carlo). If the posterior has a simple form, or even only a few dimensions, then you could imagine performing a grid-based mapping of the posterior distribution. But we certainly don't have that; each of our pulsars will have its own ephemeris and noise model, while the GW signal model will have its own set of parameters. The alternative is through random sampling techniques. With a set of random samples, $\{\vec{x}_i\}_N$, drawn from a posterior distribution, $p(\vec{x}|d,\mathcal{H})$, we can perform Monte Carlo integration over arbitrary (multivariate) functions, $f(\vec{x})$:
\begin{equation}
    \int f(\vec{x})p(\vec{x}|d,\mathcal{H})d^n x\approx{\frac{1}{N}}\displaystyle\sum_{i=1}^Nf(\vec{x}_i).
\end{equation}
It follows that marginalized posterior probability distributions can easily be obtained by simply binning the random samples in the relevant parameter subset.

So the case is made: we need an efficient path to explore high dimensional distributions, perform random draws, and to perform numerical integrals that correspond to marginalization. Iterative Markov chain Monte Carlo (MCMC) methods are the most often used in PTA GW searches. These make use of the \textit{ergodic theory of Markoc chains}, which states that regardless of the initial state of parameter exploration, after a sufficiently large number of iterative steps one will converge towards sampling of the target stationary distribution of interest.

\section{Metropolis algorithms}

\begin{algorithm}
    \caption{A typical Metroplis-Hastings algorithm} \label{alg1}
  \begin{algorithmic}[1]
    \STATE \textbf{Initialization} $\vec{x}_{(0)}\sim q(\vec{x})$
    \FOR{$i=1,2,\ldots$}
      \STATE Propose: \\\hspace{30pt}$\vec{y}\sim q(\vec{y}|\vec{x}_{i-1})$
      \STATE Acceptance probability: \\ \hspace{30pt}$\alpha(\vec{y}|\vec{x}_{i-1}) = \mathrm{min}\left\{ 1,\frac{p(\vec{y}) p(d|\vec{y})}{p(\vec{x}_{i-1})p(d|\vec{x}_{i-1})} \times \frac{q(\vec{x}_{i-1}|\vec{y})}{q(\vec{y}|\vec{x}_{i-1})} \right\}$
      \STATE $u \sim \mathrm{Uniform}(0,1)$ 
      \IF{$u<\alpha$}
        \STATE Accept the proposal: $\vec{x}_i \leftarrow \vec{y}$
      \ELSE
        \STATE Reject the proposal: $\vec{x}_i \leftarrow \vec{x}_{i-1}$
      \ENDIF
      \STATE $i = i + 1$
    \ENDFOR
  \end{algorithmic}
\end{algorithm}

Metropolis-Hastings sampling is an acceptance-rejection based method of drawing quasi-independent random samples from a target distribution. Moves within parameter space are proposed and then either probabilistically accepted or rejected based on the Metropolis-Hastings ratio \citep{hastings1970monte}. A standard Metropolis-Hastings algorithm is shown in Algorithm \ref{alg1}. We draw an initial vector of parameters, $\vec{x}_0$, from the assigned \textit{prior} distribution. At each subsequent iteration, $i$, a new trial point, $\vec{y}$, is drawn from a \textit{proposal distribution}, $q(\vec{y}|\vec{x}_{i-1})$ and the Metropolis-Hastings ratio is evaluated,
\begin{equation}
    \mathrm{MH}=\frac{p(\vec{y}) p(d|\vec{y})}{p(\vec{x}_{i-1})p(d|\vec{x}_{i-1})} \times \frac{q(\vec{x}_{i-1}|\vec{y})}{q(\vec{y}|\vec{x}_{i-1})}.
\end{equation}

With this MH value we now need a way to assess whether to accept the proposed point or reject it. We compare MH against a random draw, $u$, from a uniform distribution, $u\in U[0,1]$. If $u<$ MH then the move to the new point is accepted and we set $\vec{x}_i = \vec{y}$. If $u>$ MH, the move is rejected and we set $\vec{x}_i = \vec{x}_{i-1}$. There is usually an initial period of exploration where the chain of points is simply roaming around trying to locate regions of high probability in what could be a high-dimensional and potentially multi-modal distribution. This early sequence is usually referred to as the \textit{burn-in}, and discarded from any subsequent usage of the chain. Once the burn-in is over the chain will then be sampling from the target posterior distribution. Over a sufficiently large number of iterations, the chain of visited points corresponds to a collection of (quasi-)independent random draws that, through Monte Carlo integration, can be used for parameter estimation, computing quantiles, and even estimating detection significance.  

Including the proposal weighting ratio is crucial in ensuring \textit{detailed balance}, where the transitions in parameters between locations is proportional to the ratio of the posterior densities at those locations. Your exploration will perform best when the proposal distribution is as close as possible to the actual target distribution. To this end, it's possible and completely fine to perform pilot MCMC analyses in order to construct \textit{empirical parameter proposal distributions} for subsequent analyses that improve mixing. Remember that the proposal distribution is not the prior, so you are not double counting here. Just make sure in this case that proposal weightings are correctly implemented in the Metropolis-Hastings ratio. But by far the most common choice for the proposal distribution is that it be symmetric and centered on the current point, in which case we don't need to keep track of the proposal weight factors because that ratio is $1$. The simplest choice for such a symmetric proposal distribution is a multi-variate Gaussian distribution with mean equal to the current parameter vector, and a variance tuned to ensure good \textit{mixing}, i.e., the chain is actually exploring the parameter landscape and not simply getting stuck. We'll see more about how to choose this proposal width soon. 

\begin{figure}
\centering
	\includegraphics[width=0.32\columnwidth]{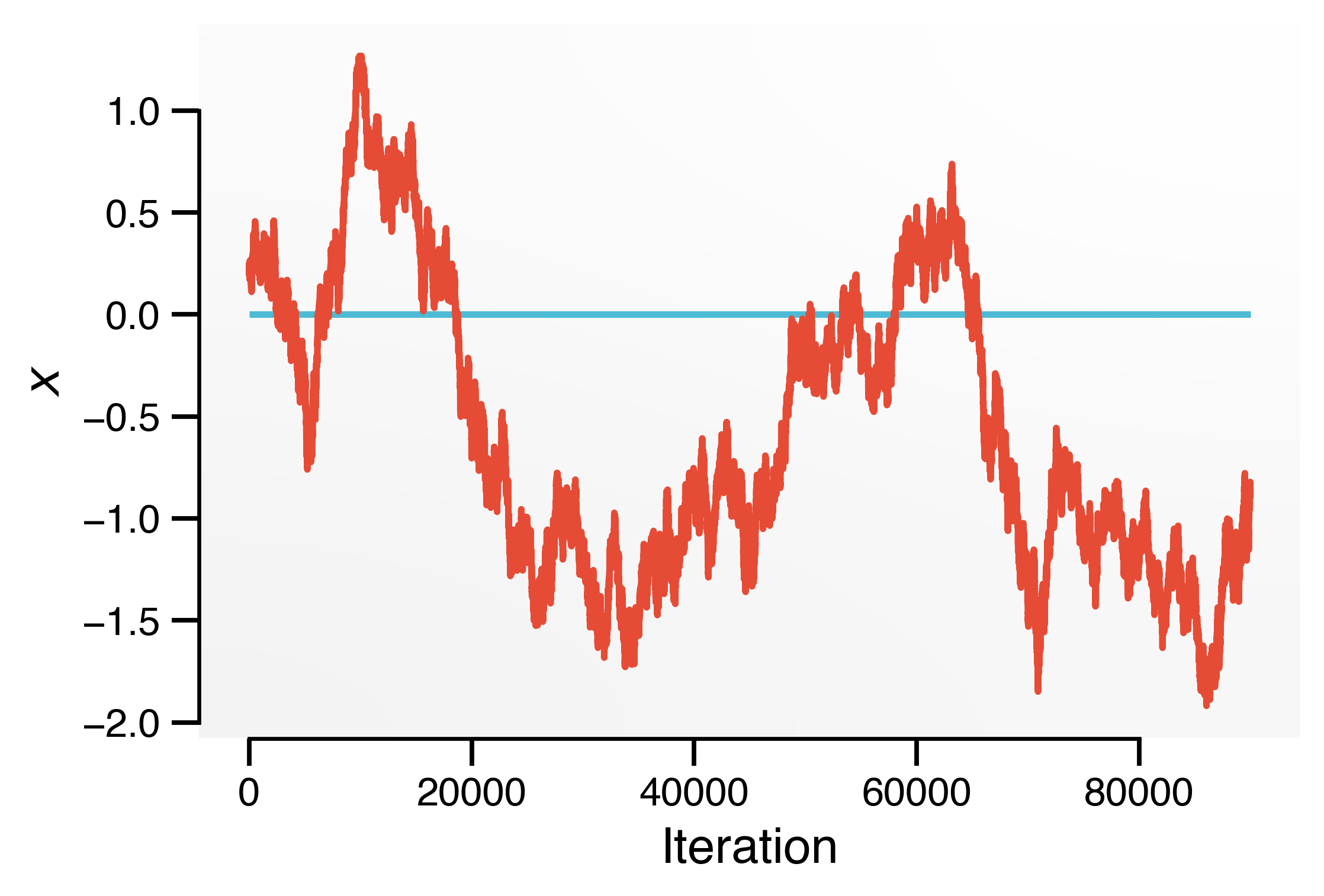}
	\includegraphics[width=0.32\columnwidth]{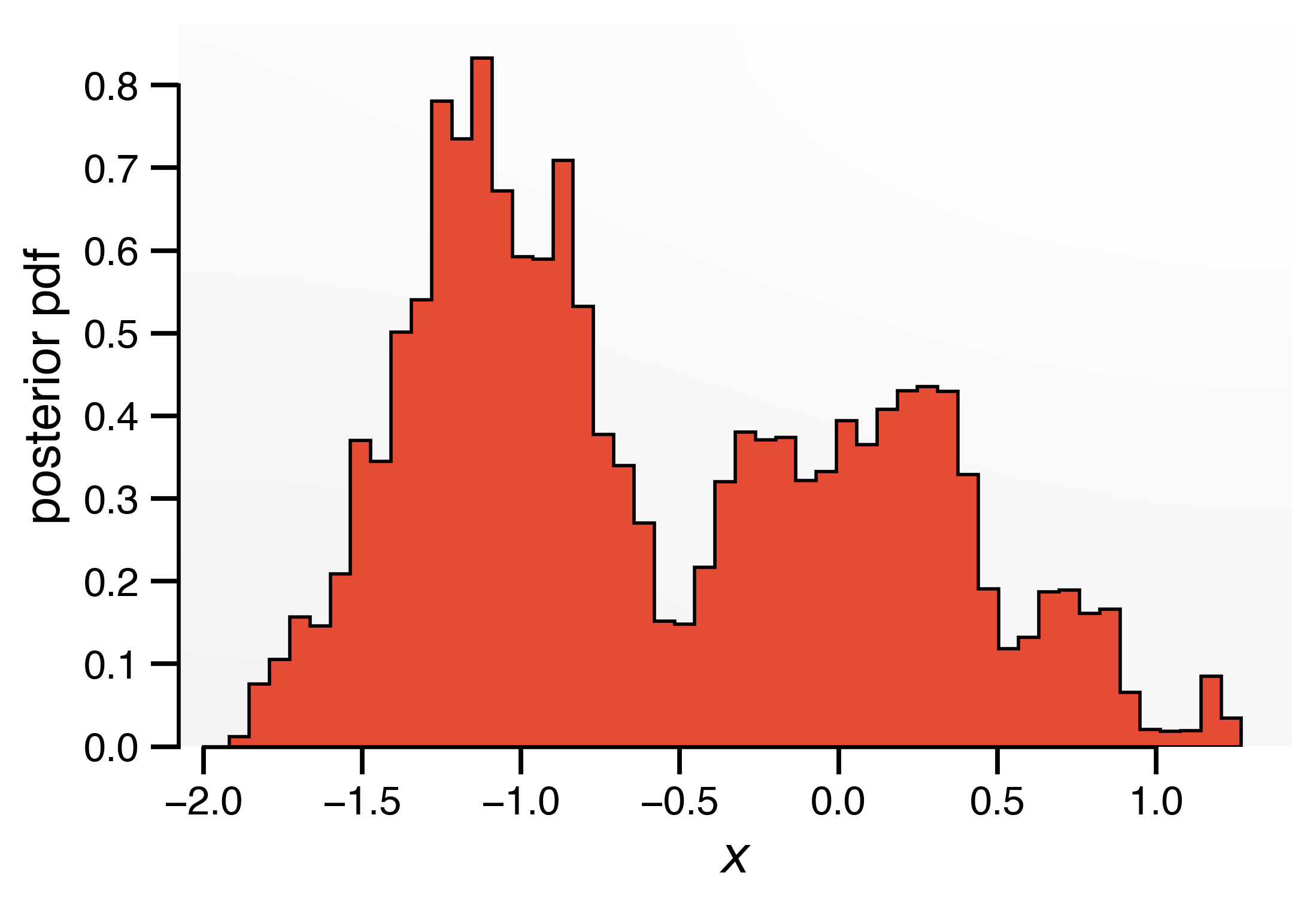}
	\includegraphics[width=0.32\columnwidth]{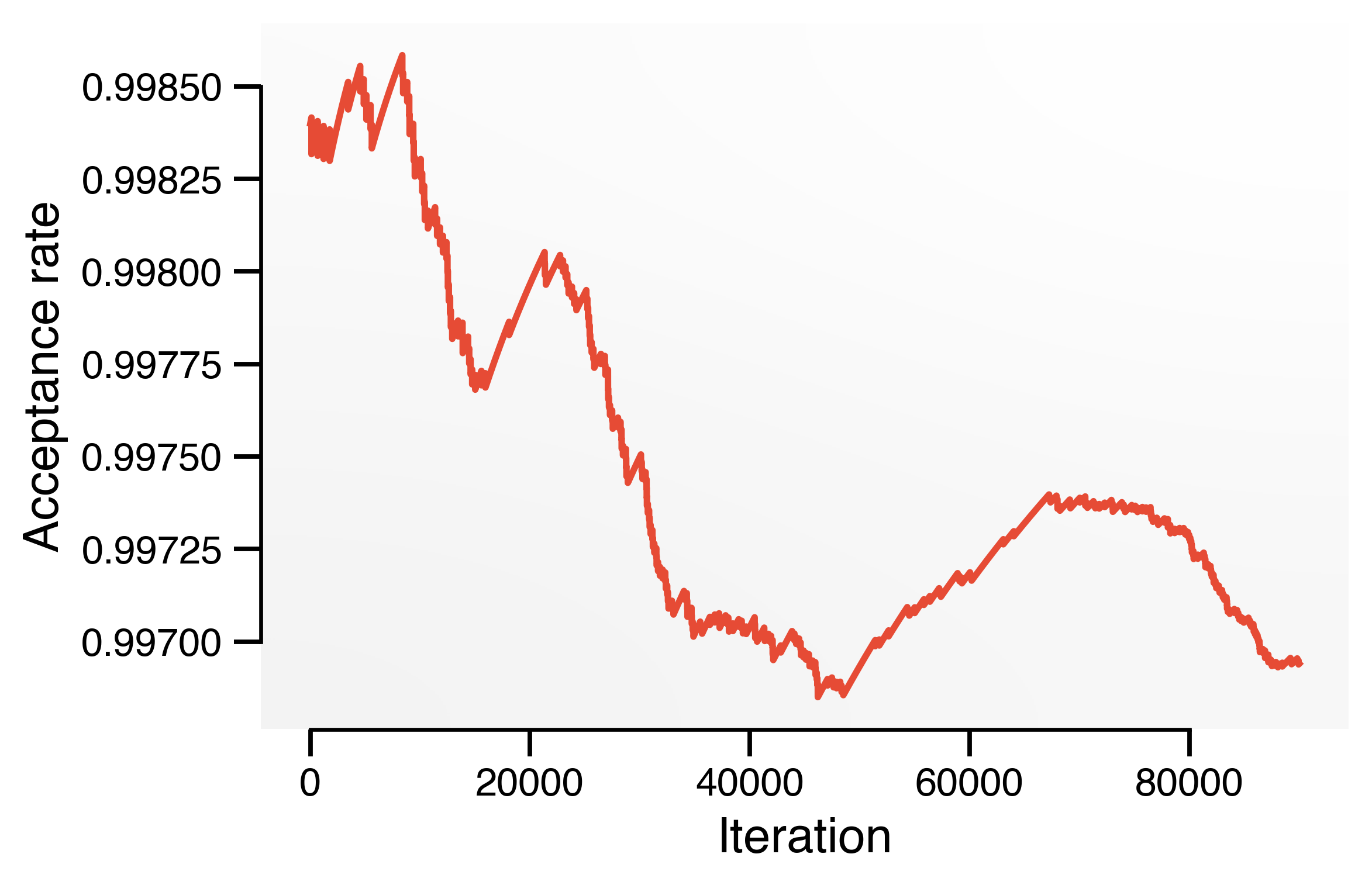} \\

	\includegraphics[width=0.32\columnwidth]{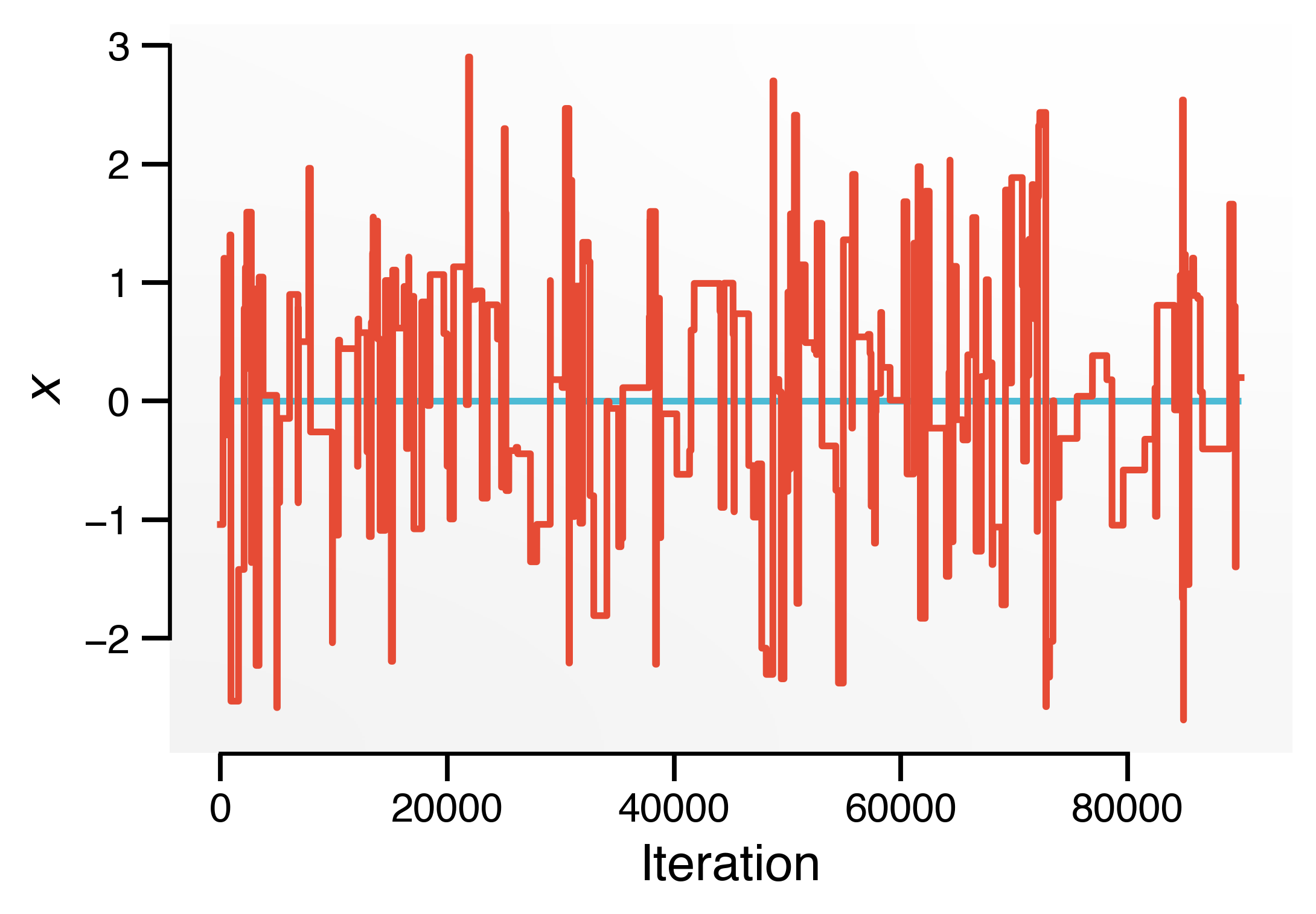}
	\includegraphics[width=0.32\columnwidth]{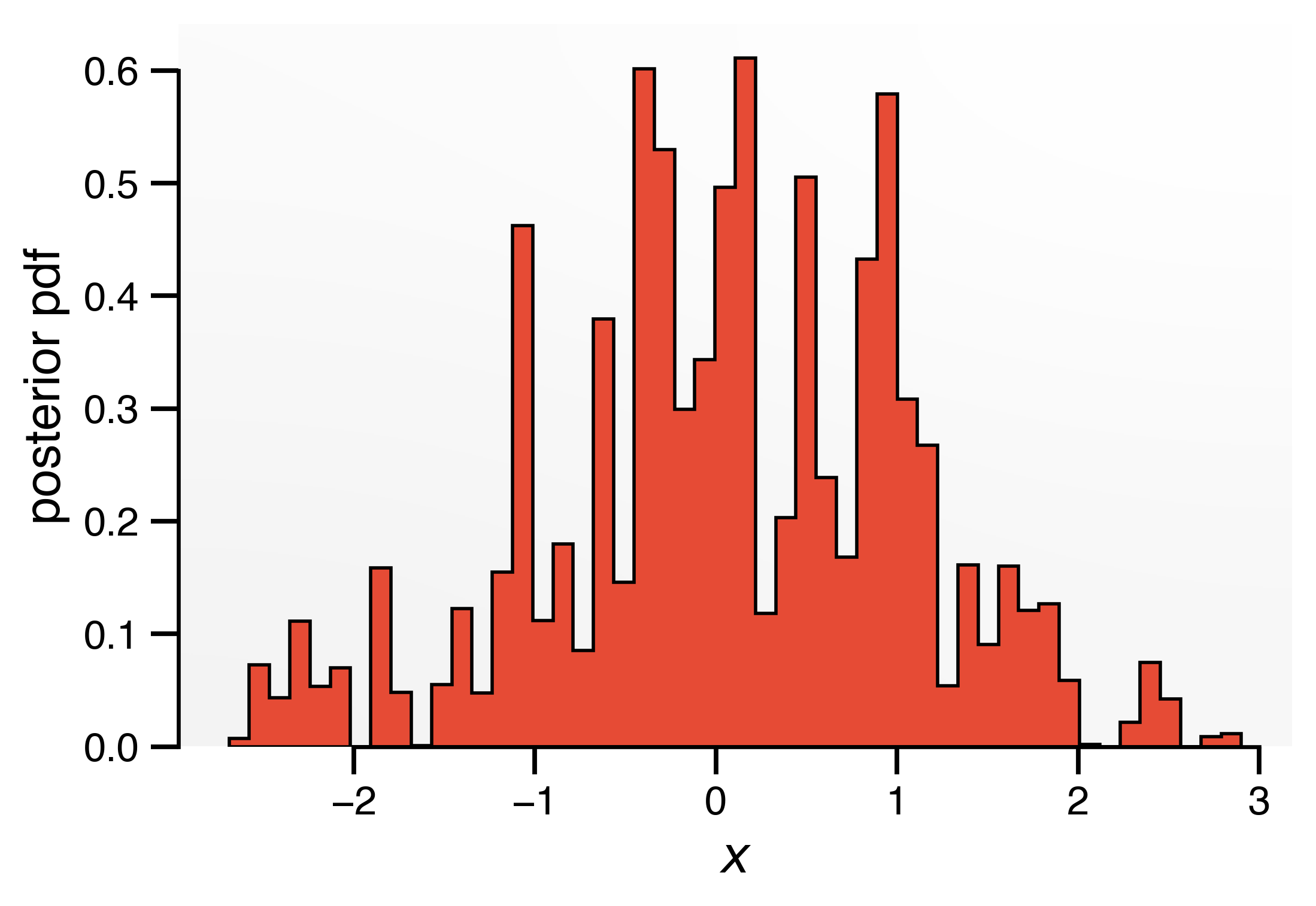}
	\includegraphics[width=0.32\columnwidth]{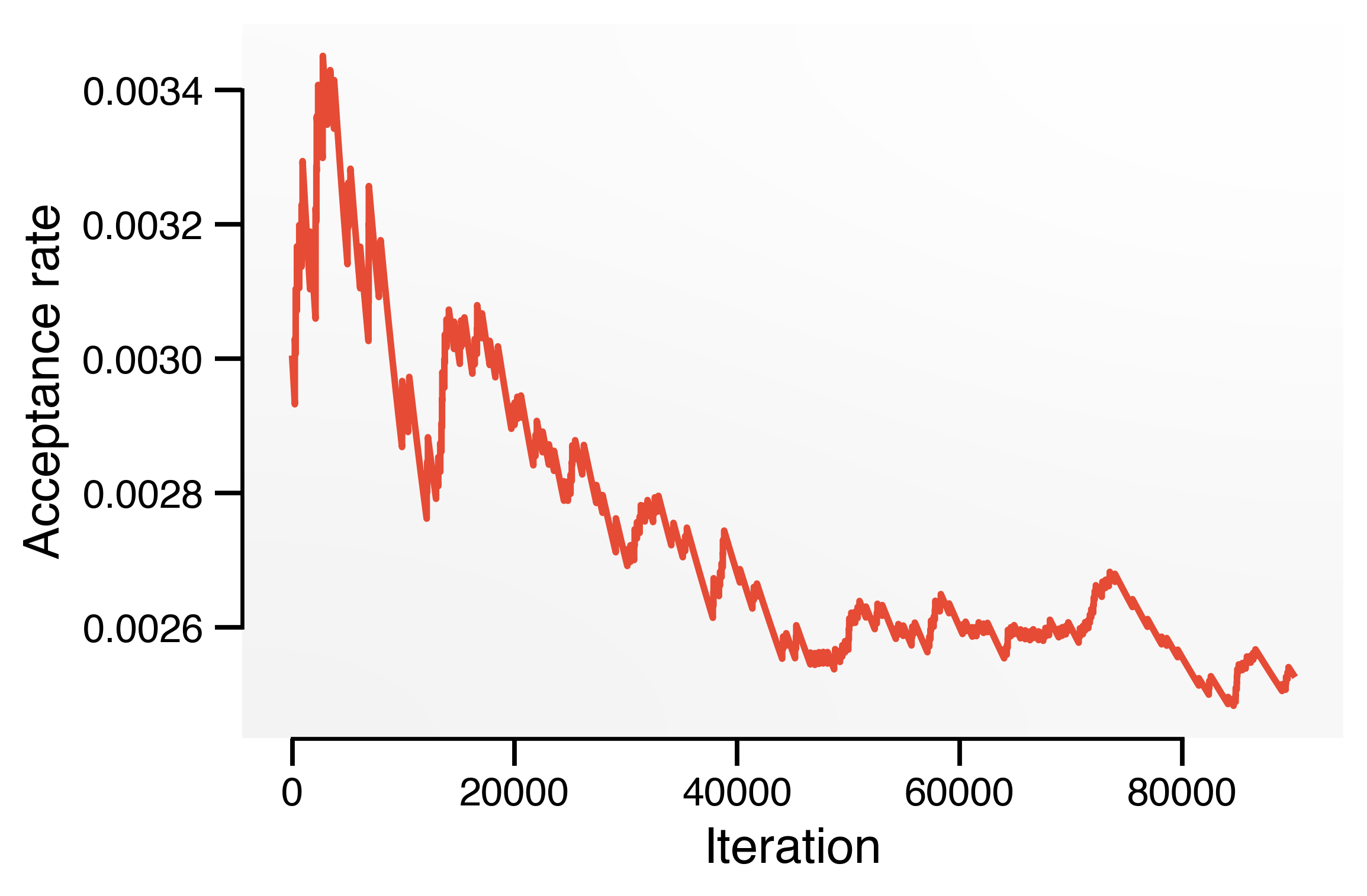} \\
    
	\includegraphics[width=0.32\columnwidth]{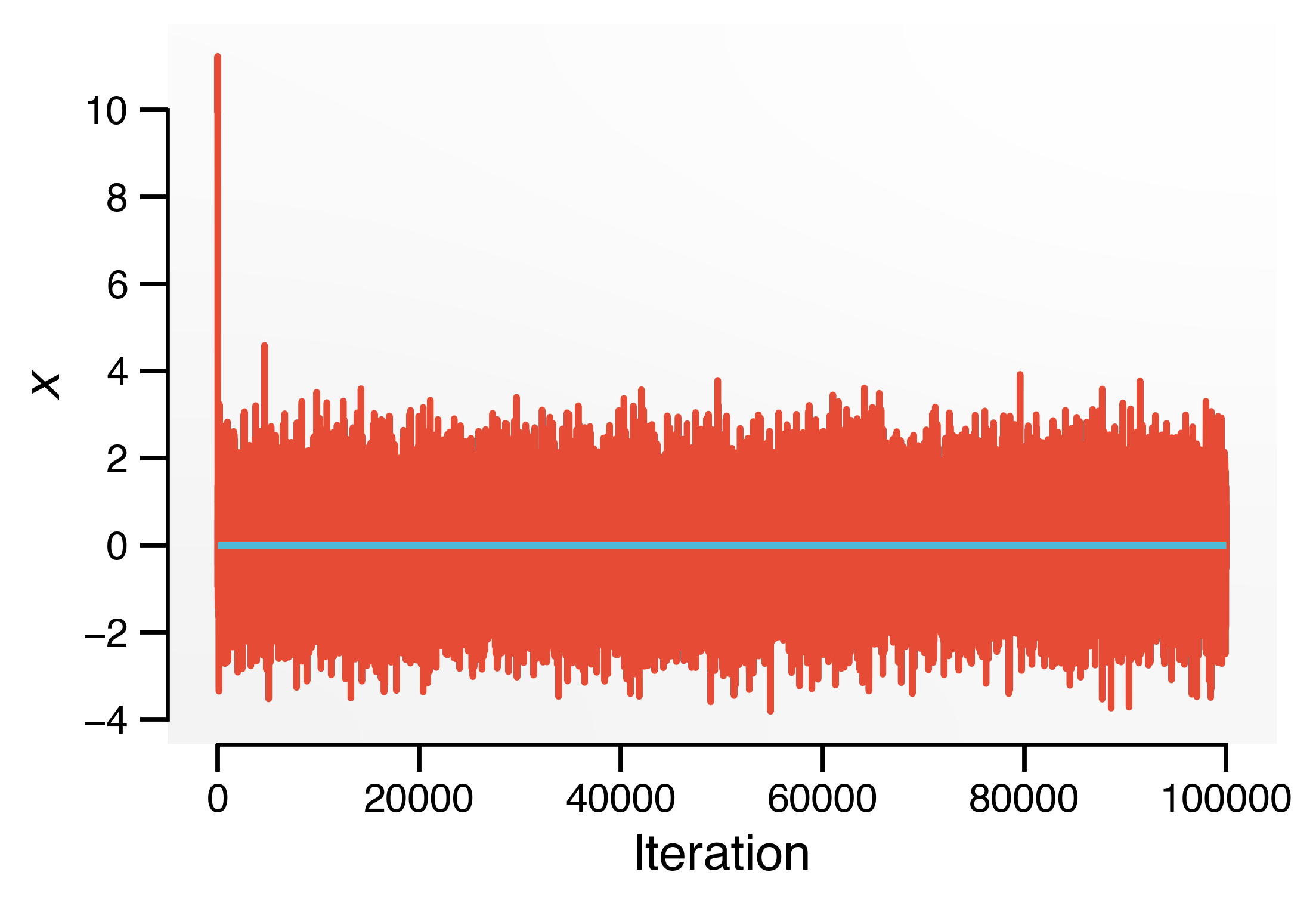}
	\includegraphics[width=0.32\columnwidth]{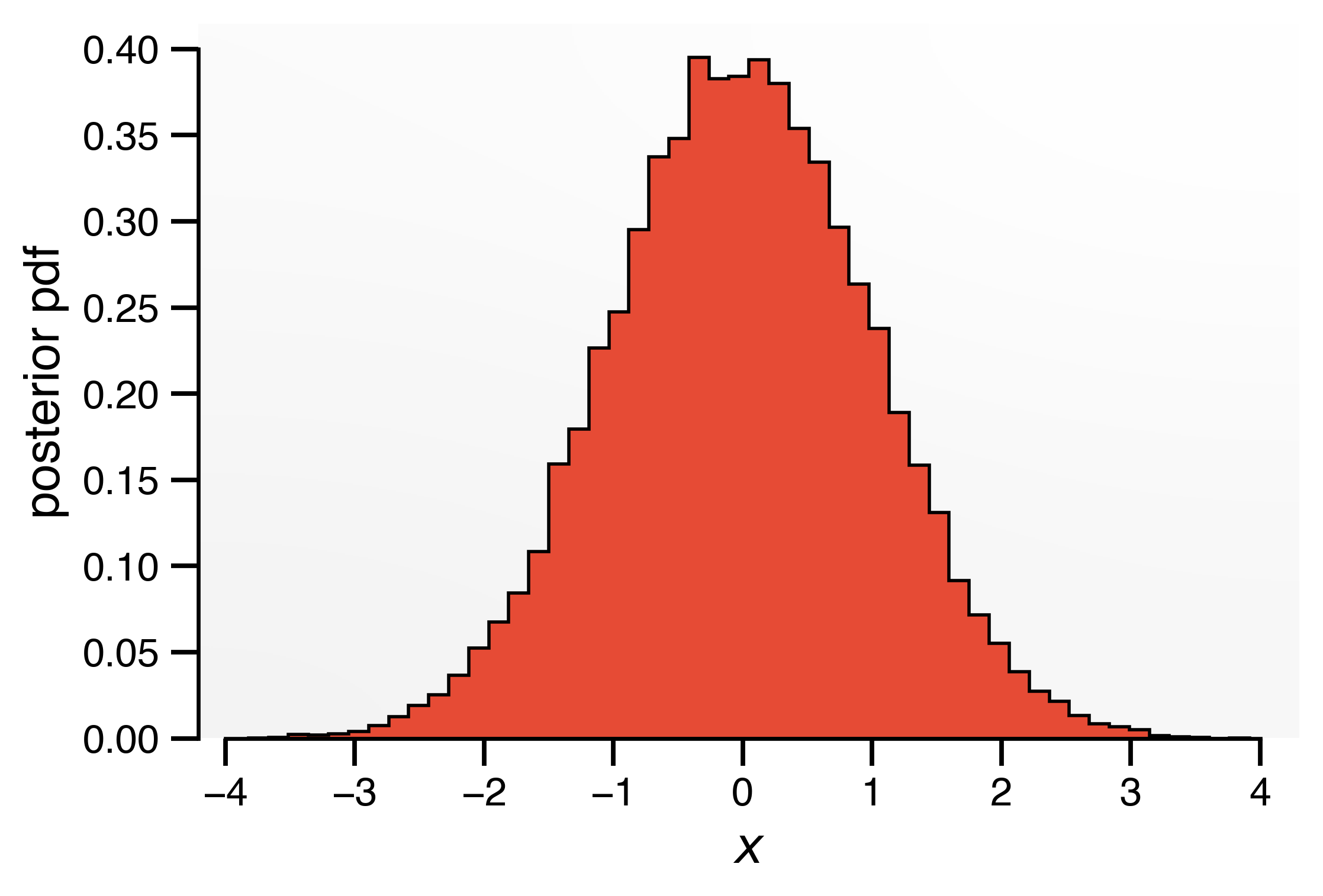}
	\includegraphics[width=0.32\columnwidth]{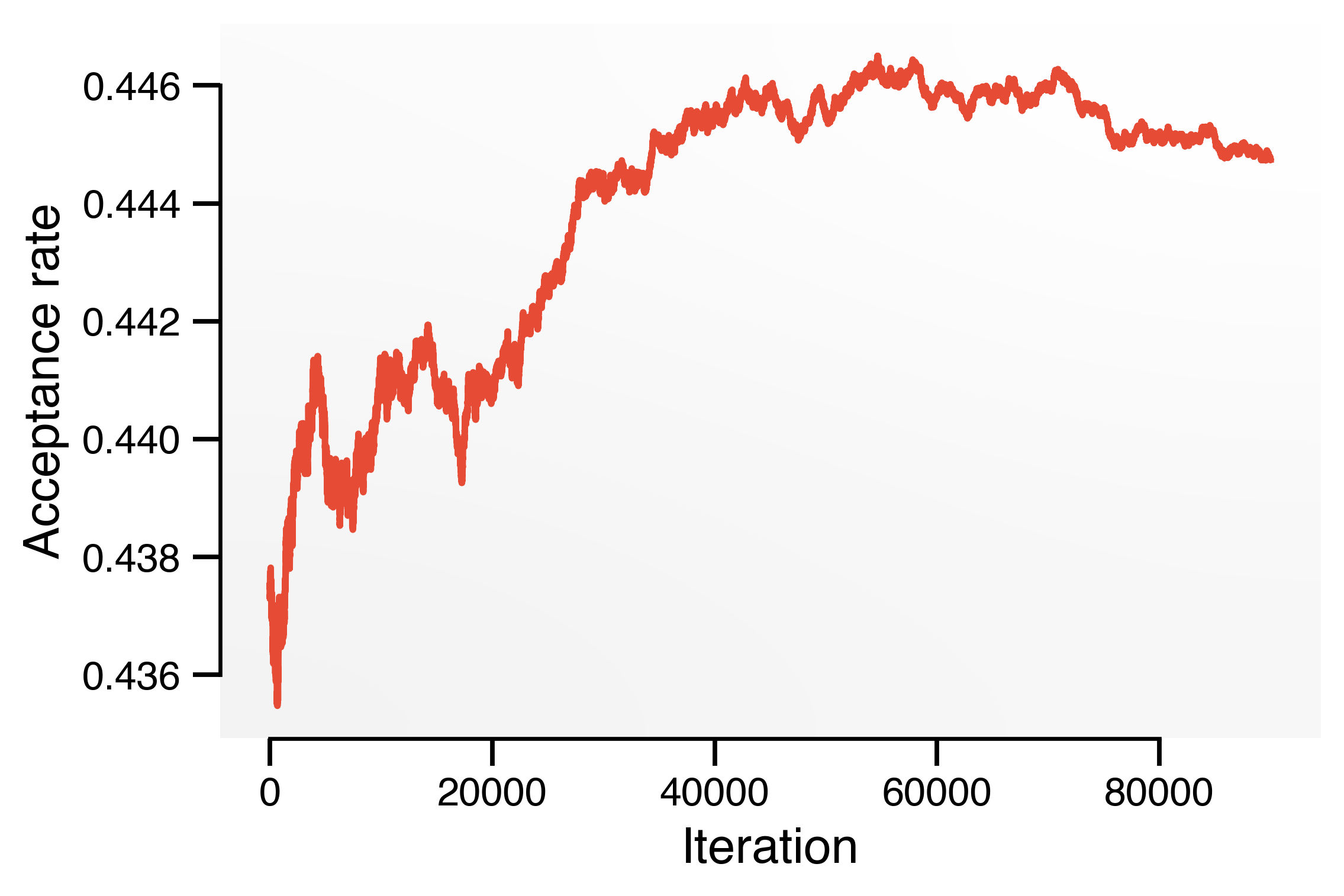}
	
    \caption{Each new MCMC problem should be initially diagnosed with some basic visual checks. Here we search for the mean of a zero-mean unit-variance Gaussian distribution. Each row corresponds to a different MCMC analysis of the same data. The top row has a Gaussian proposal width of $\sigma=0.01$, the middle row has $\sigma=500$, while the bottom row has an optimum proposal width of $\sigma=2.38$. The left panels show the \textit{trace-plot} of the parameter position versus MCMC iteration. The middle panels show the histogram of the MCMC chain (corresponding to the \textit{$1$-dimensional marginalized posterior distribution} of the parameter). The right panel shows the \textit{acceptance rate} of proposed points. A short period of \textit{burn-in} is shown in the lower left panel, with the chain starting far from the target region but quickly converging toward it. Adapted from \href{https://jellis18.github.io/post/2018-01-02-mcmc-part1}{https://jellis18.github.io/post/2018-01-02-mcmc-part1}.}
    \label{fig:mccm_chain_inspect}
\end{figure}

The basic visual chain inspections that should be performed for every MCMC problem are shown in Fig.~\ref{fig:mccm_chain_inspect}, corresponding to making \textit{trace-plots} to check chain mixing, histograms to check posterior recovery, and \textit{acceptance rate} monitoring to assess whether the width of the proposal distribution is a problematic factor to the chain's mixing. These are discussed in further detail below.

\subsection{How long to sample?} 

There are several diagnostics you can use to assess whether you have proceeded through a sufficient number of MCMC iterations. First off, always make \textit{trace-plots}! These are plots of parameter value (or even the log-likelihood value) versus MCMC iteration (e.g., left column of Fig.~\ref{fig:mccm_chain_inspect}). Not only will you be able to judge when to cut off the early burn-in stage, but you will also see whether your chain has settled down to carving out the same region(s) of parameter space. The latter is important, since your chain should be stationary, i.e., statistics computed from one chunk should be equivalent within sampling errors to the statistics of another chunk. 

Trace-plots are also an important visual way of judging whether your chain \textit{autocorrelation length} is too long, or whether your proposal acceptance rate is sub-optimal. The former is important because we perform numerical integrals using independent random draws from the target distribution. If the chain is simply wandering around in the same small region then it may be correlated over long timescales, requiring many iterations to fully explore the target distribution. Likewise, if the proposal distribution is attempting large jumps across parameter space to regions of low probability then the chain may not move at all, again leading to high autocorrelation lengths. There are two accepted practices for dealing with autocorrelation lengths larger than a few iterations: $(i)$ you can \textit{thin} your chain at the end of sampling by the computed autocorrelation length, ensuring you are using independent random samples; $(ii)$ simply run your chain for longer, using brute force to overcome the temporary local clumpiness of correlated samples. 

Another important statistic for assessing sampling convergence is the \textit{Gelman-Rubin R statistic} \citep{gelman_rubin92}. This determines convergence from $N$ independently launched MCMC explorations that initially started from a distribution of points that are ``over-dispersed'' relative to the variance of the target distribution. When the output from the $N$ chains are indistinguishable, the exploration has converged to the target distribution. The $R$-statistic compares \textit{in-chain} parameter variance to \textit{inter-chain} parameter variance. The parameter variance from the concatenation of chains is $\hat\sigma^2 = [(n-1)W + B] / n$, where $W$ is the mean of in-chain variances, $n$ is the number of MCMC iterations, and $B/n$ is the inter-chain variance. Upon convergence, $\hat\sigma^2$ and $W$ should be unbiased. But pre-convergence, $\hat\sigma^2$ should under-estimate the variance since parameter mixing has not been sufficient, while $W$ should over-estimate the variance by reflecting the over-dispersal of the initial chain starting points. The $R$ statistic assumes the target distribution is Gaussian, and has the form
\begin{equation}
    R = \left\{[(d+3)\hat{V}]/[(d+1)W]\right\}^{1/2},
\end{equation}
where $d=2\hat{V}^2/\mathrm{Var}(\hat{V})$, and $\hat{V} = \hat\sigma^2 + B/(nN)$, and $R\rightarrow 1$ as the chains converge.

\subsection{How to propose new parameters?} 

Judging sub-optimal proposal widths is related to the issue of autocorrelation lengths, since too large or too small proposed jumps will lead to poor chain exploration. You will see this in your trace-plot, and most importantly through in-flight \textit{acceptance rate} diagnostics. The ideal acceptance rate of proposed points should be somewhere between $\sim20-60\%$ (e.g., bottom right panel of Fig.~\ref{fig:mccm_chain_inspect}). Any smaller than this range runs the risk of the chain hardly moving around because the jumps are too big and correspond to regions of low probability (e.g., middle right panel of Fig.~\ref{fig:mccm_chain_inspect}). Any larger than this range and your chain is likely just performing tiny perturbations away from a region of local high probability (e.g., top right panel of Fig.~\ref{fig:mccm_chain_inspect}). 

To cut a long story short, you can find the optimum proposal width empirically through many trial MCMCs, or you can let the proposal tune itself. This self-tuning procedure is known as \textit{Adaptive Metropolis MCMC}, which, along with several other alternative proposal schemes, is the subject of the next several sub-sections.

\subsubsection{Adaptive Metropolis (AM)}

In AM \citep{haario2001adaptive} you use the empirically-estimated parameter covariance matrix to tune the width of the multi-variate Gaussian proposal distribution. Not only this, but tuning is updated during the sampling in order to reach optimal mixing. In practice this means that one uses the entire past history of the chain up until the current point to estimate the parameter covariance matrix\footnote{If one uses a recent chunk of the chain rather than the entire past history, then it can be shown that the chain is no longer ergodic \citep{haario1999adaptive}.}, scaling this covariance matrix by $\alpha=2.38^2/n_\mathrm{dim}$ to reach the optimal $25\%$ proposal acceptance rate for a target Gaussian distribution. One subtlety here is that by using more than just the most recent point to tune the sampling, our chain is no longer Markovian. This is easily resolved by allowing the chain to pass through a proposal tuning stage using AM, after which the proposal covariance matrix is frozen so that the chain is Markovian then on. 

The practical AM strategy is to factorize the chain-estimated parameter covariance matrix, $C$, using a Cholesky decomposition such that $C = LL^T$. A random draw from this multivariate Gaussian centered on the current point, $\vec{x}_{i-1}$, is given by $\vec{y} = \vec{x}_{i-1} + \alpha^{1/2} L\vec{u}$, where $u$ is an $n_\mathrm{dim}$ vector of random draws from a zero-mean unit-variance Gaussian. To reduce computational book-keeping time, the parameter covariance matrix can be updated periodically rather than after every single MCMC iteration.

\subsubsection{Single Component Adaptive Metropolis (SCAM)} 

With high-dimensional model parameter spaces, or even target posterior distributions with significant covariances amongst some parameters, the AM method may suffer from low acceptance rates. One method that addresses this is a variant on AM called Single Component Adaptive Metropolis (SCAM) \citep{haario2005componentwise}. This involves jumping along selected eigenvectors (or principal axes) of the parameter covariance matrix, which is equivalent to jumping in only one \textit{uncorrelated} parameter at a time. 

We consider the parameter covariance matrix again, except this time we perform an eigendecomposition on it, $C=D\Lambda D^T$, where $D$ is a unitary matrix with eigenvectors as columns, and $\Lambda=\mathrm{diag}(\sigma_\Lambda^2)$ is a diagonal matrix of eigenvalues. A SCAM jump corresponds to a zero-mean unit-variance jump in a randomly chosen uncorrelated parameter, equivalent to jumping along a vector of correlated parameters. A proposal draw is given by $\vec{y} = \vec{x}_{i-1} + 2.38 D_ju_j$, where $D_j$ is a randomly chosen column of D corresponding to the $j^\mathrm{th}$ eigenvector of $C$, and $u_j\sim\mathcal{N}(0,(\sigma^j_\Lambda)^2)$. 

\subsubsection{Differential Evolution (DE)}

Another popular proposal scheme is DE \citep{ter2006markov}, which is a simple genetic algorithm that treats the past history of the chain up until the current point as a \textit{population}. In DE, you choose two random points from the chain's history to construct a difference vector along which the chain can jump. A DE proposal draw is given by $\vec{y} = \vec{x}_{i-1} +  \beta(\vec{x}_{r_1} - \vec{x}_{r_2})$, where $\vec{x}_{r_{1,2}}$ are parameter vectors from two randomly chosen points in the past history of the chain, and $\beta$ is a scaling factor that is usually set to be the same as the AM scaling factor, $\alpha=2.38^2 / n_\mathrm{dim}$. 

\subsubsection{The Full Proposal Cocktail}

Real world MCMC should use a cocktail of proposal schemes, aimed at ensuring convergence to the target posterior distribution with minimal burn-in, optimal acceptance rate, and as short an autocorrelation length as possible. At each MCMC iteration the proposed parameter location can be drawn according to a weighted list of schemes, involving $(i)$ AM, $(ii)$ SCAM, $(iii)$ DE, $(iv)$ empirical proposal distributions, and finally $(iv)$ draws from the parameter prior distribution. The final prior-draw scheme allows for occasional large departures from regions of high likelihood, ensuring that we are exploring the full parameter landscape well, and avoiding the possibility of getting stuck in local maxima. Really, you can use any reasonable distribution you like to propose points-- your only constraint is to ensure that detailed balance is maintained through the relevant proposal weightings in the Metropolis-Hastings ratio.

\section{Gibbs sampling}

\begin{algorithm}
    \caption{A typical Gibbs algorithm} \label{alg2}
  \begin{algorithmic}[1]
    \STATE \textbf{Initialization} $x^{(0)}\sim p(x)$
    \FOR{$i=1,2,\ldots$}
      \STATE $x^{(i)}_1 \sim p(X_1=x_1| X_2=x^{(i-1)}_2, X_3=x^{(i-1)}_3,\ldots,X_{n_\mathrm{dim}}=x^{(i-1)}_{n_\mathrm{dim}})$
      \STATE $x^{(i)}_2 \sim p(X_2=x_2| X_1=x^{(i)}_1, X_3=x^{(i-1)}_3,\ldots,X_{n_\mathrm{dim}}=x^{(i-1)}_{n_\mathrm{dim}})$
      \STATE \vdots
      \STATE $x^{(i)}_{n_\mathrm{dim}} \sim p(X_{n_\mathrm{dim}}=x_{n_\mathrm{dim}}| X_1=x^{(i)}_1, X_2=x^{(i)}_2, X_3=x^{(i)}_3,\ldots)$
      \STATE $i = i + 1$
    \ENDFOR
  \end{algorithmic}
\end{algorithm}

Gibbs sampling \citep{geman1984stochastic} is an MCMC method that avoids acceptance-rejection techniques, and instead involves sweeping through each parameter (or block of parameters) to draw from their \textit{conditional} probability distributions, with all other parameters fixed to their current values. After sampling for a sufficiently large number of Gibbs steps, the principles of MCMC guarantee that this process of sequential conditional probability draws converges to the joint posterior distribution of the overall model parameter space. A standard Gibbs algorithm is shown in Algorithm \ref{alg2}.

What are the benefits of Gibbs sampling? Well, by drawing directly from the posterior conditionals, the auto-correlation length can be exceptionally small, with minimal burn-in. It's also fast; sequential draws directly from the parameter posterior conditionals means that we are not rejecting any points. 

What are the drawbacks of Gibbs sampling? You need to know the form of the conditional probability distributions for each parameter (or parameter blocks), and crucially how to draw samples from it. This can be a non-trivial problem, so typically a lot of effort is placed in manipulating the form of the posterior to find a conditional that is a standard probability distribution. This is where \textit{conjugate priors} become really handy; these are parameter priors for which the the posterior lies in the same family of distributions as the prior. For example, imagine that we have a Gaussian likelihood function with mean $\mu$ and variance $\sigma^2$ parameters. The conjugate prior on $\mu$ when $\sigma^2$ is known (as is assumed in Gibbs when sweeping through each parameter) is a Gaussian distribution, which means that the conditional distribution on $\mu$ is simply a Gaussian. Similarly, the conjugate prior on $\sigma^2$ when $\mu$ is known is an inverse gamma distribution. 

All is not lost if a parameter's conjugate prior is not known, or if it is not easy to directly sample from its conditional distribution. You can get inventive by embedding a short Metropolis-Hastings block within the Gibbs algorithm. For example, if there are parameters for which you can not directly draw from the conditional, then your Gibbs step for that parameter could be a short Metropolis-Hastings MCMC run. The goal is to run this until you have drawn a single quasi-independent random sample from the parameter posterior conditional distribution, with all other parameters fixed. You can then proceed through the remainder of your Gibbs steps as normal. 

\section{Evidence evaluation \& model selection}

As mentioned earlier, evaluating the Bayesian evidence is fraught with issues of ensuring adequate sampling of the prior volume, possible multi-modalities, and issues of computational power. There are some practical strategies with which to tackle it though, which I now briefly discuss. In what follows I will change notation such that model parameters are now labeled $\theta$, which can be a vector.

\subsection{Harmonic mean estimator}

It is possible to use an MCMC chain to directly compute the Bayesian evidence, $\mathcal{Z}$ \citep{10.2307/2346025}. Remember that Bayes' theorem can be phrased as
\begin{equation}
    \frac{p(\theta)}{\mathcal{Z}} = \frac{p(\theta|d)}{p(d|\theta)}. 
\end{equation}
Integrating both sides over $\theta$ gives
\begin{equation}
    \frac{1}{\mathcal{Z}} = \int \frac{p(\theta|d)}{p(d|\theta)}\, d\theta,
\end{equation}
since the prior is normalized to one, and $\mathcal{Z}$ is a constant. This means that with a sufficiently converged MCMC exploration of a model parameter space, we can use the MCMC samples to perform Monte Carlo integration to deliver the harmonic mean estimator of the evidence:
\begin{equation}
    \mathcal{Z} = \left\langle \frac{1}{p(d|\theta_i)} \right\rangle_{i=1,\ldots,N}^{-1},
\end{equation}
where the angled brackets denote the expectation over the MCMC samples, $\{\theta_1,\ldots,\theta_N\}$. This estimator is notoriously unreliable though, and is almost never employed for precision evidence calculation or model selection.

\subsection{Information criterion proxies}

There are several computationally-cheap proxies for the evidence that are based on Taylor expansions of the log likelihood function around its maximum.

\subsubsection{Bayesian Information Criterion (BIC)}

Also known as the Schwarz Information Criterion \citep{schwarz1978estimating}, the BIC is derived through an expansion of the log-likelihood function up to quadratic order around its maximum, followed by integrating over the model parameters to compute the evidence. In the limit of large numbers of observations, $N$, that exceed the number of model parameters, $k$, the BIC of model $\mathcal{H}$ can be written as
\begin{equation}
    \mathrm{BIC}(\mathcal{H}|d) \equiv -2\ln \mathcal{Z}_\mathcal{H} \approx k\ln N - 2\ln p(d|\hat{\theta}_\mathrm{MLE})
\end{equation}
where the second term after the approximation equality is the log-likelihood evaluated at the maximum likelihood parameters, $\hat{\theta}_\mathrm{MLE}$.

For a Gaussian likelihood, the BIC is linearly related to the $\chi^2$ value, but incorporates an additional level of parsimony to penalize model complexity based on the number of parameters. In model selection, the goal is to achieve the lowest BIC value possible among the ensemble of tested models. The rule of thumb with $\Delta\mathrm{BIC}$ values is that $0-2$ is not worth mentioning, $2-6$ is positive, $6-10$ is strong, and $>10$ is very strong evidence against the model with the higher BIC \citep{kass1995bayes}. 

\subsubsection{Akaike Information Criterion (AIC)}

The AIC is another information criterion based on similar approximations as the BIC, but penalizes model complexity less severely \citep{akaike1974new}:
\begin{equation}
    \mathrm{AIC}(\mathcal{H}|d) = 2k - 2\ln p(d|\hat{\theta}_\mathrm{MLE}).
\end{equation}
In fact, there is an additional form of the AIC that is corrected for small sample sizes \citep{liddle2007information}:
\begin{equation}
    \mathrm{AICc}(\mathcal{H}|d) = \mathrm{AIC}(\mathcal{H}|d) + \frac{2k(k+1)}{N-k-1} = 2k + \frac{2k(k+1)}{N-k-1} - 2\ln p(d|\hat{\theta}_\mathrm{MLE}).
\end{equation}

The model selection aim with AIC is the same as with BIC, where we try to find the model that minimizes the AIC value. However, note that both the AIC and BIC penalize model complexity based on the number of parameters, regardless of whether the parameter is constrained or not. This is different from the full Bayesian evidence, where parameters are only penalized if they are constrained by the data (but are nevertheless superfluous); parameters that are unconstrained by the data will simply produce a constant likelihood, integrating against their prior and canceling out in a ratio of model evidences.   

\subsection{Thermodynamic integration}

This evidence calculation technique employs \textit{parallel tempering}, which is a method of launching many MCMC chains of varying ``temperature'', $T$, designed to aggressively search parameter space and avoid trapping of chains in local likelihood maxima \citep{gelman1998simulating,ogata1989monte}. A chain's temperature denotes the degree to which the likelihood contrast is smoothed, where a $T=\infty$ chain is essentially exploring the prior space. Each chain has a different target distribution, $p(\theta|d,\beta) = p(\theta)p(d|\theta)^\beta$, where $\beta=1/T\in[0,1]$. Higher temperature chains are more capable of easily exploring regions far from the likelihood maximum. A multi-temperature Hastings step is used to ensure inter-chain mixing and rapid localization of the global maximum, where the multi-$T$ Hastings ratio is
\begin{equation}
    H_{i \rightarrow j} = \frac{p(d|\theta_i,\beta_j)p(d|\theta_j,\beta_i)}{p(d|\theta_i,\beta_i)p(d|\theta_j,\beta_j)}.
\end{equation}

The evidence for a chain with inverse temperature $\beta$ is \citep[see, e.g.,][and references therein]{2009PhRvD..80f3007L}
\begin{equation}
    \mathcal{Z}_\beta = \int p(\theta)p(d|\theta)^\beta\, d\theta,
\end{equation}
such that
\begin{align}
    \ln\mathcal{Z} &= \int_0^1 \frac{\partial\ln\mathcal{Z}}{\partial\beta}\,d\beta \nonumber\\
    &= \int_0^1\,d\beta \int \frac{p(\theta)p(d|\theta)^\beta}{\mathcal{Z_\beta}} \ln{p(d|\theta)}\,d\theta \nonumber\\
    &= \int_0^1 \langle\ln{p(d|\theta)}\rangle_\beta\,d\beta,
\end{align}
where angled brackets denote the expectation over the MCMC chain with inverse temperature $\beta$. This is an exact method for evidence evaluation, in the sense that the limitations on the evidence precision are through practical considerations like sampling efficiency, and the design of the temperature spacings (sometimes called the \textit{temperature ladder}). The choice of temperature-ladder spacing and maximum temperature is problem- and dimension-dependent; for an example of the ladder construction for individual SMBBH searches with PTAs, see Ref.~\citep{2013CQGra..30v4004E,2014ApJ...794..141A}, and for more general GW-search diagnostics, see Ref.~\citep{2010PhRvD..82j3007L}.  

\subsection{Nested sampling}

The most ubiquitous implementation of this Monte Carlo technique (originally proposed by Skilling \citep{skilling2004}) is the \textsc{MultiNest} algorithm (see Ref.~\citep{2008MNRAS.384..449F,2009MNRAS.398.1601F,2019OJAp....2E..10F}). We discuss this algorithm in the following, but note that there are many other variants \citep[e.g.,][]{2015MNRAS.453.4384H,2017arXiv170508978P,2019arXiv190911873S,2020MNRAS.493.3132S}. A model's parameter space is populated  with ``live'' points drawn from the prior. These points climb through nested contours of increasing likelihood, where at each iteration the points are ranked by likelihood such that the lowest ranked point is replaced by a higher likelihood substitute. The latter step poses the biggest difficulty-- drawing new points from the prior volume provides a steadily decreasing acceptance rate, since at later iterations the live-set occupies a smaller volume of the prior space. \textsc{MultiNest} uses an ellipsoidal rejection-sampling technique, where the current live-set is enclosed by (possibly overlapping) ellipsoids, and a new point drawn uniformly from the enclosed region. This technique successfully copes with multimodal distributions and parameter spaces with strong, curving degeneracies.

The multi-dimensional evidence integral is calculated by transforming to a one-dimensional integral that is easily numerically evaluated. The prior volume, $X$ is defined as
\begin{equation}
    dX = p(\theta)d^n\theta,
\end{equation}
such that
\begin{equation}
    X(\lambda) = \int_{\mathcal{L}(\theta)>\lambda}p(\theta)d^n\theta,
\end{equation}
where the integral extends over the region of the $n$-dimensional parameter space contained within the iso-likelihood contour $\mathcal{L}(\theta)=\lambda$. Thus the evidence integral can be written as
\begin{equation}
    \mathcal{Z} = \int p(d|\theta)p(\theta)\,d^n\theta = \int_0^1\mathcal{L}(X)\,dX,
\end{equation}
where $\mathcal{L}(X)$ is a monotonically decreasing function of $X$. Ordering the $X$ values allows the evidence to be approximated numerically using the trapezium rule
\begin{equation}
    \mathcal{Z} = \displaystyle\sum_{i=1}^M\mathcal{L}_iw_i,
\end{equation}
where the weights, $w_i$, are given by $w_i = \left(X_{i-1}-X_{i+1}\right)/2$.

Although designed for evidence evaluation, the final live-set and all discarded points can be collected and assigned probability weights to compute the posterior probability of each point. These points can be used to deduce posterior integrals as in other MCMC methods.

\subsection{Savage-Dickey density ratio}

We consider the case of nested models, $\mathcal{H}_1$, $\mathcal{H}_2$ that share many common parameters, $\theta$. However one of the models, $\mathcal{H}_1$ includes a signal that can be switched on and off by some parameter, $A$. Identifying this parameter (or corner of parameter space) that nulls the signal is the key to implementing
this Savage-Dickey approach \citep{dickey1971}, such that
\begin{equation} \label{eq:nested_mods}
    p(d|A=0,\theta; \mathcal{H}_1) = p(d|\theta; \mathcal{H}_2).
\end{equation}
Let's work out the posterior density for $A=0$:
\begin{align}
    p(A=0|d;\mathcal{H}_1) &= \int p(A=0,\theta|d;\mathcal{H}_1)\,d^n\theta \\\nonumber
    &= \int \frac{p(d|A=0,\theta;\mathcal{H}_1)p(A=0)p(\theta)}{p(d|\mathcal{H}_1)}\,d^n\theta \nonumber\\
    &= \int \frac{p(d|\theta;\mathcal{H}_2)p(A=0)p(\theta)}{p(d|\mathcal{H}_1)}\,d^n\theta \nonumber\\
    &= \frac{p(A=0)}{p(d|\mathcal{H}_1)} \int p(d|\theta;\mathcal{H}_2)p(\theta)\,d^n\theta \nonumber\\
    &= \frac{p(d|\mathcal{H}_2)}{p(d|\mathcal{H}_1)}p(A=0),
\end{align}
where we have used Bayes' Theorem on the second line, and Eq.~\ref{eq:nested_mods} on the third line. We can see immediately that the Bayes factor between model $1$ and model $2$ is
\begin{equation}
    \mathcal{B}_{12} = \frac{p(d|\mathcal{H}_1)}{p(d|\mathcal{H}_2)} = \frac{p(A=0)}{p(A=0|d;\mathcal{H}_1)},
\end{equation}
corresponding to the ratio of the prior to marginal posterior densities of $A=0$. If the data is informative of a non-zero signal being present then the posterior support at $A=0$ will be less than the prior support, such that $\mathcal{B}_{12}>1$, as desired.

The Savage-Dickey density ratio has been used extensively in PTA searches for GWs, where the amplitude of a fixed spectral-index GWB is the relevant on/off signal switch. For practical purposes, the position of zero signal is the lower sampling limit of the GWB strain amplitude, $\log_{10}A_\mathrm{low} = -18$. This limit is chosen to make the lowest
GWB signal level much smaller than the typical intrinsic noise levels in the pulsars. The recovered Bayes factor is between a model with intrinsic pulsar noise plus a GWB versus a model with noise alone. This does not immediately give a measure of the significance of GW-induced inter-pulsar correlations, which is the most sought-after statistic for PTA searches.

\subsection{Product space sampling}

In this approach we treat model selection as a parameter estimation problem \citep{carlin1995bayesian,godsill2001relationship,2018ApJ...859...47A,2020PhRvD.102h4039T}. We define a hypermodel, $\mathcal{H}_*$, whose parameter space is the concatenation of all sub-model spaces under consideration, along with an additional model tag, $n$. This model tag is discrete, but in the process of sampling we can simply apply appropriate model behaviour within some bounded region of a continuously sampled variable. 

At a given iteration in the sampling process we evaluate the model tag's position that indicates the ``active'' sub-model to be used for the likelihood evaluation. The hypermodel parameter space, $\mathbf{\theta_*}$, is filtered to find the relevant parameters of the active sub-model, which are then passed to the likelihood function. The parameters of the inactive sub-models do not contribute to, and are not constrained by, the active likelihood function. As sampling proceeds, the model tag will vary between all sub-models such that the relative fraction of iterations spent in each sub-model provides an estimate of the posterior odds ratio. 

Let's write the marginalized posterior distribution of the model tag from the output of MCMC sampling:
\begin{align}
    p(n|d;\mathcal{H}_*) &= \int p(\theta_*,n|d;\mathcal{H}_*)d\mathbf{\theta} \nonumber\\
    &= \frac{1}{\mathcal{Z}_*} \int p(d|\theta_*,n;\mathcal{H}_*)p(\theta_*,n|\mathcal{H}_*)d\theta,
\end{align}
where $\mathcal{Z}_*$ is the hypermodel evidence. For a given $n$ the hypermodel parameter space is partitioned into active, $\theta_n$, and inactive, $\theta_{\bar{n}}$, parameters, $\theta_* = \{\theta_n,\theta_{\bar{n}}\}$, where the likelihood $p(d|\theta_*,n;\mathcal{H}_*)$ is independent of the inactive parameters. The prior term can then be factorized:
\begin{equation}
    p(\theta_*,n|\mathcal{H}_*) = p(\theta_n|\mathcal{H}_n)p(\theta_{\bar{n}}|\mathcal{H}_{\bar{n}})p(n|\mathcal{H}_*),
\end{equation}
such that
\begin{align}
    p(n|d,\mathcal{H}_*) &= \frac{p(n|\mathcal{H}_*)}{\mathcal{Z}_*} \int p(d|\theta_n;\mathcal{H}_n) p(\theta_n|\mathcal{H}_n) d\theta_n \nonumber\\
    &= \frac{p(n|\mathcal{H}_*)}{\mathcal{Z}_*} \mathcal{Z}_n,
\end{align}
where $\mathcal{Z}_n$ is the evidence for sub-model $n$. We have already marginalized over inactive parameters since they only appear in the prior term $p(\mathbf{\theta}_{\bar{n}}|\mathcal{H}_{\bar{n}})$, which integrates to one. Thus the posterior odds ratio between two models is given by:
\begin{equation}
    \mathcal{O}_{12} = \frac{p(n_1|\mathcal{H}_*)\mathcal{Z}_1}{p(n_2|\mathcal{H}_*)\mathcal{Z}_2} = \frac{p(n_1|d;\mathcal{H}_*)}{p(n_2|d;\mathcal{H}_*)},
\end{equation}
where the hypermodel evidence cancels in this ratio of the two sub-models. 

We can refine this method further by designing an iterative scheme, where pilot runs provide an initial estimate of $\tilde{\mathcal{O}}_{12}$, which is then followed by focused runs for an improved estimate. In a focused run, we weight model $1$ by $1 / (1 + \tilde{\mathcal{O}}_{12})$ and model $2$ by $\tilde{\mathcal{O}}_{12} / (1 + \tilde{\mathcal{O}}_{12})$. This improves mixing and exploration by reducing the posterior contrast across the model landscape. The posterior odds ratio from the focused run is then re-weighted by the pilot run estimate to provide a more accurate value of $\mathcal{O}_{12}$.  
The product-space estimator (sometimes called a \textit{hypermodel analysis} in PTA circles) of the posterior odds ratio is simple to implement, applicable to high dimensional parameter spaces, and allows direct model comparison.

Product space sampling is being used within PTA data analysis to deduce bespoke noise models for the time-series of each pulsar, and to assess the odds of our data showing inter-pulsar correlations matching the Hellings \& Downs curve. The implementation can be found in {\texttt {enterprise$\_$extensions}}\footnote{\href{https://github.com/nanograv/enterprise_extensions}{https://github.com/nanograv/enterprise$\_$extensions}}.

\bibliographystyle{unsrt_new}
\bibliography{refs}

%% file: 07.tex
\chapter{The PTA Likelihood} \label{chap:pta_likelihood}
\epigraph{\textit{``It's a kind of magic.''}}{Connor MacLeod, ``Highlander''}

Constructing the PTA likelihood begins with an understanding of the dominant influences on the pulse TOAs. To leading order, this is of course the deterministic pulsar timing ephemeris, depending on the pulse period, spindown rate, astrometric effects, radio-frequency dependent delays due to propagation through the ionized interstellar medium, etc. Pulsar-timing astronomers diligently construct these timing ephemerides over repeated observations, accruing more evidence for marginal effects that may be related to binary orbital effects, such as e.g., Shapiro delay. The result of such painstaking work over years and years is a best-fit timing ephemeris with respect to leading-order noise processes in the TOAs, such as radiometer noise (really just pulse shape template-fitting errors), low-frequency/long-timescale noise, etc. 

The search for GWs is in the \textit{timing residuals}, literally the remnant time-series upon subtracting the best-fit timing ephemeris from the raw observed TOAs. Anything left after this subtraction should merely be noise and GWs. ``But wait!'', I hear you holler, with righteous indignation; ``What if the original timing-ephemeris fit accidentally removed some of the GW signal? Won't this bias or hinder the search?'' This is of course true, so let's take a dive into the pulsar-timing data model. 

Throughout the following it will be convenient to consider both sampling frequencies of a pulsars' time-series data, $f$, and the radio frequencies at which the TOAs are observed, $\nu$. Unless otherwise stated, ``frequency'' will be taken to mean the sampling frequency of the time-series, which is often synonymous with the frequency of GWs being probed. A final note that a detailed study of the computational evaluation time of each part of the PTA likelihood will not be given in the following; readers interested in that should see the comprehensive overview given in Ref.~\citep{2014PhRvD..90j4012V}.

\section{The Pulsar-timing Data Model}

A pulsar's TOAs can be written as a sum of deterministic and stochastic parts
\begin{equation}
    \vec{t}_\mathrm{TOA} = \vec{t}_\mathrm{det} + \vec{t}_\mathrm{stoch},
\end{equation}
where $\vec{t}_\mathrm{TOA}$ is the vector of pulse arrival times of length $N$, and $\vec{t}_{\mathrm{det}/\mathrm{stoch}}$ are the corresponding deterministic and stochastic components. We model all stochastic contributions as Gaussian random processes; there are important reasons for this that will be seen in detail later, but essentially this means that all statistical properties can be encapsulated in the first and second moments of the stochastic time-series, allowing us to use some of the lovely properties of Gaussian distributions and their integrals when formulating the PTA likelihood\footnote{For a discussion of modifying the PTA data model in the cases of non-Gaussianity or non-stationarity, see \citep{2014MNRAS.444.3863L} and \citep{2016PhRvD..93h4048E}, respectively.}.

\subsection{Timing ephemeris} \label{sec:tmodel}

\begin{figure}
	\includegraphics[width=\columnwidth]{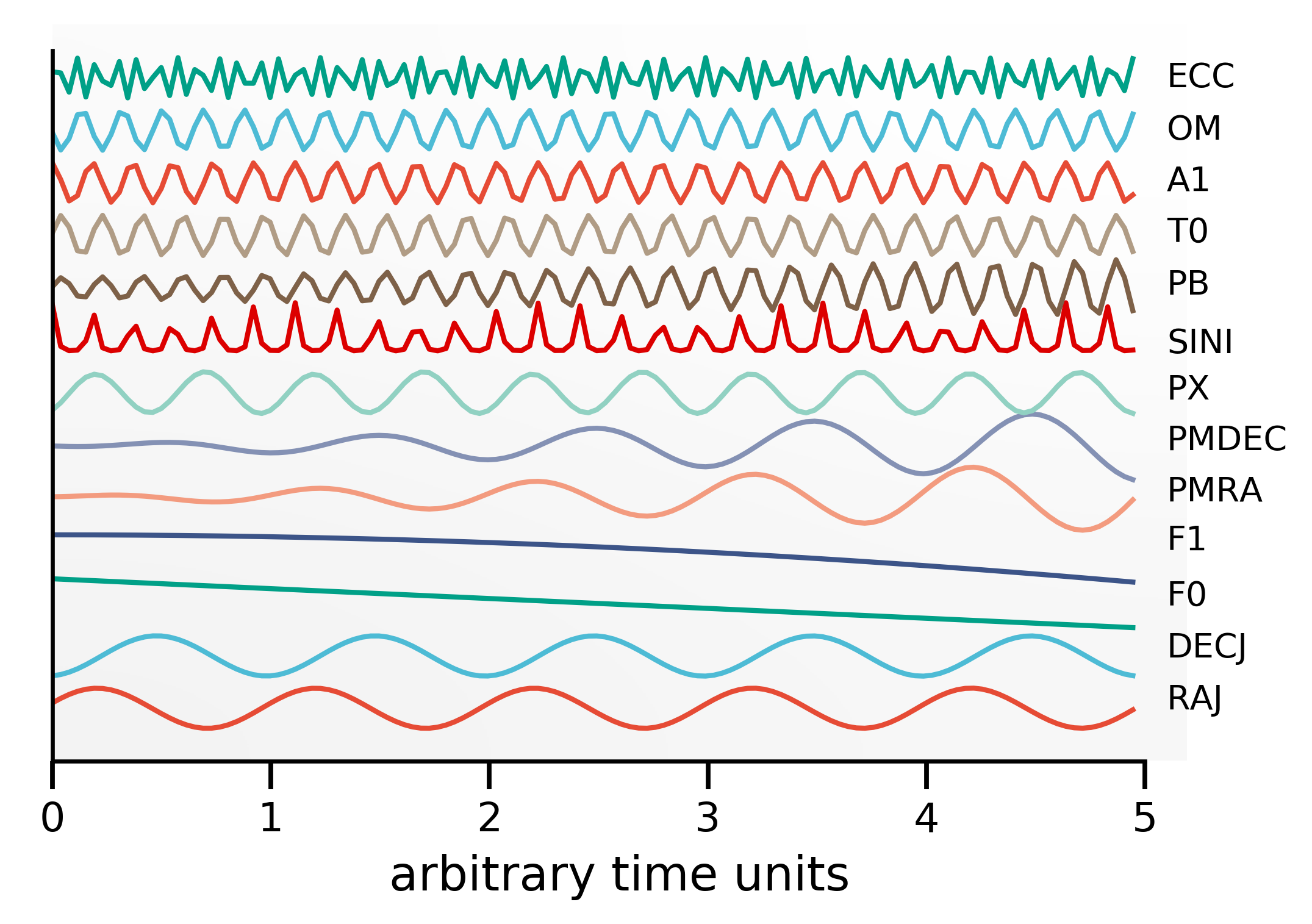}
    \caption{The columns of $\mathbf{M}$ in Eq.~\ref{eq:tm_design} are treated as a basis on which to expand timing-ephemeris perturbations in terms of different parameters. The time-dependent behavior of these basis vectors are shown here. Some timing-ephemeris offsets are clearly (quasi-)periodic in time, such as those due to positional and proper motion uncertainties. Adapted from \href{https://github.com/nanograv/pulsar_timing_school}{https://github.com/nanograv/pulsar$\_$timing$\_$school}.}
    \label{fig:tm_basis}
\end{figure}

Assuming that we have some estimate of the $m$ timing-ephemeris parameters from the Herculean efforts of pulsar-timing astronomers, $\vec{\beta}_0$, we can form the pulsar-timing residuals, $\vec{\delta t}$
\begin{equation}
    \vec{\delta t} \equiv \vec{t}_\mathrm{TOA} - \vec{t}_\mathrm{det}(\vec{\beta}_0).
\end{equation}
Ideally, we would like to vary the timing-ephemeris parameters at the same time as we search for GWs, thus ensuring that we don't erroneously remove some of the signal through this initial fitting procedure. I'll explain later how we can do this explicitly, but for now we're going to try to be smart about this. Realistically, any GW-induced perturbation to the TOAs will be miniscule, swamped by the determinstic effects and noise processes. Therefore we can safely assume that the difference between the initial best-fit parameters $\vec{\beta}_0$ and those we would get from the full analysis that includes GWs, $\vec{\beta}_\mathrm{full}$, is very, very small. We can then linearize the deterministic timing ephemeris around the initial parameters, allowing us to search for small linear departures alongside noise processes and GW signals. For a single observation this can be written as
\begin{equation}
    t_{\mathrm{det},i}(\vec{\beta}) = t_{\mathrm{det},i}(\vec{\beta}_0) + \left[\sum_j \frac{\partial t_{\mathrm{det},i}}{\partial \beta_j}\bigg|_{\vec{\beta}_0} \times(\beta_j-\beta_{0,j})\right],
\end{equation}
and for the full deterministic vector we write this compactly as
\begin{equation}
    \vec{t}_{\mathrm{det}}(\vec{\beta}) = \vec{t}_{\mathrm{det}}(\vec{\beta}_0) + \mathbf{M}\vec{\epsilon},
\end{equation}
where $\mathbf{M}$ is an $(N \times m)$ matrix of partial derivatives of the TOAs with respect to each timing-ephemeris parameter (evaluated at the initial fitting solution, and referred to as the \textit{design matrix}), and $\vec{\epsilon}$ is a vector of linear parameter offsets from the initial fitting solution. For a simple timing ephemeris that involves a constant offset and quadratic spindown, the design matrix takes the form
\begin{equation} \label{eq:tm_design}
    \mathbf{M} = \begin{pmatrix}
                    1 & t_1 & t_1^2 \\
                    1 & t_2 & t_2^2 \\
                    \vdots & \vdots & \vdots \\
                    1 & t_N & t_N^2
                 \end{pmatrix}.
\end{equation}
An example of the time-dependent behavior of these quadratic spindown terms, as well as other astrometric and pulsar-binary terms, are shown in Fig.~\ref{fig:tm_basis}.

The design matrix typically contains columns with disparate dynamic ranges of the elements. This could lead to numerical instabilities when evaluating some of the quantities introduced later. We can easily resolve this by creating a stabilized version of $\bm{M}$ where each column is divided by its $L^2$ norm. Alternatively, we can perform an SVD (singular value decomposition) on $\bm{M}$ such that $\bm{M}=\bm{U}\bm{S}\bm{V}^\mathrm{T}$. The matrix $\bm{U}=[\bm{G}_c\,\,\bm{G}]$ can be partitioned into column spaces corresponding to the first $m$ non-singular components, $\bm{G}_c$ (an ($N\times m$) matrix), and the singular null-space components, $\bm{G}$ (an $(N\times(N-m))$ matrix). The design matrix can then be replaced with $\bm{G}_c$, or the $L^2$-stabilized $\bm{M}$, in all subsequent inference. However care should be taken to store all quantities associated with stabilizing $\bm{M}$ in order to transform the inferred timing-ephemeris deviations back to their physical values (if relevant). 

\textbf{Prior:} The timing ephemeris parameters are very well-constrained by pulsar-timing observations; their inference is likelihood dominated. Hence we usually place an improper uniform prior on these small linear departures, $\vec{\epsilon}$, which is equivalent to a zero-mean Gaussian prior of infinite variance:
\begin{equation}
    p(\vec{\epsilon}) = N(\mathrm{mean}=\vec{0},\mathrm{variance}=\vec\infty).
\end{equation}
Introducing infinities seems like a dangerous prospect, but as we discuss later, we never actually have to worry about this in practice \citep{2014PhRvD..90j4012V}.

\subsection{Achromatic low-frequency processes}

\begin{figure}
	\includegraphics[width=\columnwidth]{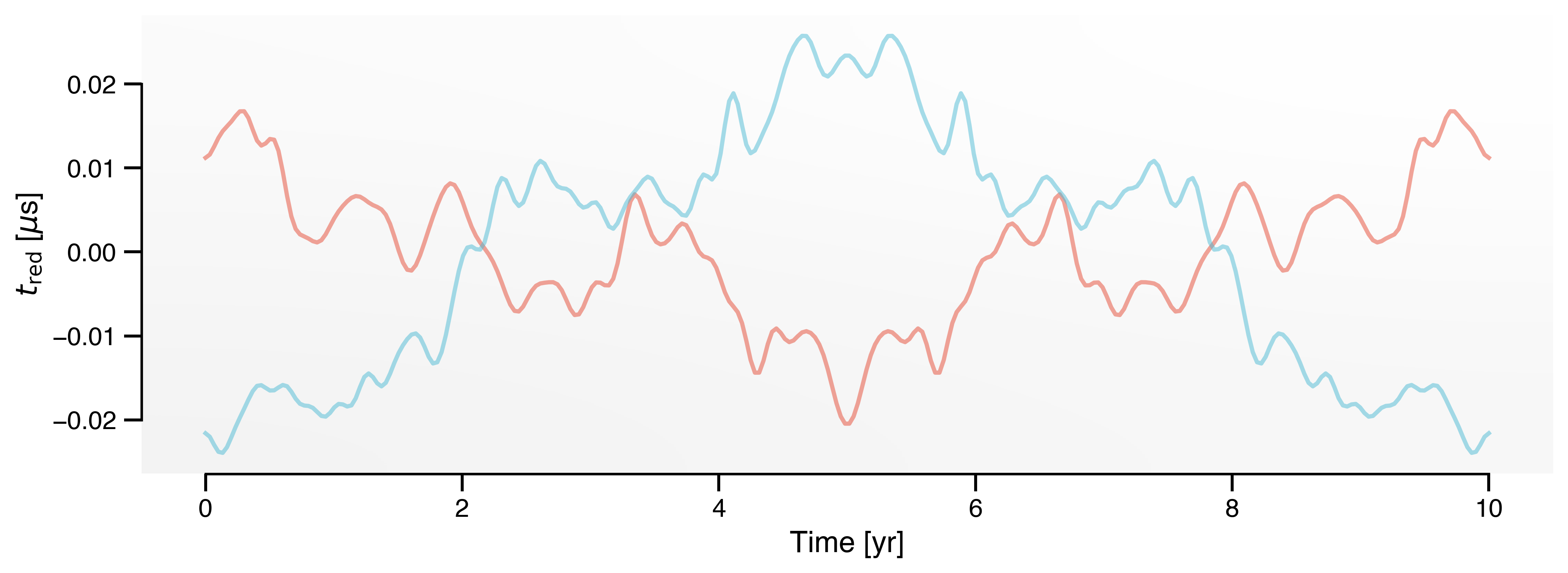}
    \caption{Example time-domain realizations of red noise, with $A=10^{-15}$ and $\gamma=2.5$. Adapted from \href{https://github.com/nanograv/pulsar_timing_school}{https://github.com/nanograv/pulsar$\_$timing$\_$school}.}
    \label{fig:red_noise}
\end{figure}

Millisecond pulsars are very stable rotators, but they're not perfect. It is well known \citep[e.g.,][]{2010ApJ...725.1607S} that pulsars can suffer from intrinsic instabilities that cause quasi-random-walk behaviour in pulsar pulse phase, period, or spindown rate. The result appears as long-timescale noise processes, often called \textit{red noise} since the noise power is predominately at low sampling frequencies of the timing residuals\footnote{To understand why we call things ``white noise'', ``red noise'', etc., consider visible light. White light has equal power in all components of visible light, just as white noise has equal power across all sampling frequencies. By contrast, red light has an excess of lower frequency red components, as does red noise.}. Some example time-domain red noise realizations are shown in Fig.~\ref{fig:red_noise}. We characterize this intrinsic red noise (and indeed any low-frequency process that has no radio-frequency dependence, including the GWB) in the timing residuals using a Fourier basis. Consider the Fourier series for an arbitrary function of time, $f(t)$:
\begin{equation}
    f(t) = \frac{1}{2}a_0 + \sum_{k=1}^\infty a_k\sin(2\pi kt/T) + \sum_{k=1}^\infty b_k\cos(2\pi kt/T),
\end{equation}
where $a_0,a_k,b_k$ are Fourier coefficients, and $k$ indexes the harmonics of the base sampling frequency $1/T$ that is the inverse of the data span. 

In PTA analysis we lose all sensitivity to constant offsets such as $a_0$ by fitting for constant phase offsets. We are also only interested in describing low-frequency behavior, so we can truncate the harmonic sum to some reasonable frequency, $N_f$. What remains of this expression can be written in more compact form for the vector of delays induced by low-frequency processes as
\begin{equation}
    \vec{t}_\mathrm{red} = \mathbf{F}\vec{a},
\end{equation}
where $\vec{a} = (a_1,b_1,a_2,b_2,\ldots,a_{N_f},b_{N_f})^\mathrm{T}$ and $\mathbf{F}$ is the $(N \times 2N_f)$ \textit{Fourier design matrix} that corresponds to columns of alternating sines and cosines for each frequency evaluated at the different observation times
\begin{equation}
    \mathbf{F} = \begin{pmatrix} \sin(2\pi t_1/T) \!\!&\!\! \cos(2\pi t_1/T) \!\!&\!\! \cdots \!\!&\!\! \sin(2\pi N_f t_1/T) \!\!&\!\! \cos(2\pi N_f t_1/T)\\
    \sin(2\pi t_2/T) \!\!&\!\! \cos(2\pi t_2/T) \!\!&\!\! \cdots \!\!&\!\! \sin(2\pi N_f t_2/T) \!\!&\!\! \cos(2\pi N_f t_2/T)\\
    \vdots \!\!&\!\! \vdots \!\!&\!\! \vdots \!\!&\!\! \vdots \!\!&\!\! \vdots \\
    \sin(2\pi t_N/T) \!\!&\!\! \cos(2\pi t_N/T) \!\!&\!\! \cdots \!\!&\!\! \sin(2\pi N_f t_N/T) \!\!&\!\! \cos(2\pi N_f t_N/T)\\
    \end{pmatrix}.
\end{equation}
Lower frequency red-noise effects are highly covariant with the quadratic spindown of the timing ephemeris, so modeling below $1/T$ is not necessary. We call this limitation to low-frequency sensitivity the ``quadratic wall''. However, some techniques have been developed to improve the modeling of very red processes by including frequencies below $1/T$ \citep{2015MNRAS.446.1170V}.

\textbf{Prior:} We model these achromatic red noise terms as zero-mean Gaussian random processes
\begin{equation}
    p(\vec{a}|\vec{\eta}) = \frac{\exp\left(-\frac{1}{2}\vec{a}^\mathrm{T}\bm{\phi}^{-1}\vec{a}\right)}{\sqrt{\mathrm{det}(2\pi\bm{\phi})}},
\end{equation}
where $\langle \vec{a}\,\vec{a}^\mathrm{T}\rangle = \bm{\phi}$, and $\vec{\eta}$ are hyper-parameters of the Gaussian variance on these processes. The covariance matrix of the Fourier coefficients will include all potential achromatic low-frequency processes, which necessarily also includes the GWB and any other spatially-correlated noise processes. Hence:
\begin{equation} \label{eq:phi_prior}
    [\phi]_{(ak)(bj)} = \Gamma_{ab}\rho_k\delta_{kj} + \kappa_{ak}\delta_{kj}\delta_{ab},
\end{equation}
where $(a,b)$ index over pulsars, $(k,j)$ index over sampling frequencies of the timing residuals, and $\Gamma_{ab}$ is the GWB overlap reduction function coefficient for pulsars $(a,b)$ (or the equivalent cross-correlation coefficient for spatially-correlated noise processes).  The terms $\rho_k$ / $\kappa_{ak}$ are related to the power spectral density (PSD) of the timing delay, $S(f)$ (units of time$^3$), induced by the GWB (or spatially-correlated noise process) and intrinsic per-pulsar red noise, respectively, such that $\rho(f) = S(f)\Delta f$, and $\Delta f=1/T$ (and likewise for $\kappa_a(f)$).

There is alot of flexibility in how we can model the PSD of these processes. The default assumption is for a power-law GWB and/or intrinsic per-pulsar red noise, such that
\begin{equation}
    \rho(f) = \frac{h_c(f)^2}{12 \pi^2 f^3}\frac{1}{T} = \frac{A^2}{12\pi^2}\frac{1}{T}\left(\frac{f}{1\,\mathrm{yr}^{-1}}\right)^{-\gamma}\,\,\mathrm{yr}^2,
\end{equation}
where $h_c(f)=A(f/1\,\mathrm{yr}^{-1})^\alpha$ is the GWB characteristic strain spectrum, $\gamma\equiv3-2\alpha$, and
\begin{equation}
    \kappa_a(f) = \frac{A_a^2}{12\pi^2}\frac{1}{T}\left(\frac{f}{1\,\mathrm{yr}^{-1}}\right)^{-\gamma_a}\,\,\mathrm{yr}^2.
\end{equation}
In these parametrizations the power-law amplitude, $A$, is referenced to a frequency of an inverse year. There are many other parametrized spectrum representations that are in common usage, including: a turnover spectrum to encapsulate the influence of non-GW final-parsec dynamical influences on SMBBH evolution \citep{Sampson2015}; spectra with high-frequency knees to account for SMMBH population finiteness \citep{svc08}; astrophysics-driven spectral parametrizations \citep{2017MNRAS.468..404C,2017MNRAS.470.1738C,2019MNRAS.488..401C,2017PhRvL.118r1102T}; t-process spectra that allow for some degree of fuzziness around a power-law \citep{2019ApJ...880..116A}; and finally the free-spectrum model that is agnostic to the spectral shape, allowing complete per-frequency flexibility in the modeling \citep{2013PhRvD..87j4021L}. In all of these we care more about recovering the hyper-parameters, $\vec{\eta}$ (e.g., $A$, $\gamma$), than the actual Fourier coefficients of the process.

\subsection{Chromatic low-frequency processes}

\begin{figure}
	\includegraphics[width=\columnwidth]{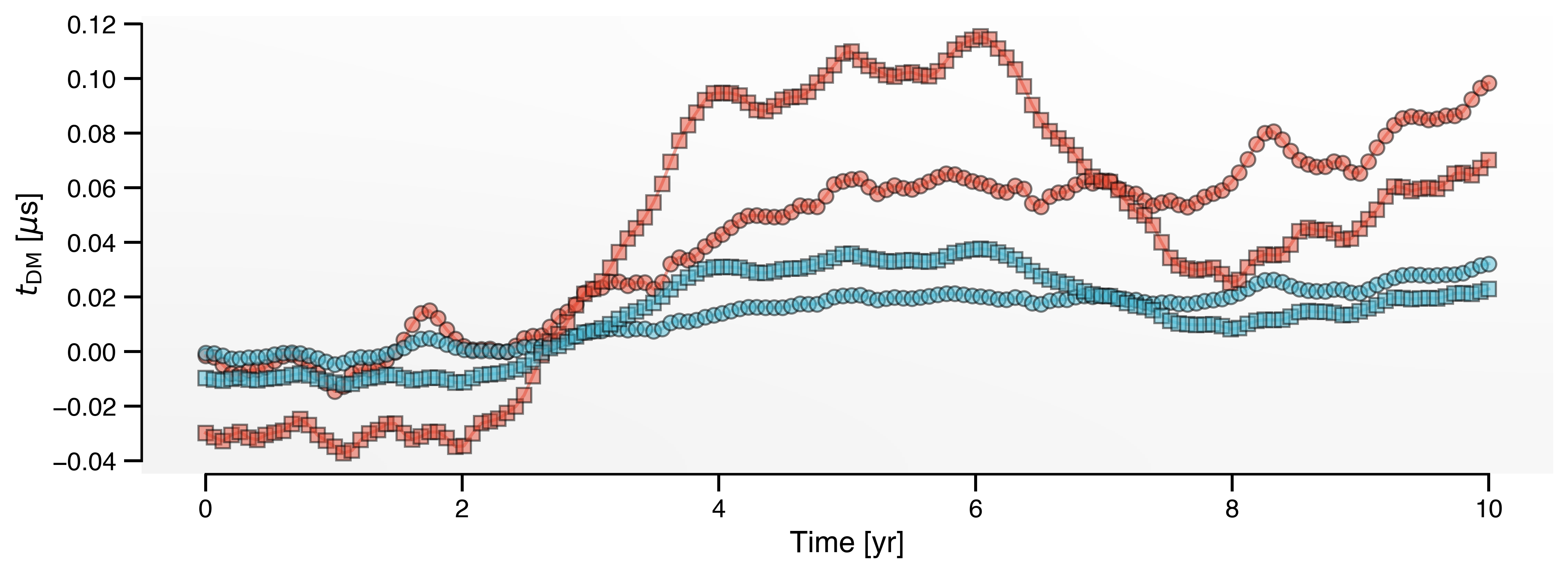}
    \caption{Example time-domain realizations of chromatic delays induced by dispersion-measure variations, with $A=10^{-15}$ and $\gamma=2.5$. There are two realizations of delays for two different radio-frequency components (indicated by the line colors). Adapted from \href{https://github.com/nanograv/pulsar_timing_school}{https://github.com/nanograv/pulsar$\_$timing$\_$school}.}
    \label{fig:dm_noise}
\end{figure}

Pulsars emit radio pulses that propagate through interstellar space on the way to our observatories on Earth. This space is filled with ionized material, leading to radio-frequency dependent (i.e., chromatic) propagation delays for different components of the pulse. Components at lower radio frequencies will experience greater delays than those at higher frequencies. The leading-order chromatic effect is \textit{dispersion}, where the group delay of a given radio-frequency component, $\nu_\mathrm{obs}$, is $t_\mathrm{delay} = (\mathrm{DM}/K)(1/\nu_\mathrm{obs}^2)$, where the dispersion constant is defined to be $K = 2.41\times 10^{-16}\,\,\mathrm{Hz}^{-2}\mathrm{cm}^{-3}\mathrm{pc}\,\mathrm{s}^{-1}$, and the \textit{dispersion measure}, DM, is defined to be the line-of-sight integrated column density of free electrons
\begin{equation}
    \mathrm{DM} = \int_0^L n_e\,\,dl.
\end{equation}
Leading-order dispersive effects are modeled in the timing ephemeris, however variations in the electron column density can induce chromatic red noise in the timing residuals. Example time-domain realizations of chromatic delays induced by DM variations are shown in Fig.~\ref{fig:dm_noise}. There are variety of techniques used to model these DM variations \citep[e.g.,][]{2014MNRAS.441.2831L,2013MNRAS.429.2161K}, but we focus the discussion here on the two most prominent. 

\begin{enumerate}[leftmargin=*,wide,labelwidth=!,labelindent=0pt]
    \setlength\itemsep{0pt}
    \item[\textbf{DMX:}] This model is used by the NANOGrav Collaboration, where DM variations are treated as an epoch-by-epoch offset in DM. The DM offset is fit alongside all other timing-ephemeris parameters as a purely deterministic effect. 
    \item[\textbf{DM GP:}] This model treats DM variations as a chromatic red-noise (Gaussian) process. This is written as
    \begin{equation} \label{eq:dmgp}
        \vec{t}_\mathrm{DM} = \mathbf{F}_\mathrm{DM}\vec{a}_\mathrm{DM},
    \end{equation}
    where each element of $\mathbf{F}_\mathrm{DM}$ is related to that of $\mathbf{F}$ via
    \begin{equation}
        F_\mathrm{DM}(f_k,t_i) =  F(f_k,t_i)\times 1/(K\nu_{\mathrm{obs},i}^2),
    \end{equation}
    and $\nu_{\mathrm{obs},i}$ is the observed radio frequency of the $i^\mathrm{th}$ TOA. Unlike achromatic red noise, this DM GP has no quadratic function of the time-series to act as a proxy for low-frequency power. Hence two additional parameters (DM1 and DM2) are added to the deterministic timing ephemeris, corresponding to coefficients of chromatic linear and quadratic trends in the TOAs. Note also that, while we have assumed a Fourier basis for the DM variations here, one could also consider a coarse-grained time-domain basis to allow other kinds of GP kernels to model quasi-periodic features and even radio-frequency-band evolution of DM variations\footnote{{\tt The Kernel Cookbook: \href{https://www.cs.toronto.edu/~duvenaud/cookbook}{https://www.cs.toronto.edu/~duvenaud/cookbook}}}.  
\end{enumerate}

Dispersion may be the leading-order chromatic effect on the TOAs, but is by no means the only one. There are a litany of other chromatic influences on pulse arrival times as they propagate through the ionized interstellar medium \citep{2010arXiv1010.3785C}, such as turbulence-induced pulse broadening, which can lead to $\sim 1/\nu_\mathrm{obs}^{4.4}$ dependent delays. These influences are easily modeled through additional terms like Eq.~\ref{eq:dmgp}. 

\textbf{Prior:} Under the DMX model, DM variations are part of the timing ephemeris, and treated accordingly in the prior assumption. When modeling DM variations as a Gaussian process, the prior treatment is very similar to achromatic processes, where $\langle \vec{a}_\mathrm{DM}\,\vec{a}_\mathrm{DM}^\mathrm{T} \rangle = \bm{\phi}_\mathrm{DM}$, and
\begin{equation}
    [\phi_\mathrm{DM}]_{(ak)(bj)} =  \lambda_{ak}\delta_{kj}\delta_{ab},
\end{equation}
where $\lambda(f) = S_\mathrm{DM}(f)\Delta f$. The PSD of DM variations can be modeled with all of the flexibility of achromatic processes, but typically a power-law is adopted with a minor modification in the definition:
\begin{equation}
    \lambda_a(f) = \frac{A^2_{\mathrm{DM,a}}}{T} \left(\frac{f}{1\,\mathrm{yr}^{-1}}\right)^{-\gamma_{\mathrm{DM},a}}\,\,\mathrm{yr}^2.
\end{equation}

\subsection{White noise}

\begin{figure}
    \centering
	\includegraphics[width=0.5\columnwidth]{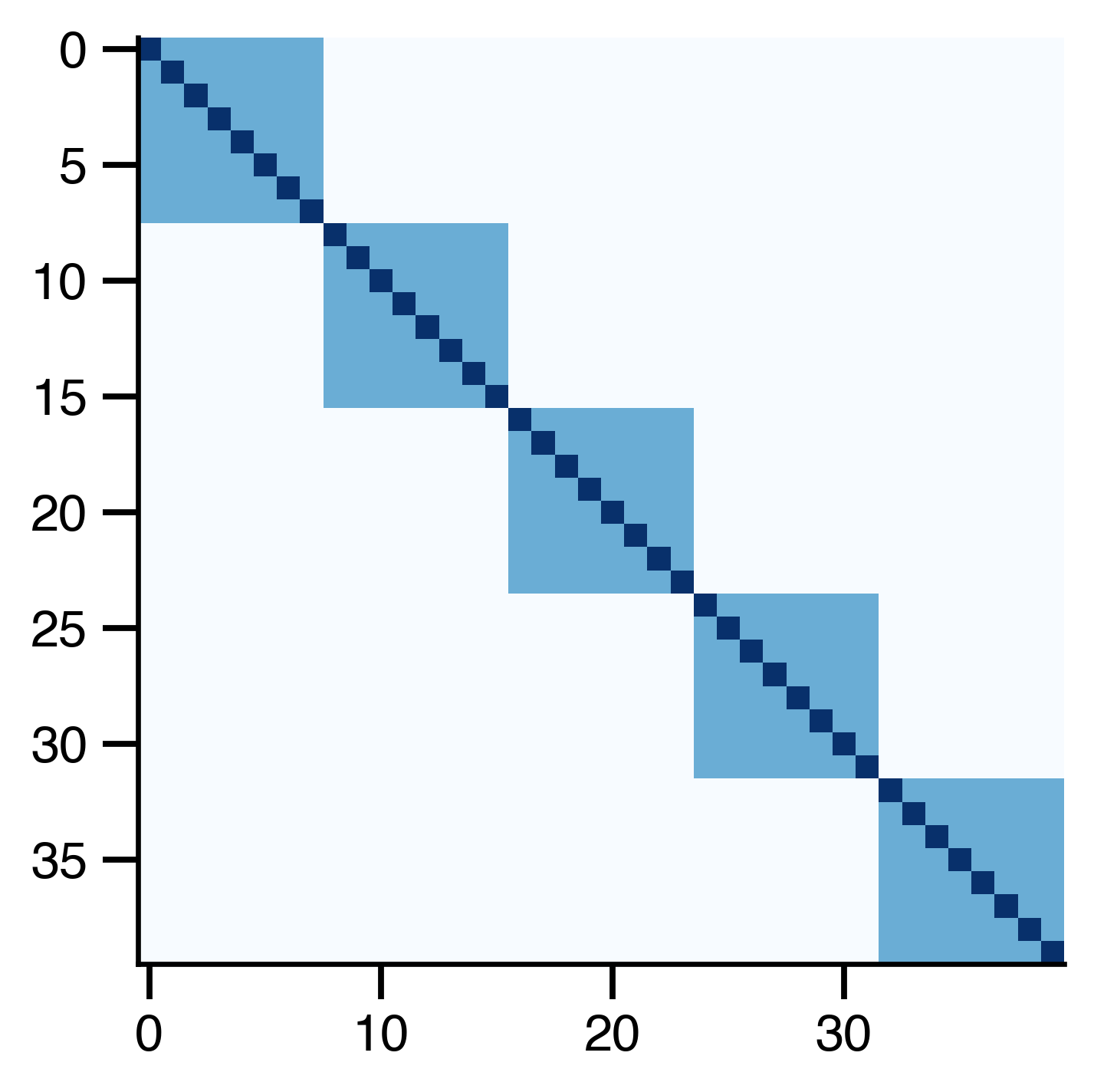}
    \caption{Example time-domain covariance matrix of white noise. Noise from EFAC-scaled TOA uncertainties and EQUAD contributions are uncorrelated with other observations, manifesting as a diagonal covariance matrix (dark blue). Jitter/ECORR is uncorrelated between different epochs, but fully correlated within epochs, giving a block-diagonal covariance matrix (light blue). Adapted from \href{https://github.com/nanograv/pulsar_timing_school}{https://github.com/nanograv/pulsar$\_$timing$\_$school}.}
    \label{fig:white_noise}
\end{figure}

White noise has a flat power spectral density across all sampling frequencies. It therefore does not exhibit long-timescale trends in the residual time-series. An example time-domain covariance matrix from the different white noise sources mentioned below is shown in Fig.~\ref{fig:white_noise}. White noise also exhibits no inter-pulsar correlations. There are three main sources of white noise in PTA analysis:
\begin{enumerate}[leftmargin=*,wide,labelwidth=!,labelindent=0pt]
    \item[\textbf{Radiometer noise \& EFAC:}] All pulses observed within a given data-taking epoch ($\sim20-30$ minutes) are de-dispersed and folded over the pulsar's spin period, which is then fit to a long-timescale averaged pulse template to compute the actual TOA. The standard TOA uncertainties from radiometer noise are merely template-fitting uncertainties. But not all sources of fitting uncertainty may be propagated into the final quoted TOA uncertainty, leading us to introduce a correction factor, or \textit{Extra FACtor (EFAC)} that acts as a multiplicative correction to the uncertainties. We apply these EFACs to unique combinations of telescope receivers and backends (a given unique combination is denoted a ``system'' \citep{2016MNRAS.458.2161L})
    \begin{equation}
        \langle n_{i,\mu} n_{j,\nu} \rangle = F_\mu^2\sigma_i^2\delta_{ij}\delta_{\mu\nu},
    \end{equation}
    where $n_{i,\mu}$ is the timing delay induced by white noise at observation $i$ in receiver-backend system $\mu$; $\sigma_i$ is the TOA uncertainty for observation $i$; and $F_\mu$ is the EFAC for $\mu$. EFACs that are significantly different from unity would raise suspicions of data quality. 
    \item[\textbf{EQUAD:}] There may be additional white noise that is completely separate from radiometer noise, from e.g., instrumental effects. To account for additional non-multiplicative corrections to the TOA uncertainties, we include an \textit{Extra QUADrature (EQUAD)} noise term. This EQUAD adds in quadrature to the term above, which now becomes
    \begin{equation}
        \langle n_{i,\mu} n_{j,\nu} \rangle = F_\mu^2\sigma_i^2\delta_{ij}\delta_{\mu\nu} + Q_\mu^2\delta_{ij}\delta_{\mu\nu},
    \end{equation}
    and also applies to unique instrument systems.
    \item[\textbf{Pulse phase jitter \& ECORR:}] Within a given observing \textit{epoch}, a finite number of pulses are being folded and fit to a standard pulse profile template. This leads to an effect known as \textit{pulse phase jitter} that contributes an additional white noise term; essentially the folding of a finite number of pulses from epoch to epoch leaves some residual shape fluctuation with respect to the profile template. While this depends on the number of pulses recorded in a given observing epoch, this is typically not changing much over the span of the pulsar's dataset, and so it is again just discriminated by unique receiver-backend systems. What's more, in a given observing epoch NANOGrav typically records many near-simultaneous TOAs across neighboring radio-frequency bands. Given that these correspond to the same folded pulses, the pulse phase jitter will be fully correlated across these bands. This gives us \textit{Extra Correlated (ECORR)} white noise, which is uncorrelated between different observing epochs, but fully correlated between different bands within the same epoch.
    
    ECORR can be treated in two different ways within our pipelines. An important piece of initial book-keeping involves making sure that our TOAs are sorted by epoch. We can then either,
    \begin{enumerate}
        \item Model ECORR using a low-rank basis (similar to the Fourier basis for low-frequency processes, but now on a basis of epochs), such that
        \begin{equation}
            \vec{n}_\mathrm{ECORR} = \bm{U}\vec{j},
        \end{equation}
        where $\vec{j}$ is a jitter vector of length $N_\mathrm{epoch}$, and $\bm{U}$ is an $(N\times N_\mathrm{epoch})$ matrix that has entry values of $1$ when a TOA row lies within an epoch column, and zero everywhere else, e.g.,
        \begin{equation}
            \bm{U} = \underbrace{
                        \begin{pmatrix}
                        1 & & & \\
                        \vdots & & & \\
                        1 & & & \\
                          & 1 & & \\
                          & \vdots & & \\
                          & 1 & & \\
                          & & \ddots & \\
                          & & & 1 \\
                          & & & \vdots \\
                          & & & 1
                      \end{pmatrix}
                      }_{N_\mathrm{epoch}}.
        \end{equation}
        $\bm{U}$ is sometimes referred to as the \textit{epoch exploder matrix}, given that it explodes epoch assignments out to the full rank of TOAs. 
        
        \textbf{Prior:} We treat jitter/ECORR in this sense as a Gaussian process that is fully correlated within an epoch but uncorrelated across different epochs. As such our prior is
        \begin{equation}
            p(\vec{j}|\vec{J}) = \frac{\exp\left(-\frac{1}{2}\vec{j}^\mathrm{T}\bm{\mathcal{J}}^{-1}\vec{j}\right)}{\sqrt{\mathrm{det}(2\pi\bm{\mathcal{J}})}},
        \end{equation}
        where $\langle\vec{j}\,\vec{j}^\mathrm{T}\rangle=\bm{\mathcal{J}}$, and $\vec{J}$ is a vector of jitter rms values for distinct receiver-backend systems that parametrize the jitter covariance matrix $\bm{\mathcal{J}}$.
        \item Model ECORR alongside other white noise processes like radiometer noise and EQUAD. This is a tempting approach, because we are not actually interested in what $\vec{j}$ is, in the same way that we are not interested in what the time-series of other white-noise timing delays are. Jitter/ECORR can be packaged alongside EFAC and EQUAD as simply another white noise term
        \begin{equation}
            \langle n^J_{i,\mu} n^J_{j,\nu} \rangle = J_\mu^2\delta_{e(i)e(j)}\delta_{\mu\nu},
         \end{equation}
        where $e(i)$ indexes the epochs of each observation such that $\delta_{e(i)e(j)}$ is zero unless TOAs lie within the same epoch, and as before $\mu$ indexes each receiver-backend system.
    \end{enumerate}
\end{enumerate}

\section{The Likelihood}

We now have all the main ingredients with which to piece together the PTA likelihood. To keep things relatively straightforward when introducing this likelihood, we model the residuals as being only the sum of contributions from linear timing ephemeris deviations (containing the DMX model of DM variations), achromatic low-frequency processes (including a GWB), and various white noise processes\footnote{We will consider other deterministic processes later, such as chromatic features in individual pulsars, and resolvable individual SMBBH signals across the entire PTA.}:  
\begin{equation}
    \vec{\delta t} = \bm{M}\vec{\epsilon} + \bm{F}\vec{a} + \vec{n},
\end{equation}
where $\vec{n}$ contains jitter noise. We form a new vector of noise- and signal-mitigated timing residuals that acts as a model-dependent estimate of $\vec{n}$, such that
\begin{equation} \label{eq:whitened_res}
    \vec{r} = \vec{\delta t} - \bm{M}\vec{\epsilon} - \bm{F}\vec{a}.
\end{equation}
Since $\vec{r}$ acts as our estimate of Gaussian white noise, its likelihood is simple:
\begin{equation} \label{eq:like0}
    p(\vec{r}|\vec{\epsilon},\vec{a},\ldots,\vec{\eta}) = \frac{\exp\left(-\frac{1}{2}\vec{r}^\mathrm{T}\bm{N}^{-1}\vec{r}\right)}{\sqrt{\mathrm{det}(2\pi\bm{N})}},
\end{equation}
where $\vec{\eta}$ are hyper-parameters that (for now) just appear as EFACs, EQUADs, ECORRs etc., in the white noise covariance matrix, $\bm{N}$. 

For convenience, we now group all rank-reduced processes (e.g., linear timing-ephemeris deviations, low-frequency processes) together, such that
\begin{equation}
    \vec{r} = \vec{\delta t} - \bm{T}\vec{b},
\end{equation}
where $\bm{T} = [\bm{M}\,\, \bm{F}]$, and $\vec{b} = \left[\begin{matrix}\vec{\epsilon} \\ \vec{a}\end{matrix}\right]$. This kind of compact representation is easily expanded with additional model compoenents, whether they be the Gaussian process model of DM variations ($\bm{F}_\mathrm{DM}\vec{a}_\mathrm{DM}$), or the epoch-expanded model of jitter/ECORR ($\bm{U}\vec{j}$). All you need to do is concatenate the relevant basis matrices onto $\bm{T}$, and likewise the coefficient vectors onto $\vec{b}$.\footnote{There is a notable alternative to the $\bm{T}$-matrix method, called the $\bm{G}$-matrix method \citep{2013MNRAS.428.1147V}, where $\bm{G}$ is the null space of the timing ephemeris design matrix, $\bm{M}$ (see Sec.~\ref{sec:tmodel}). This method is appropriate for when you don't care about the timing ephemeris at all, and simply want to marginalize over it. When constructing a statistic or writing a likelihood, you just replace all vectors $\vec{x}$ by $\bm{G}^\mathrm{T}\vec{x}$, and all matrices $\bm{X}$ by $\bm{G}^\mathrm{T}\bm{X}\bm{G}$. In so doing, you are projecting these quantities into the null space of the timing ephemeris, which can be shown to be mathematically equivalent to marginalization \citep{2014PhRvD..90j4012V} (see also \href{https://gwic.ligo.org/assets/docs/theses/taylor_thesis.pdf}{https://gwic.ligo.org/assets/docs/theses/taylor$\_$thesis.pdf}).}

What now? Well, we have a likelihood, and in principle we could proceed with inference on our pulsar-timing residuals using just Eq.~\ref{eq:like0}. That would be a grave mistake though; we would likely learn very little about all these various coefficient vectors in $\vec{b}$ or the white noise parameters. That's because we have forgotten to include a proper prior on $\vec{b}$. At the moment the prior is implicitly uniform over the unbounded range $[-\infty,+\infty]$. This is appropriate for the timing-ephemeris deviations, since the inference on those coefficients will be data-dominated. But all other processes are exceptionally small, requiring proper control with well-chosen priors. Looking back to the various descriptions of each process contributing to the pulse time-series, the answer is straightforward: we assume these are are zero-mean random Gaussian processes with parametrized variances. The prior on $\vec{b}$ is then
\begin{equation} \label{eq:bprior}
    p(\vec{b}|\vec{\eta}) = \frac{\exp\left(-\frac{1}{2}\vec{b}^\mathrm{T}\bm{B}^{-1}\vec{b}\right)}{\sqrt{\mathrm{det}(2\pi\bm{B})}},
\end{equation}
where
\begin{equation}
    \bm{B} = \begin{pmatrix}\bm{\infty} & \bm{0} \\
                            \bm{0} & \bm{\phi} \end{pmatrix},
\end{equation}
and $\vec{\eta}$ are parameters that control the behaviour of $\bm{B}\equiv\bm{B}(\vec{\eta})$. However, given that these are parameters of a prior, we call these \textit{hyper-parameters}. Once again, this prior matrix $\bm{B}$ is easily expanded to include more blocks like $\bm{\phi}_\mathrm{DM}$ or $\bm{\mathcal{J}}$. We'll see next how we can piece together the likelihood and this prior to get the full hierarchical PTA likelihood. Finally, don't worry about that infinity block! As we'll see soon, we only ever have to deal with the inverse of $\bm{B}$, making this infinity drop to zero. 

\subsection{Full hierarchical likelihood}

The joint probability density of the process coefficients and their variance hyper-parameters can be written as a chain of conditional probabilities
\begin{equation}
    p(\vec{b},\vec{\eta}|\vec{\delta t}) \propto p(\vec{\delta t}|\vec{b}) \times p(\vec{b}|\vec{\eta}) \times p(\vec{\eta}),
\end{equation}
where $p(\vec{\eta})$ is the prior density on the hyper-parameters. The first two terms in this chain give the hierarchical PTA likelihood, $p(\vec{\delta t}|\vec{b},\vec{\eta}) = p(\vec{\delta t}|\vec{b}) \times p(\vec{b}|\vec{\eta})$. Let's see what this looks like when written out in full:
\begin{equation} \label{eq:like1}
    p(\vec{\delta t}|\vec{b},\vec{\eta}) = \frac{\exp\left(-\frac{1}{2}(\vec{\delta t}-\bm{T}\vec{b})^\mathrm{T}\bm{N}^{-1}(\vec{\delta t}-\bm{T}\vec{b})\right)}{\sqrt{\mathrm{det}(2\pi\bm{N})}} \times \frac{\exp\left(-\frac{1}{2}\vec{b}^\mathrm{T}\bm{B}^{-1}\vec{b}\right)}{\sqrt{\mathrm{det}(2\pi\bm{B})}}.
\end{equation}

For a single pulsar, this doesn't look too scary, and in fact it can be pretty quick to evaluate on modern laptops. The speed with which we can evaluate this actually depends on how jitter has been treated. If jitter is packaged along with other rank-reduced processes in $\bm{T}\vec{b}$, then $\bm{N}$ is a diagonal matrix that is trivial to invert and whose determinant is simply the product of diagonal elements. Likewise, for a single pulsar $\bm{B}$ is a relatively small and easily-invertable diagonal matrix, such that $\bm{B}^{-1}=\mathrm{diag}(\bm{0}, \bm{\phi}^{-1})$ and the diagonal matrix $\bm{\phi}$ is trivial to invert. Using the fact that the inverse of a matrix determinant is equivalent to the determinant of the matrix's inverse, we see that the infinity block poses no practical problems. 

What about when jitter is instead modeled inside $\bm{N}$? This white noise covariance matrix then becomes \textit{block diagonal}, with a block for each epoch. The inverse is simply another block diagonal matrix formed of each inverted block. Let's consider each block to be of the form $\bm{N}_e = \bm{D}_e + J_e^2 \vec{u}_e \vec{u}_e^\mathrm{T}$, where $D_e$ is the usual diagonal part containing radiometer noise, EFAC, and EQUAD contributions; $J_e$ is the jitter rms in this epoch; and $u_e = (1,1,\ldots,1)^\mathrm{T}$ contains as many entries as there are TOAs in this epoch. The inverse of this block is efficiently computed using the \textit{Sherman-Morrison formula} \citep{sherman1950}:
\begin{equation}
    \bm{N}_e^{-1} = \bm{D}_e^{-1} - \frac{\bm{D}_e^{-1} \vec{u}_e \vec{u}_e^\mathrm{T} \bm{D}_e^{-1}}{J_e^{-2} + \vec{u}_e^\mathrm{T}\bm{N}_e^{-1}\vec{u}_e}.
\end{equation}
Note that the denominator of the right-most term is a scalar quantity. The Sherman-Morrison formula is a special case of the \textit{Woodbury matrix identity} that we will see more of soon. 

For a full PTA, Eq.~\ref{eq:like1} is modified to become a product over the $p(\vec{\delta t}|\vec{b})$ terms, and controlled by a prior on $\vec{b}$ that allows for inter-pulsar correlations induced by a GWB or systematic noise processes:
\begin{equation} \label{eq:like2}
    p(\{\vec{\delta t}\}|\{\vec{b}\},\vec{\eta}) = \left[\prod_{a=1}^{N_p}\, p(\vec{\delta t}_a|\vec{b}_a)\right] \times p(\{\vec{b}\}|\vec{\eta}),
\end{equation}
where $p(\{\vec{b}\}|\vec{\eta})$ has the same form as Eq.~\ref{eq:bprior} with $\vec{b}$ replaced by the concatenation of coefficient vectors over all pulsars, and with the $\bm{B}$ matrix becoming a matrix of $N_p\times N_p$ blocks for each pair of pulsars. For cross-pairings of pulsars, the usual infinity block becomes zero since the pulsar timing ephemerides are uncorrelated, and the $\bm{\phi}$ block contains only contributions from inter-pulsar correlated processes (see Eq.~\ref{eq:phi_prior}). $\bm{B}$ is now a large band diagonal matrix, which still allows for straightforward and efficient computation of its inverse and determinant. We do so by temporarily reordering the matrix into an $N_b\times N_b$ matrix of blocks for each $\vec{b}$ coefficient; this makes $\bm{B}$ a block diagonal matrix that we can easily invert one block at a time, where upon inverting each block we slot the elements back into their original band diagonal structure to return $\bm{B}^{-1}$. 

Equation \ref{eq:like2} has a \textit{very} large parameter space to explore! For each pulsar there are timing ephemeris deviations, intrinsic red noise coefficients and hyper-parameters, EFACs, EQUADs, and ECORRs, plus potential chromatic coefficients plus hyper-parameters, jitter vectors, and any other parameters that are used to model features within the residual time series. Add to that the coefficients and hyper-parameters of inter-pulsar correlated processes like the GWB, and we have a parameter space that encroaches on $\sim1000$ dimensions or more for currently-sized PTAs of $\sim 50$ pulsars. Unguided Metropolis-Hastings--based MCMC is going to have big difficulties in sampling from this parameter space, for reasons that include the necessary burn-in time, autocorrelation lengths, and the well-known \textit{Neal's funnel} conundrum of trying to sample coefficient parameters from distributions whose variances are also being updated \citep{10.2307/3448413}. However, recent advances in Gibbs sampling (where parameter blocks are updated through sequential conditional distribution draws) \citep{2014PhRvD..90j4012V}, and Hamiltonian MCMC (where likelihood gradient information is used to guide the chain exploration) \citep{2013PhRvD..87j4021L,2017MNRAS.466.4954V} have made sampling from this hierarchical likelihood tractable. 

\subsection{Marginalized likelihood}

Let's ask ourselves what we really care about most of the time in our noise or GW analyses. Is it the particular coefficients that describe the one realization of noise or GW signal that we see in the residuals, or is it the statistical properties of these processes? We're dealing with stochastic processes, so the vast majority of the time it is the latter. Hence, we need not trouble ourselves with numerically sampling from this huge parameter space of $\vec{b}$ coefficients plus hyper-parameters, when all we care about are the hyper-parameters. We can use the wonderful integration properties of chained Gaussian distributions to \textit{analytically} marginalize over these $\vec{b}$ coefficients, leaving a marginalized likelihood that depends only on the hyper-parameters \citep{2013PhRvD..87j4021L,2013MNRAS.428.1147V}. This has the form
\begin{equation} \label{eq:marg_like}
    p(\{\vec{\delta t}\}|\vec{\eta}) = \int p(\{\vec{\delta t}\}|\{\vec{b}\},\vec{\eta})\,d^{N_p}\vec{b} = \frac{\exp\left(-\frac{1}{2}\vec{\delta t}^\mathrm{T}\bm{C}^{-1}\vec{\delta t}\right)}{\sqrt{\mathrm{det}(2\pi\bm{C})}},
\end{equation}
where $\bm{C} = \bm{N} + \bm{T}\bm{B}\bm{T}^\mathrm{T}$. Now, to get to this compact form, we have implicitly used the very powerful \textit{Woodbury matrix identity} \citep{woodbury1950inverting}. This simplifies the inversion of the large dense matrix $\bm{C}$ such that
\begin{equation} \label{eq:woodbury}
    \bm{C}^{-1} = (\bm{N} + \bm{T}\bm{B}\bm{T}^\mathrm{T})^{-1} = \bm{N}^{-1} - \bm{N}^{-1}\bm{T}\bm{\Sigma}^{-1}\bm{T}^\mathrm{T}\bm{N}^{-1},
\end{equation}
where $\bm{\Sigma}=\bm{B}^{-1} + \bm{T}^\mathrm{T}\bm{N}^{-1}\bm{T}$. This may not look simpler, but it is! It's also much faster to compute. Performing a Cholesky decomposition \citep{cholesky2005resolution} to invert a symmetric positive-definite matrix like $\bm{C}$ usually scales as $\mathcal{O}(N_p^3N_\mathrm{TOA}^3)$.\footnote{A Cholesky decomposition factors a symmetric positive-definite matrix into $\bm{X}=\bm{L}\bm{L}^\mathrm{T}$ such that $\bm{L}$ is a lower triangular matrix, making equation solving and inversion simpler to evaluate. Furthermore, the determinant of $\bm{X}$ is simply the product of the squares of the diagonal elements of $\bm{L}$.} But in Eq.~\ref{eq:woodbury}, the bottleneck operation is $\bm{\Sigma}^{-1}$, which still requires a Cholesky inversion, but which now scales as $\mathcal{O}(N_p^3 N_b^3)$. The number of $\vec{b}$ coefficients for each pulsar is typically much less than the number of TOAs, rendering this a significant acceleration. The determinant of $\bm{C}$ is also made more tractable by the Woodbury identity, such that $\mathrm{det}(\bm{C}) = \mathrm{det}(\bm{N})\mathrm{det}(\bm{B})\mathrm{det}(\bm{\Sigma})$. 

It's important to note what the form of $\bm{T}\bm{B}\bm{T}^\mathrm{T}$ actually means for the rank-reduced processes. If we simply look at the achromatic process sector of this dense matrix to inspect the covariance between times $t$ and $(t+\tau)$ in pulsars $a$ and $b$, we get
\begin{equation} \label{eq:disc_wk}
    [\bm{T}\bm{B}\bm{T}^\mathrm{T}]_{(ab),\tau} = \sum_k^{N_f}\,[\bm{\phi}]_{ab}\cos(2\pi k\tau/T),
\end{equation}
where $[\bm{\phi}]_{ab}=\Gamma_{ab}\rho/T$, $\Gamma_{ab}$ is the cross-correlation coefficient between pulsars, $\rho$ is the PSD of the process, and $T$ is the observational timing baseline. In this form we see that Eq.~\ref{eq:disc_wk} is simply the discretized form of the \textit{Wiener-Khinchin theorem} \citep{wiener1930generalized,khintchine1934} to translate the PSD of a stochastic process into its temporal covariance. 

Equation \ref{eq:woodbury} is how we practically implement the form of Eq.~\ref{eq:marg_like} within production-level GW search pipelines, the most prominent of which is {\texttt {ENTERPRISE}} (\texttt{E}nhanced \texttt{N}umerical \texttt{T}oolbox \texttt{E}nabling a \texttt{R}obust \texttt{P}ulsa\texttt{R} \texttt{I}nference \texttt{S}uit\texttt{E})\footnote{\texttt{enterprise: \href{https://github.com/nanograv/enterprise}{https://github.com/nanograv/enterprise}}}. We use this marginalized likelihood with all of the numerical Bayesian techniques outlined in Chapter~\ref{chap:nbayes}, including parameter estimation of GW signals and noise processes, upper limits on signal parameters\footnote{In Bayesian inference the upper limit is just the parameter value at the required percentile of the $1$D-marginalized parameter posterior distribution.}, and model selection. Given that we expect the stochastic GWB will be the first manifestation of GW detection in our PTAs, the model selection of primary current interest is between a spatially-correlated GWB signal and that of a spatially-uncorrelated common-spectrum red process. The only difference between these models is the presence of GWB-induced inter-pulsar correlations as described by the Hellings \& Downs function, since these cross-correlations are the undeniable fingerprint of the influence of GWs on our pulsar-timing experiment.

\begin{figure}
	\includegraphics[width=\columnwidth]{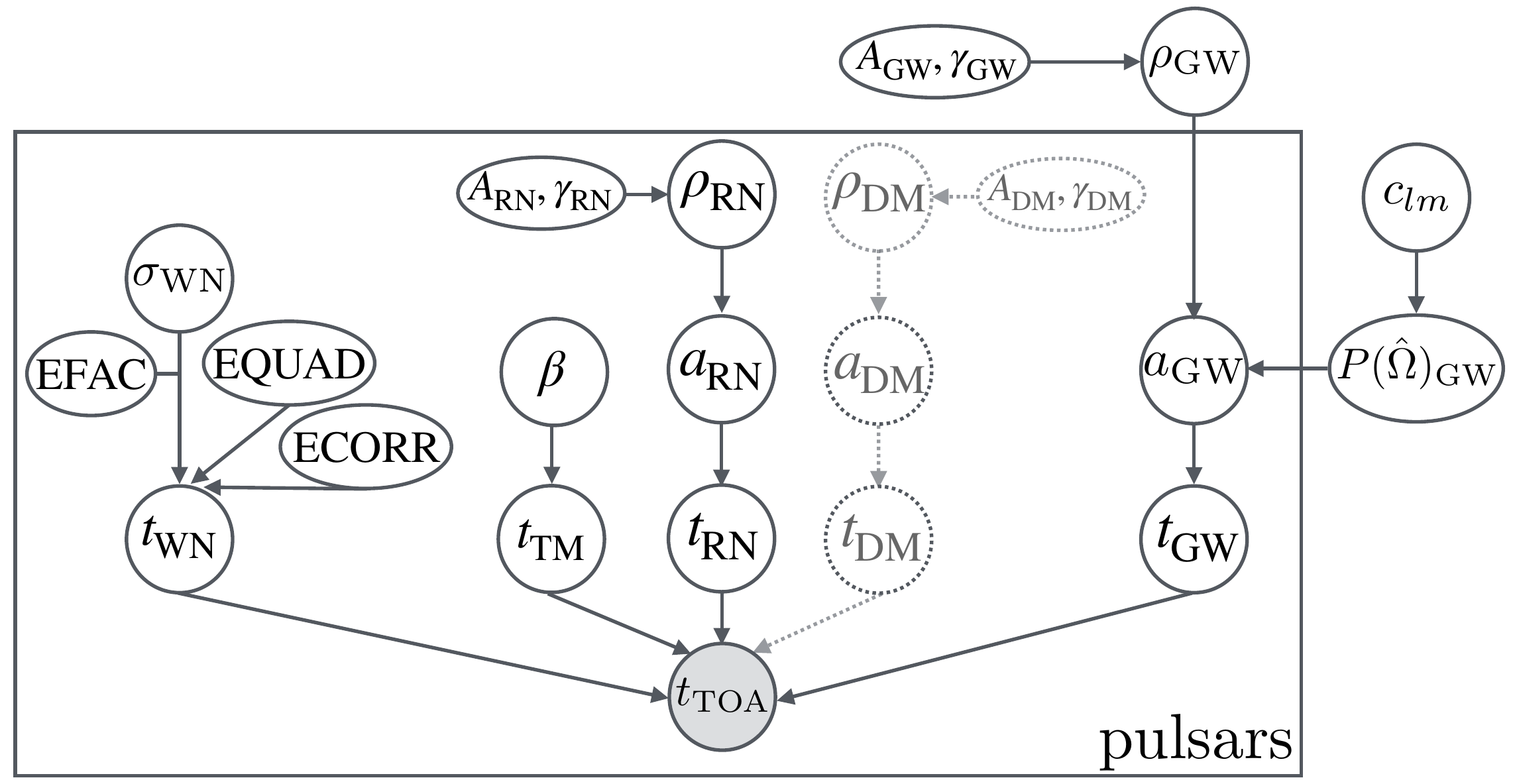}
    \caption{Probabilistic graphical model (or Bayesian network) for a PTA search for a SGWB. Arrows indicate the chain of conditional statistical dependencies. All processes inside the box are intrinsic to each pulsar, implying that the joint probability distribution of data is given by the product of probabilities over all pulsars. Arrows outside the box indicate priors on inter-pulsar correlated processes, which in this case is the frequency and angular spectral behaviour of the SGWB. Dashed arrows and circles modify this model to treat DM variations as a random Gaussian process, rather than part of the timing ephemeris. Additional deterministic per-pulsar (transient chromatic features) or common signals (Solar System ephemeris perturbations and individual GW signals) can be trivially added to this framework.}
    \label{fig:pta_pgm}
\end{figure}

A graphical representation of a PTA search for the SGWB is shown in Fig.~\ref{fig:pta_pgm} as a Bayesian network. This network illustrates the chain of conditional statistical dependencies of all processes that constitute our model of the data.

\subsection{Modeling deterministic signals}

Thus far, we have talked about modeling stochastic processes, either signal or noise. However, we also want to be able to model signals for which we have a deterministic description of the time-domain behavior, e.g., some of the sources mentioned in Chapter~\ref{chap:sources}--- individually resolvable SMBBH signals; GW bursts; Solar System ephemeris perturbations; or even a full non-linear pulsar timing ephemeris analysis. Fortunately, it is incredibly easy to slot this kind of signal description into the PTA likelihood framework. We recall that the likelihood is meant to be a probabilistic description of the random Gaussian components of the residual time series. Way back in Eq.~\ref{eq:whitened_res}, we created a time-series of our best estimate of the Gaussian white noise in our data. To model additional deterministic processes, we simply replace $\vec{r}$ in Eq.~\ref{eq:whitened_res} with $\vec{r}\rightarrow\vec{r} - \vec{s}(\vec{\theta})$, where $\vec{s}(\theta)$ is a deterministic signal function that depends on parameters $\vec{\theta}$. For the marginalized likelihood of Eq.~\ref{eq:marg_like}, this just means we make the following replacement
\begin{equation}
    \vec{\delta t}\rightarrow\vec{\delta t} - \vec{s}(\vec{\theta}).
\end{equation}

\section{Likelihood-based Statistics} \label{sec:likestats}

The full form of the PTA likelihood is a sufficient statistic for all kinds of deterministic signal and stochastic process searches. However there are many statistics that can be derived from this PTA likelihood for the purpose of more limited searches or frequentist constraints. The following are some of the most used.

\subsection{GWB statistics}

\subsubsection{Optimal statistic}

The optimal statistic (OS) is a frequentist estimator of the amplitude of an isotropic stochastic GWB \citep{2009PhRvD..79h4030A,2013ApJ...762...94D,2015PhRvD..91d4048C}. It can be derived by maximizing the PTA likelihood under a first-order expansion around the Hellings \& Downs inter-pulsar correlation coefficients \citep{2013ApJ...769...63E}. Given that $\bm{C}=\langle\vec{\delta t}\vec{\delta t}^\mathrm{T}\rangle$ is the time-domain covariance of residuals between arbitrary pairs of pulsars, let us introduce the following labels for the following specific combinations:
\begin{align}
    \bm{P}_a &= \bm{C}_{aa}, \nonumber\\
    \bm{S}_{ab} &= A_\mathrm{GWB}^2\bm{\tilde{S}}_{ab} = \bm{C}_{ab}.
\end{align}
The first-order--expanded and marginalized PTA likelihood function can be written as (see Ref.~\citep{2013ApJ...769...63E} for full details)
\begin{equation}
    \ln p(\{\vec{\delta t}\}|\vec{\eta}) \approx -\frac{1}{2}\left[\sum_{a=1}^{N_p} (\mathrm{Tr}\ln\bm{P}_a + \vec{\delta t}_a^\mathrm{T}\bm{P}_a^{-1}\vec{\delta t}_a) - \sum_{a=1}^{N_p}\sum_{b>a}^{N_p}\vec{\delta t}_a^\mathrm{T}\bm{P}_a^{-1}\bm{S}_{ab}\bm{P}_b^{-1}\vec{\delta t}_b \right].
\end{equation}
If we have performed a program of noise estimation for all pulsars, then the autocovariance matrices can all be assumed known, allowing us to form the log-likelihood ratio: $\ln\Lambda = \ln p(\{\vec{\delta t}\}|\vec{\eta},\mathcal{H}_\mathrm{gw}) - \ln p(\{\vec{\delta t}\}|\vec{\eta},\mathcal{H}_\mathrm{noise})$.\footnote{The models being compared here are a GWB versus a spatially-uncorrelated common process, giving a statistic that acts as a frequentist proxy for the Bayesian odds of GWB-induced cross-correlations.} With all pulsar noise parameters known, we further assume the GWB characteristic strain spectrum takes its default power-law form with the exponent for a SMBBH population, $\alpha=-2/3$. This leaves only one parameter to be estimated: the amplitude of the characteristic strain spectrum at a frequency of $1/\mathrm{yr}$, $A_\mathrm{GWB}$. Evaluating this log-likelihood ratio gives
\begin{equation}
    \ln\Lambda = \frac{A_\mathrm{GWB}^2}{2} \sum_{a=1}^{N_p}\sum_{b>a}^{N_p} \vec{\delta t}_a^\mathrm{T}\bm{P}_a^{-1}\bm{\tilde{S}}_{ab}\bm{P}_b^{-1}\vec{\delta t}_b.
\end{equation}
It can also be shown that
\begin{equation}
    \left\langle\sum_{a=1}^{N_p}\sum_{b>a}^{N_p} \vec{\delta t}_a^\mathrm{T}\bm{P}_a^{-1}\bm{\tilde{S}}_{ab}\bm{P}_b^{-1}\vec{\delta t}_b\right\rangle = A_\mathrm{GWB}^2 \sum_{a=1}^{N_p}\sum_{b>a}^{N_p} \mathrm{Tr}\left[ \bm{P}_a^{-1}\bm{\tilde{S}}_{ab}\bm{P}_b^{-1}\bm{\tilde{S}}_{ba} \right].
\end{equation}
Therefore we can write the optimal estimator of the GWB amplitude as
\begin{equation}
    \hat{A}^2 = \frac{\sum_{a=1}^{N_p}\sum_{b>a}^{N_p} \vec{\delta t}_a^\mathrm{T}\bm{P}_a^{-1}\bm{\tilde{S}}_{ab}\bm{P}_b^{-1}\vec{\delta t}_b}{\sum_{a=1}^{N_p}\sum_{b>a}^{N_p} \mathrm{Tr}\left[ \bm{P}_a^{-1}\bm{\tilde{S}}_{ab}\bm{P}_b^{-1}\bm{\tilde{S}}_{ba} \right]},
\end{equation}
thereby ensuring $\langle\hat{A}^2\rangle = A_\mathrm{GWB}^2$. In the weak-signal or noise-only regime, $\langle\hat{A}^2\rangle=0$ and the standard deviation of this estimator is
\begin{equation}
    \sigma_0 = \left(\sum_{a=1}^{N_p}\sum_{b>a}^{N_p} \mathrm{Tr}\left[ \bm{P}_a^{-1}\bm{\tilde{S}}_{ab}\bm{P}_b^{-1}\bm{\tilde{S}}_{ba} \right]\right)^{-1/2}.
\end{equation}
Using this, we can define a signal-to-noise ratio (SNR), corresponding to the number of standard deviations away from zero that the measured statistic is found to be:
\begin{equation}
    \hat{\rho} = \frac{\hat{A}^2}{\sigma_0} = \frac{\sum_{a=1}^{N_p}\sum_{b>a}^{N_p} \vec{\delta t}_a^\mathrm{T}\bm{P}_a^{-1}\bm{\tilde{S}}_{ab}\bm{P}_b^{-1}\vec{\delta t}_b}{\left(\sum_{a=1}^{N_p}\sum_{b>a}^{N_p} \mathrm{Tr}\left[ \bm{P}_a^{-1}\bm{\tilde{S}}_{ab}\bm{P}_b^{-1}\bm{\tilde{S}}_{ba} \right]\right)^{1/2}},
\end{equation}
which has expectation value
\begin{equation}
    \langle\rho\rangle = A_\mathrm{GWB}^2\left(\sum_{a=1}^{N_p}\sum_{b>a}^{N_p} \mathrm{Tr}\left[ \bm{P}_a^{-1}\bm{\tilde{S}}_{ab}\bm{P}_b^{-1}\bm{\tilde{S}}_{ba} \right]\right)^{1/2}.
\end{equation}
Hence
\begin{equation} \label{eq:lambda_to_rho}
    \langle\ln\Lambda\rangle = \langle\rho\rangle^2/2.
\end{equation}
The various procedures under which the OS can be used to define false alarm probabilities and detection probabilities can be found in detail in Refs.~\citep{2009PhRvD..79h4030A,2013ApJ...762...94D,2015PhRvD..91d4048C}. Note that the OS estimator was initially developed for the weak signal regime, and can be biased for the intermediate signal regime and beyond. However, its utility can be extended by hybridizing it with Bayesian principles-- the noise-marginalized OS \citep{2018PhRvD..98d4003V} averages the OS over noise parameters drawn from an MCMC chain that has been run on a fixed-$\alpha$ PTA GWB search. The OS can be used to explore how the GWB SNR scales with various PTA configuration variables, like timing precision, number of pulsars, GWB amplitude, etc. \citep{2013CQGra..30v4015S,2016PhRvD..94l3003V}.

It is also possible to take a broader view of the OS as a maximum likelihood estimator, which uses pairwise correlation measurements as the data with heterogeneous uncertainties. In this view, we are free to fit models of inter-pulsar correlation signatures to the measured pairwise correlations as we choose. The pairwise correlations and their uncertainties are \citep{2015PhRvD..91d4048C}
\begin{align}
    \rho_{ab} &= \frac{\vec{\delta t}_a^\mathrm{T}\bm{P}_a^{-1}\bm{\hat{S}}_{ab}\bm{P}_b^{-1}\vec{\delta t}_b}{\mathrm{Tr}\left[ \bm{P}_a^{-1}\bm{\hat{S}}_{ab}\bm{P}_b^{-1}\bm{\hat{S}}_{ba} \right]}, \nonumber\\
    \sigma_{ab} &= \left(\mathrm{Tr}\left[ \bm{P}_a^{-1}\bm{\hat{S}}_{ab}\bm{P}_b^{-1}\bm{\hat{S}}_{ba} \right]\right)^{-1/2},
\end{align}
where now $\bm{S}_{ab} = A_\mathrm{GWB}^2\Gamma_{ab}\bm{\hat{S}}_{ab}$, and $\Gamma_{ab}$ is the ORF or inter-pulsar correlation signature. We can often describe $\Gamma_{ab}$ as a \textit{linear model}, e.g., in terms of pixel or multipole power for anisotropic modeling \citep{2009PhRvD..80l2002T,2013PhRvD..88h4001T,2013PhRvD..88f2005M,2020PhRvD.102h4039T}, in terms of a Legendre or Chebyshev series for an agnostic description \citep{2014PhRvD..90h2001G,2015MNRAS.453.2576L,2019ApJ...876...55R}, or in terms of multiple processes that may involve a SGWB and systematics \citep{2016MNRAS.455.4339T}. This simply means that we can write $\Gamma_{ab} = \sum_k c_k X_{ab,k}$, or alternatively in vector notation $\vec\Gamma = \bm{X}\vec{c}$, such that $\bm{X}$ is a $(N_\mathrm{pairs}\times N_\mathrm{features})$ design matrix of the basis functions evaluated for the different pulsar pairs. Treating the $\{\rho_{ab}, \sigma_{ab}\}$ in vector form, and assuming $\sigma_{ab}$ are Gaussian uncertainties, we can use linear regression to deduce maximum-likelihood model estimators with an associated uncertainty covariance matrix:
\begin{equation}
    \vec{c}_\mathrm{ML} = \left(\bm{X}^T \bm{C}^{-1} \bm{X} \right)^{-1} \bm{X}^T \bm{C}^{-1}\vec\rho, \quad \Sigma_c = \left(\bm{X}^T \bm{C}^{-1} \bm{X} \right)^{-1},
\end{equation}
where $C$ is a diagonal matrix of squared $\sigma_{ab}$ values. The diagonal elements of $\Sigma_c$ give the variance of the $\vec{c}_\mathrm{ML}$ estimates, while the off-diagonal elements describe the parameter covariances. 

\subsubsection{Bridging the Bayesian odds ratio and the frequentist optimal statistic}

Under certain circumstances it is possible to relate Bayesian model selection to frequentist hypothesis testing \citep{2017LRR....20....2R}. When data is informative such that the likelihood is strongly peaked, the Bayesian evidence can be computed under the \textit{Laplace approximation}, such that
\begin{equation}
    Z_\mathcal{H}\equiv\int\,d\theta\, p(d|\theta,\mathcal{H})p(\theta|\mathcal{H})\approx p(\theta_\mathrm{ML}|\mathcal{H})\Delta V_\mathcal{H}/V_\mathcal{H},
\end{equation} 
where $p(d|\theta_\mathrm{ML},\mathcal{H})$ maximizes the likelihood with parameters $\theta$ given data $d$ under model $\mathcal{H}$. $\Delta V_\mathcal{H}/V_\mathcal{H}$ measures the compactness of the parameter space volume occupied by the likelihood with respect to the total prior volume, incorporating the Bayesian notion of model parsimony. Taking the ratio of Bayesian evidences between two models, labeled 1 and 2, and assuming equal prior odds, allows the Bayesian odds ratio, $\mathcal{O}_{12}$, to be written as
\begin{equation}
    \ln\mathcal{O}_{12}\approx\ln\Lambda_\mathrm{ML}(d) + \ln\left[ (\Delta V_1/V_1)/(\Delta V_2/V_2) \right],
\end{equation}
where $\Lambda_\mathrm{ML}(d)$ is the maximum likelihood ratio. The relevant maximum likelihood statistic for PTA GWB detection is the \textit{optimal statistic}, which as we have seen is a noise-weighted two-point correlation statistic between all unique pulsar pairs, comparing models with and without spatial correlations between pulsars \cite{2009PhRvD..79h4030A,2013ApJ...762...94D,2015PhRvD..91d4048C}. From Eq.~\ref{eq:lambda_to_rho}, the SNR of such GWB-induced correlations can be written as $\langle\ln\Lambda_\mathrm{ML}\rangle=\langle\rho\rangle^2/2$. While the likelihood may be marginally more compact under the model with Hellings \& Downs correlations, there is no difference in parameter dimensionality; hence we ignore the likelihood compactness terms. The key relationship between the Bayesian odds ratio in favor of Hellings \& Downs correlations and the frequentist SNR of such correlations can then be written as
\begin{equation} \label{eq:laplace}
    \ln\mathcal{O}_\mathrm{HD} \approx \rho^2 / 2.
\end{equation}

\subsection{Individual binary statistics}

\subsubsection{$\mathcal{F}_e$ statistic}

The $\mathcal{F}_e$ statistic is a maximum likelihood estimator of the sky-location and orbital frequency of an individually-resolvable monochromatic (i.e. circular) SMBBH signal in PTA data. The assumed signal model includes only the \textit{Earth term} of timing delays. This concept was originally developed in the context of detecting continuous GWs from neutron stars \citep{1998PhRvD..58f3001J,2005PhRvD..72f3006C}, then later adapted for PTA continuous GW searches \citep{2012PhRvD..85d4034B,2012ApJ...756..175E}. Developing this statistic requires us to re-arrange the signal model of induced timing delays into a format where we can maximize over the coefficients of a set of time-dependent basis functions. We use a parenthetical notation to denote the noise-weighted inner product of two vectors, such that $(\vec{x}|\vec{y}) = \vec{x}^\mathrm{T} C^{-1} \vec{y}$, where $C$ is a covariance matrix of noise in the observed vectors $\{\vec{x},\vec{y}\}$. The log-likelihood ratio between the signal$+$noise model and the noise-only model can be written as
\begin{align}
    \ln\Lambda &= \ln\left( \frac{p(\vec{\delta t}|\vec{s})}{p(\vec{\delta t}|\vec{0})} \right) \nonumber\\
    &= -\frac{1}{2}(\vec{\delta t}-\vec{s}|\vec{\delta t}-\vec{s}) + \frac{1}{2}(\vec{\delta t}|\vec{\delta t}) \nonumber\\
    &= (\vec{\delta t}|\vec{s}) - \frac{1}{2}(\vec{s}|\vec{s}).
\end{align}
Using our earlier introduced waveform model for an individually resolvable SMBBH (see Chap.~\ref{chap:sources}), we re-write the Earth-term signal model in terms of coefficients of \textit{extrinsic} variables ($\zeta,\iota,\Phi_0,\psi$) and basis functions of \textit{intrinsic} variables $(\theta,\phi,\omega_0)$ \citep{1998PhRvD..58f3001J}
\begin{equation}
    \vec{s}(t,\hat\Omega) = \sum_{k=1}^4 a_k(\zeta,\iota,\Phi_0,\psi)\vec{A}^k(t,\theta,\phi,\omega_0), 
\end{equation}
where $\vec{s}$ and $\vec{A}^k$ are concatenated vectors of time-series for all pulsars, such that
\begin{equation}
    \vec{s} = \left[ \begin{matrix} s_1 \\ s_2 \\ \vdots \\ s_{N_p} \end{matrix} \right], \qquad\vec{A}^k = \left[ \begin{matrix} A^k_1 \\ A^k_2 \\ \vdots \\ A^k_{N_p} \end{matrix} \right],
\end{equation}
\begin{align}
    A^1_a &= F^+_a(\hat\Omega)\omega(t)^{-1/2}\sin(2\Phi(t)), \nonumber\\
    A^2_a &= F^+_a(\hat\Omega)\omega(t)^{-1/2}\cos(2\Phi(\vec{t})), \nonumber\\
    A^3_a &= F^\times_a(\hat\Omega)\omega(t)^{-1/2}\sin(2\Phi(t)), \nonumber\\
    A^4_a &= F^\times_a(\hat\Omega)\omega(t)^{-1/2}\cos(2\Phi(t)),
\end{align}
\begin{align} \label{eq:fe_coeffs}
    a_1 &= \zeta[(1+\cos^2\iota)\cos2\Phi_0\cos2\psi +2\cos\iota\sin2\Phi_0\sin2\psi], \nonumber\\
    a_2 &= -\zeta[(1+\cos^2\iota)\sin2\Phi_0\cos2\psi -2\cos\iota\cos2\Phi_0\sin2\psi], \nonumber\\
    a_3 &= \zeta[(1+\cos^2\iota)\cos2\Phi_0\sin2\psi -2\cos\iota\sin2\Phi_0\cos2\psi], \nonumber\\
    a_4 &= -\zeta[(1+\cos^2\iota)\sin2\Phi_0\sin2\psi +2\cos\iota\cos2\Phi_0\cos2\psi].
\end{align}
The binary is assumed to be slowly evolving over the course of the pulsar observation baseline, such that $\omega(t)\approx\omega_0$ and $\Phi(t)\approx\omega_0t$.
The log-likelihood can then be written as
\begin{equation}
    \ln\Lambda = a_k\bm{N}^k - \frac{1}{2}\bm{M}^{kl}a_ka_l,
\end{equation}
where $\bm{N}^k = (\vec{\delta t}|\vec{A}^k)$ and $\bm{M}^{kl} = (\vec{A}^k|\vec{A}^l)$. Summation convention is used to evaluate terms with paired superscript and subscript indices. Maximizing this log-likelihood over the four coefficients, $a_k$, yields their maximum likelihood estimates
\begin{equation}
    a^\mathrm{ML}_k = \bm{M}_{kl}\bm{N}^l,
\end{equation}
where $\bm{M}_{kl} = (\bm{M}^{kl})^{-1}$. These coefficients can be plugged back into the log-likelihood to yield the $\mathcal{F}_e$ statistic
\begin{equation}
    2\mathcal{F}_e = \bm{N}^k\bm{M}_{kl}\bm{N}^l.
\end{equation}
This statistic can then be mapped over different orbital frequencies and sky locations, or globally maximized to find the best-fit values of these parameters. The form of this statistic's false alarm probability, detection probability, and procedures under which it is used are detailed in Ref.~\citep{2012PhRvD..85d4034B,2012ApJ...756..175E}. While only the sky location and orbital frequency are explicitly searched over, it is possible to estimate the extrinsic parameters from the maximum likelihood amplitude coefficeints \citep{2007CQGra..24.5729C,2012ApJ...756..175E}. Note also that a generalized version of the $\mathcal{F}_e$-statistic applicable to arbitrary binary eccentricities has also been developed \citep{2016ApJ...817...70T}, where the space of intrinsic parameters additionally includes binary eccentricity, $e$, and the initial mean anomaly, $l_0=2\pi t_0/P$, where $P$ is the orbital period.

\subsubsection{$\mathcal{F}_p$ statistic}

As a maximum-likelihood estimator for individually-resolvable monochromatic SMBBH signals in PTA data, the $\mathcal{F}_p$ statistic is similar to the $\mathcal{F}_e$ statistic, except that it also accounts for the \textit{pulsar term} in the GW-induced timing delays. This is sometimes also known as the \textit{incoherent} $\mathcal{F}$ statistic because it involves summing over squares of data quantities, rather than in the $\mathcal{F}_e$ statistic where one squares the sum of data quantities. A central assumption used here when introducing the pulsar term is that the source evolution remains slow enough that the Earth-term orbital frequency, $\omega_0$, and pulsar-term orbital frequency, $\omega_p$, are identical (or at least indistinguishable within the PTA resolution). We can perform a Taylor expansion of the pulsar-term frequency such that
\begin{align}
    \omega(t_p) &= \omega_0\left[1 - \frac{256}{5}\left(\frac{G\mathcal{M}}{c^3}\right)^{5/3}\omega_0^{8/3}t_p\right]^{-3/8}, \nonumber\\
    &\approx \omega_0\left[1 + \frac{96}{5}\left(\frac{G\mathcal{M}}{c^3}\right)^{5/3}\omega_0^{8/3}(t_e - L_p(1+\hat\Omega\cdot\hat{p})\right],
\end{align}
where $\mathcal{M}$ is the binary chirp mass, $L_p$ is the pulsar distance, $\hat\Omega\equiv -(\sin\theta\cos\phi,\sin\theta\sin\phi,\cos\theta)$ is the GW propagation direction for GWs originating from sky-location $(\theta,\phi)$, and $\hat{p}$ is the pulsar unit-vector direction on the sky. The condition that $\omega_p = \omega(t_p)\approx\omega_0$ requires
\begin{equation}
    \omega_0 \ll \left[\frac{5}{96}\left(\frac{c^3}{G\mathcal{M}}\right)^{5/3}\left|T-L(1+\hat\Omega\cdot\hat{p})\right|\right]^{3/8},
\end{equation}
where $T$ is the total pulsar observation time. With both Earth and pulsar terms sharing a common signal frequency, we can now write the GW signal in pulsar $a$ as
\begin{equation}
    s_a(t,\hat\Omega) = \sum_{k=1}^2 b_{k,a}(\zeta,\iota,\Phi_0,\psi,\tilde\Phi_a,\theta,\phi)B^k_a(t,\omega_0),
\end{equation}
where
\begin{equation}
    \tilde\Phi_a = \omega_0L_a(1+\hat\Omega\cdot\hat{p}_a)/c + \Phi_0,
\end{equation}
is the pulsar-term orbital phase. The amplitude and basis functions are defined as
\begin{align}
    b_{1,a} &= \zeta\left[ (1+\cos^2\iota)(F^+_a\cos2\psi + F^\times_a\sin2\psi)(\cos2\Phi_0 - \cos2\tilde\Phi_a) \right. \nonumber\\
    &\left. + 2\cos\iota(F^+_a\sin2\psi - F^\times_a\cos2\psi)(\sin2\Phi_0 - \sin2\tilde\Phi_a) \right], \nonumber\\
    b_{2,a} &= -\zeta\left[ (1+\cos^2\iota)(F^+_a\cos2\psi + F^\times_a\sin2\psi)(\sin2\Phi_0 - \sin2\tilde\Phi_a) \right. \nonumber\\
    &\left. - 2\cos\iota(F^+_a\sin2\psi - F^\times_a\cos2\psi)(\cos2\Phi_0 - \cos2\tilde\Phi_a) \right], 
\end{align}
and
\begin{equation}
    B^1_a = \frac{1}{\omega_0^{1/3}}\sin(2\omega_0 t),\quad B^2_a = \frac{1}{\omega_0^{1/3}}\cos(2\omega_0 t).
\end{equation}
The log-likelihood ratio can then be written as
\begin{equation}
    \ln\Lambda = \sum_{a=1}^{N_p} \left[b_{k,a}P^k_a - \frac{1}{2}Q^{kl}b_{k,a}b_{l,a} \right],
\end{equation}
where $P^k_a = (\delta t_a|B^k_a)$ and $Q^{kl} = (B^k_a|B^l_a)$. Maximizing $\ln\Lambda$ over the $2N_p$ coefficients yields
\begin{equation}
    b^\mathrm{ML}_{k,a} = (Q^{kl}_a)^{-1}P^k_a,
\end{equation}
which, when inserted back into the log-likelihood ratio, give
\begin{equation}
    2\mathcal{F}_p = \sum_{a=1}^{N_p} P^k_a (Q^{kl}_a)^{-1} P^l_a. 
\end{equation}
This statistic is more often used than the $\mathcal{F}_e$ statistic because of the realism of the model that includes the pulsar term, and all factors involving the (uncertain) pulsar distances have been absorbed into amplitude coefficients that are maximized over. It can be used to examine pulsar-timing data to compute a value of the $\mathcal{F}_p$ statistic as a function of frequency, and in so doing produce sensitivity curves for indivudally-resolvable binary GW sources. The form of this statistic's false alarm probability, detection probability, and procedures under which it is used are detailed in Ref.~\citep{2012ApJ...756..175E}. Note also that an alternative to the $\mathcal{F}_p$ statistic's maximization over amplitude coefficients is to marginalize over these, or to maximize over the physical pulsar-term phase parameters, $\tilde\Phi_a$. Both schemes were explored in Ref.~\citep{2014PhRvD..90j4028T} with the former marginalization technique referred to as the $\mathcal{B}_p$ statistic. 

\bibliographystyle{unsrt_new}
\bibliography{refs}

%% file: 08.tex
\chapter{The Past, Present, \& Future Of PTAs}
\epigraph{\textit{``Fix your little problem and light this candle.''}}{Alan B.~Shepard Jr., May 5, 1961}

By the very nature of the signals they hunt, PTAs are long timescale experiments. The field has taken some time to gain traction, with the slow trickle in accumulated information being mirrored by a steady growth in interest and activity of the regional collaborations. There are three major regional PTA collaborations that each have greater than a decade of precision timing observations, all inaugurated at roughly the same time in the early $2000$'s. Ironically, at the same time as the regional PTAs formed, the most constraining limit on the SGWB came from the combination of international datasets recorded in Australia (Parkes Radio Telescope) and Puerto Rico (Arecibo Radio Telescope) \citep{2006ApJ...653.1571J}, corresponding to $1.1\times 10^{-14}$ at $f=1/\mathrm{yr}^{-1}$ with $95\%$ confidence for a SGWB with $\alpha=-2/3$. While earlier campaigns had produced limits before this \citep{1983ApJ...265L..39H,1990PhRvL..65..285S,1994ApJ...428..713K,1996PhRvD..54.5993M,2002nsps.conf..114L},\footnote{In fact, Ref.~\citep{1996PhRvD..54.5993M} from 1996 was the first to suggest a Bayesian approach.} Ref.~\citep{2006ApJ...653.1571J} was the first to be within an order of magnitude of current results. The limits quoted below will all correspond to the characteristic strain at $f=1/\mathrm{yr}^{-1}$ for a power-law with index $-2/3$, either with $95\%$ confidence for frequentist studies or $95\%$ credibility for Bayesian studies. 

The Parkes Pulsar Timing Array (PPTA) uses the $64$~m  Parkes Radio Telescope located in Parkes, New South Wales, Australia to time $26$ millisecond pulsars over a current baseline of $\sim 17$~years (since $2004$) \citep{2020PASA...37...20K}. The PPTA is distinguished amongst other PTA collaborations through its access to the southern hemisphere, allowing it to observe one of the highest quality pulsars, PSR J$0437$$-$$4715$ (although see Ref.~\citep{2021ApJ...911..137L}). Throughout its operations, the PPTA has boasted some of the tightest constraints on the amplitude of the SGWB \citep[e.g.,][]{2013Sci...342..334S}; its most recent published $95\%$ upper limit was $10^{-15}$ at $f=1/\mathrm{yr}^{-1}$ \citep{2015Sci...349.1522S}. However, this limit used only four pulsars and an outdated Solar-system ephemeris model (DE421, released in 2008 \citep{folkner2009planetary}). Several analyses have shown that relying on a small number of pulsars \citep{2016ApJ...819L...6T} and older ephemerides \citep{2018ApJ...859...47A,2020ApJ...893..112V} can lead to systematic biases in SGWB constraints. The PPTA is currently analyzing its second data release \citep{2020PASA...37...20K}. 

The European Pulsar Timing Array (EPTA) is a collaboration using five radio telescopes spread across several countries: the $94$~m-equivalent Westerbork Synthesis Radio Telescope (Westerbork, Netherlands), the $100$~m Effelsberg Radio Telescope (Bad M\"unstereifel, Germany), the $76$~m Lovell Telescope (Jodrell Bank, UK), the $94$~m Nan\c cay Radio Telescope (Sologne, France), and the $64$~m Sardinia Radio Telescope (San Basilio, Sardinia, Italy). Established in 2006, the EPTA was the first PTA collaboration to use Bayesian statistical inference to simultaneously model the intrinsic noise in each pulsar, the cross correlations between all pulsars, and the amplitude and spectral index of the SGWB's strain spectrum \citep{2011MNRAS.414.3117V}, finding a $95\%$ limit of $6\times 10^{-15}$ from five pulsars. It incorporates some of the longest observed pulsars in its datasets; in its most recent data release in 2016, it included 42 pulsars whose baselines ranged from $\sim7-18$~years until 2014 \citep{2016MNRAS.458.3341D}, of which six pulsars were used to place a limit of $3\times 10^{-15}$ \citep{2015MNRAS.453.2576L}. The collaboration is currently working on a new data release. 

The North American Nanohertz Observatory for Gravitational waves (NANOGrav), which was established in $2007$, uses the $100$~m Green Bank Telescope (Green Bank, West Virginia, USA) and the (former) $305$~m Arecibo Observatory (Arecibo, Puerto Rico, USA) to time $\sim 77$ pulsars over a baseline as long as $\sim 15$ years \citep{2019BAAS...51g.195R}. The collaboration has steadily added new pulsars to its array, where the number of analyzed pulsars in its most recent SGWB searches increased from $17$ in its $5$-year dataset (limit of $7\times 10^{-15}$) \citep{2013ApJ...762...94D}, $18$ in its $9$-year dataset (limit of $1.5\times 10^{-15}$) \citep{2016ApJ...821...13A}, $34$ in its $11$-year dataset (limit of $1.45\times 10^{-15}$ \citep{2018ApJ...859...47A}, but see also Ref.~\citep{2020ApJ...905L...6H}), and $45$ in its $12.5$~year dataset (detection of a common process with median amplitude $1.92\times 10^{-15}$) \citep{2020ApJ...905L..34A}. Notice the trend in the last few datasets: the limit appeared to decrease rapidly, saturate, then increase as a detection of a low-frequency common process was made. The saturation occurred because the NANOGrav array hit a systematic noise floor corresponding to Solar-system ephemeris precision, leading to the development of a Bayesian ephemeris model to mitigate this. Revision of the priors for intrinsic pulsar red noise in the $11$~year analysis has shown greater consistency with the $12.5$~year results \citep{2020ApJ...905L...6H}. With regards to the latter, NANOGrav has now entered a regime of overwhelming statistical evidence for a common-spectrum low-frequency process being positively supported by $10$ pulsars amongst the $45$ analyzed. This process has median amplitude $1.92\times 10^{-15}$ and $5\%-95\%$ quantiles of $1.37-2.67\times 10^{-15}$. The Bayes factor in favor of this common-spectrum process versus merely intrinsic per-pulsar noise is in excess of $10^5 : 1$. While this may potentially be the first signs of the SGWB emerging from noise \citep{2021PhRvD.103f3027R}, no evidence has yet been found for the distinctive Hellings \& Downs inter-pulsar correlation signature from an isotropic background of GWs. The collaboration is currently working on a $\sim 15$~year dataset with the number of pulsars exceeding $60$. 

What was once steady growth and advancement now feels like a torrent of progress. Emerging PTA collaborations such as the Indian PTA (InPTA) \citep{joshi2018precision}, the Chinese PTA (CPTA) \citep{lee2016prospects}, and more telescope-centered groups like CHIME \citep{2018IAUS..337..179N} and MeerTime \citep{2016mks..confE..11B}, are ramping up efforts to join the hunt. In fact, the InPTA has recently joined the International Pulsar Timing Array (IPTA), which was originally founded as a consortium of the three major PTA consortia (NANOGrav, EPTA, PPTA). The IPTA has produced two data releases so far, which usually lag several years behind when the corresponding components are released by regional PTAs. There is good reason for this; data combination is a delicate and challenging process, and where pulsars are observed in common, this requires timing solutions and noise models to be harmonized across different telescopes and processing pipelines. The first IPTA data release \citep{2016MNRAS.458.1267V} consisted of $49$ pulsars, and was built from NANOGrav's $5$~year dataset \citep{2013ApJ...762...94D}, the extended first PPTA data release \citep{2013PASA...30...17M,2016MNRAS.455.1751R}, the first EPTA data release \citep{2016MNRAS.458.3341D}, and publicly available data from Kaspi \textit{et al.} (1994) \citep{1994ApJ...428..713K} on PSRs J$1857$$+$$0943$ and J$1939$$+$$2134$ and from Zhu \textit{et al.} 2015 \citep{2015ApJ...809...41Z} on PSR J$1713$$+$$0747$. The most sensitive four pulsars were used to derive an upper limit of $1.7\times 10^{-15}$. The second IPTA data release \citep{2019MNRAS.490.4666P}, consisting of $65$ pulsars, adds to the first data release through expanding the NANOGrav component to the $9$~year data release \citep{2015ApJ...813...65N} and adding additional PPTA data that was included in Shannon \textit{et al.} \citep{2015Sci...349.1522S}. Work is ongoing to perform a full-array SGWB search within this dataset.  

So what's next? At least one PTA collaboration has detected a common-spectrum process that could be the first trumpet announcing a cresting SGWB detection. This must be confirmed (or refuted) by other regional PTAs and the IPTA. As of writing, these are all in the works. If this is an early announcement of the SGWB, then the amplitude is a little larger than recent models have suggested, although well within the prediction spread \citep{2021MNRAS.502L..99M}. Detection of the Hellings \& Downs inter-pulsar correlations should follow with only a few more years of data beyond existing baselines \citep{2013CQGra..30v4015S,2015MNRAS.451.2417R,2016ApJ...819L...6T,k+17,2021ApJ...911L..34P}, and will come packaged with $\sim 40\%$ constraints on the amplitude and spectral index of the characteristic strain spectrum \citep{2021ApJ...911L..34P}. This precision should be sufficient to test some astrophysical models of the SGWB and discriminate the origin under simple assumptions of the spectral index. After that, attention will shift to inference of spectral features such as turnovers that can indicate continued environmental coupling of the SMBHB population into the nanohertz band \citep{Sampson2015,tss17,2019MNRAS.488..401C}. 

Beyond spectral characterization, the inter-pulsar correlations encode a map of the GW sky, allowing PTAs to probe the angular structure of the SGWB signal and potentially unveil the first departures from statistical isotropy \citep{2013PhRvD..88f2005M,2013PhRvD..88h4001T,2015PhRvL.115d1101T,2020PhRvD.102h4039T}. Bright pixels or extended regions of angular power could be the first indicators of nearby or massive single sources to which the PTA is slowly becoming sensitive enough to individually resolve \citep{2020PhRvD.102h4039T}. This brings us to the next big milestone for PTAs, and a goal that is being actively searched for with only slightly less fervor than the SGWB-- individual SMBHB signals. While we think the SGWB will be detected first \citep{2015MNRAS.451.2417R}, resolution and characterization of individual SMBHB systems will confirm once and for all that they are the dominant source class for PTAs, and dangle the tantalizing opportunity for multi-messenger detections of massive black-hole binaries years before LISA will fly \citep[e.g.,][]{2019MNRAS.485.1579K}. While much of this chapter has been focused on ever-improving constraints on the SGWB, the strides in single source constraints have been huge; NANOGrav has placed constraints on the strain amplitude corresponding to $h\leq 7.3\times 10^{-15}$ at $f=8$~nHz with its $11$~year dataset, and in its most sensitive sky location has ruled out $\mathcal{M}=10^9 (10^{10}) M_\odot$ binaries closer than $120$~Mpc ($5.5$~Gpc) with $95\%$ credibility \citep{2019ApJ...880..116A}. Recent constraints from the EPTA \citep{2016MNRAS.455.1665B} and PPTA \citep{2014MNRAS.444.3709Z} are broadly similar. NANOGrav has also used its $11$~year dataset to improve constraints on a prominent binary candidate, 3C66B, finding that guiding constraints on the period of the fiducial binary that are less than an order of magnitude can in turn improve the chirp mass constraints by an order of magnitude \citep{2020ApJ...900..102A}. Furthermore, NANOGrav recently compiled a target list of $216$ massive galaxies within its sensitivity volume, placing multimessenger constraints on the secondary black-hole mass of a fiducial binary, and finding that $19$ systems could be constrained to within the same precision as SMBHB constraints in our own Milky Way \citep{2021arXiv210102716A}. These results demonstrate the power of the multimessenger goal, and the excitement for the future. PTAs may be able to detect several individual binaries by the end of the $2020$'s \citep{2015MNRAS.451.2417R,2017NatAs...1..886M,2018MNRAS.477..964K}, and if any of these have detectable lightcurve variability from accretion \citep{} or Doppler-boosting variability \citep{}, then the Rubin Observatory's Legacy Survey of Space and Time could snag the electromagnetic counterpart \citep{2019MNRAS.485.1579K}. 

Speculation further into the future runs into many unknowns, but the goals are clear. We aim to perform precision constraints on modified gravity theories through the presence of alternative GW polarization modes that may be sub-luminal; these modes will have their own distinct overlap reduction functions and lead to departures from the Hellings \& Downs curve. Beyond SMBHBs (and associated background, continuous, and burst searches), significant effort must be devoted to spectral separation of the binary signal from cosmological signals lurking beneath. These could include a primordial GW background, cosmic string signatures, imprints of a first-order cosmological phase transition, or heralds of unconstrained fundamental physics, such as the nature of dark matter. Addressing all of these important questions will require sophisticated physical models, robust statistical techniques, and powerful radio facilities to carry us into the middle of this century and beyond. The recent loss of Arecibo is a major blow, but NANOGrav is planning to shift most observations over to the Green Bank Telescope, and legacy Arecibo observations will continue to have impactful weight on future GW searches. In the longer term, the US community requires a replacement world-class radio facility (such as the DSA-$2000$ concept \citep{Hallinan2021DSA} and/or the ngVLA \citep{murphy2018science}). The successive stages of the Square Kilometre Array \citep{dewdney2009square} in South Africa and Australia will be transformative for PTA science \citep{2015aska.confE..37J}, providing a major boost to the number of pulsars and TOA precision. In China, beyond pulsar surveys and timing with the Five-hundred-meter Aperture Spherical Telescope (FAST) \citep{nan2011five,2020Innov...100053Q}, construction is underway on the $110$~m Xingjiang QTT \citep{wang2014xinjiang} and the $120$~m Jingdong Radio Telescope.

Pulsar-timing campaigns that focus on searching for gravitational waves are a mature enterprise. Though with exciting possibilities on the horizon, it feels as if our work is just beginning. There is much to discover, resolve, and strategize in order to steer our sensitivity beyond current possibilities toward future opportunities \citep{2018ApJ...868...33L,2012MNRAS.423.2642L}. PTAs offer a radically different means to GW detection, wherein a collection of dense, compact astrophysical objects are themselves an integral component of our detector. Much of this book has focused on GWs to the neglect of the extraordinary science of pulsars and the ionized interstellar medium; justice is given to those topics in other volumes by better-qualified authors. For now, I am invigorated by the state of this field and what may soon be possible. I invite the casual reader to join us on this hunt, and hope the expert reader finds this has been a useful overview of the state of the art of nanohertz GW searches with PTAs.

\bibliographystyle{unsrt_new}
\bibliography{refs}